\documentclass[12pt,a4paper]{article}
\usepackage{geometry}
\usepackage[utf8]{inputenc}

\usepackage{multicol}
\usepackage{diagbox}
\usepackage{caption}
\usepackage{subcaption}
\usepackage{jcappub}
\usepackage{bm}
\usepackage{epsfig}
\usepackage{bbold}
\usepackage{graphicx}
\usepackage{amsmath}
\usepackage{amssymb}
\usepackage{amsbsy}
\usepackage{color}
\usepackage{slashed}
\usepackage{afterpage}
\usepackage{psfrag}
\usepackage[noabbrev]{cleveref}
\usepackage{dsfont}
\usepackage{enumerate}
\usepackage[shortlabels]{enumitem}
\usepackage[utf8]{inputenc}
\usepackage{amsmath}
\usepackage[dvipsnames]{xcolor}
\usepackage{graphicx}
\usepackage{braket}
\usepackage[most]{tcolorbox}
\usepackage{empheq}

\newcommand{\beq}{\begin{equation}}
\newcommand{\eeq}{\end{equation}}
\newcommand{\vp}{\varphi}
\newcommand{\nmax}{n_{max}}
\newcommand{\VV}{{\cal V}}

\title{\begin{center}
		Soft Oscillons
\end{center}}

\vfill
\author{Fabio van Dissel}
\author{and}
\author{Oriol Pujolàs}

\affiliation{Institut de Física d’Altes Energies (IFAE) and Barcelona Institute of Science and Technology (BIST), Campus UAB, 08193 Bellaterra, Barcelona, Spain}

\makeatletter
\gdef\@fpheader{}
\makeatother

\abstract{We present a novel type of soliton dubbed soft oscillons. In contrast with conventional oscillons the soft counterparts come in a continuum of unboundedly large sizes. They are peculiar also in that the oscillation frequency is set by their size and in their high harmonic content, that leads to a peculiar core structure and pulsating emission of radiation.
Soft oscillons appear in a broad class of scalar field theories with plateau potentials and owe their existence to a simple confinement mechanism: a large field perturbation automatically acts as a cavity that traps massless modes. 
We study the classical evaporation rate, lifetime and stability under anisotropic deformations. Soft oscillons are longer lived in gapless models, where lifetimes can  reach $100$ times the initial radius or more.
}

\begin{document}

\maketitle

\newpage

\section{Introduction}
\label{sec:intro}
Oscillons are non-topological defects that arise quite generically in many massive bosonic theories due to attractive self-interactions \cite{Bogolyubsky:1976nx,Bogolyubsky:1976sc,Gleiser:1993pt,Kolb:1993hw,Copeland:1995fq}. Indeed a wide variety of models exhibit the formation of long-lived, localised states oscillating at frequencies below the mass threshold in the semi-classical limit \cite{Kasuya:2002zs,Hertzberg:2010yz,Amin:2010jq,Sfakianakis:2012bq,Amin:2013ika,Kawasaki:2015vga,Ibe:2019vyo,Fodor:2019ftc,Olle:2020qqy,Zhang:2020bec,Cyncynates:2021rtf,Levkov:2022egq,vanDissel:2023zva,Dorey:2023sjh,Blaschke:2024dlt,Zhou:2024mea}. See also Refs.~\cite{Gleiser:1993pt,Kolb:1993hw,Copeland:1995fq,Amin:2010dc,Amin:2011hj, Fonseca:2019ypl,Gleiser:2009ys, Farhi:2005rz,Graham:2006vy,Zhou:2013tsa, Fodor:2008es, Hiramatsu:2020obh, VanDissel:2020umg, Zhang:2020ntm, Antusch:2019qrr, Antusch:2017flz,Zhou:2013tsa,Hiramatsu:2020obh,Amin:2018xfe, Antusch:2017vga,Lozanov:2022yoy,Levkov:2023ncb,Lozanov:2023knf,Lozanov:2023aez,Pirvu:2023plk} for applications to early universe cosmology, dark matter \cite{Cotner:2018vug,Olle:2019kbo,Kawasaki:2019czd,Arvanitaki:2019rax,Kitajima:2020rpm,Kitajima:2021inh,Dvali:2023xfz} 
and topological defects \cite{Kolb:1993hw,Copeland:1995fq,Hindmarsh:2007jb,Braden:2015vza,Bond:2015zfa,Vaquero:2018tib,Gorghetto:2020qws,Blanco-Pillado:2020smt,Kitajima:2022jzz}.\\

While the oscillon lifetime is substantially model dependent, their size is typically set by the boson Compton radius, 
\begin{equation}\label{size}
R_o\simeq {\rm (few)} \cdot\, m^{-1}~.
\end{equation}
This can be understood as resulting from the competition between gradient and potential terms that takes place in the oscillon core. Alternatively, from the QFT perspective, oscillons correspond to many particle bound states \cite{Dvali:2011aa,Dvali:2012en,Dvali:2012gb,Dvali:2017ruz,Dvali:2017eba}, see also \cite{Eby:2014fya,Guth:2014hsa,Olle:2020qqy,vanDissel:2023zva}. Since the theory is bosonic, particles can be in the same single-particle state. For  moderately small binding energies the typical spread of this state is not far from the Compton wavelength, which reproduces \eqref{size}. \\

In this work, we aim to challenge the `universal' relation for the oscillon size  \eqref{size}.
We will show that under some conditions, a new type of oscillon state emerges, with remarkably different properties. 
In particular, we will show that well defined oscillons with a continuum of sizes 
$$
R_o \gg m^{-1}
$$
actually exist in a rather broad class of models. This is a new oscillon branch, mainly characterized by its size  $R_o$ (and low oscillation frequency  $w\sim 1/R_o \ll m $), which is why we dub them {\em soft} oscillons. \\

\begin{figure}[t]
\begin{center}
\includegraphics[width=\textwidth]{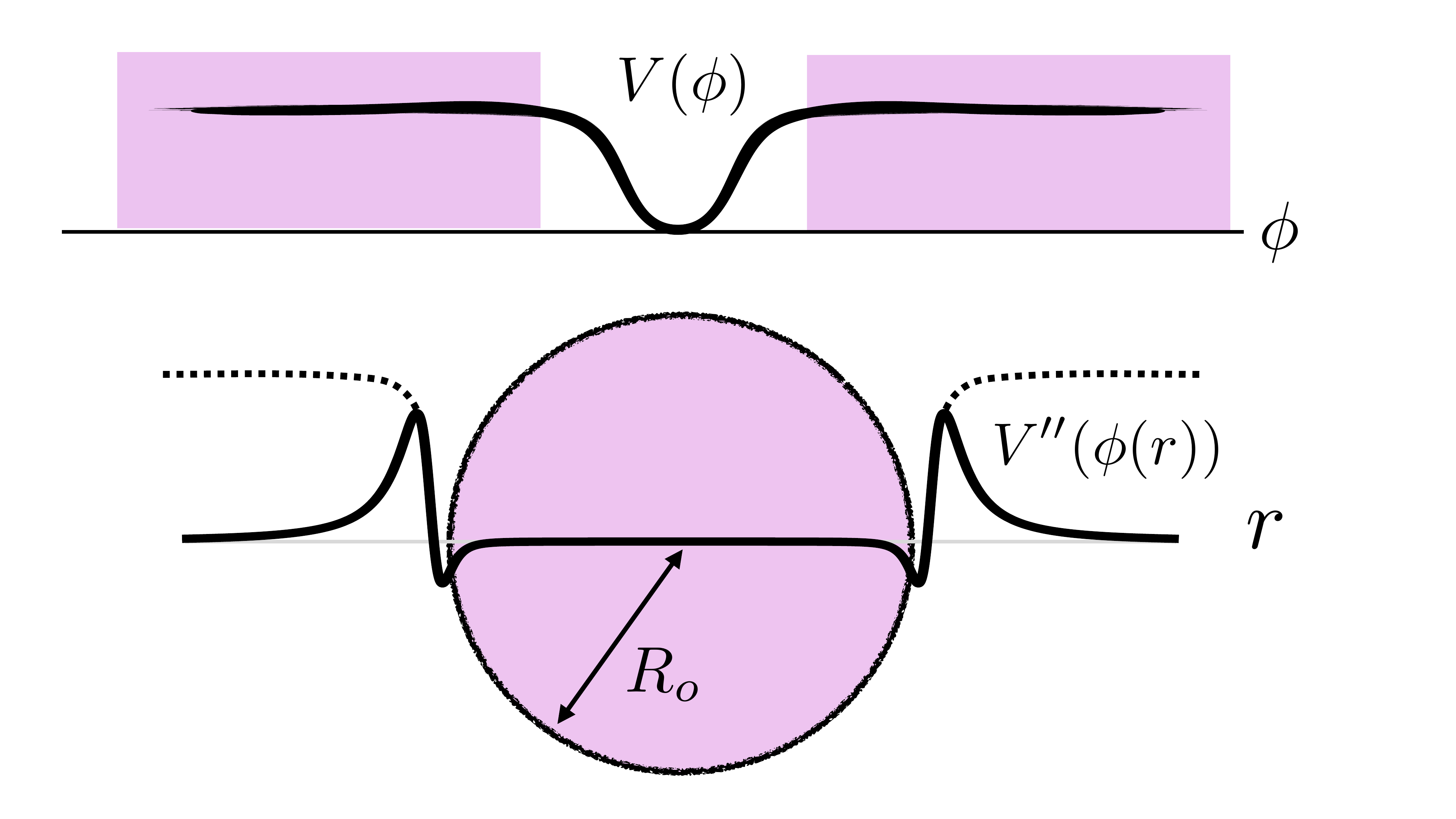}
\caption{Illustration of the {\em cavity effect} arising in scalar theories with a plateau potential $V(\phi)$ (upper panel). Given a `background' lump $\phi_o(r)$ with large field excursion, high field values correspond to the interior of the lump. The effective mass-squared of the field fluctuations  (lower panel) is given by $V''(\phi_o(r))$ and it vanishes inside the lump. 
For massive theories such as \eqref{Vmassive} (dotted line) it approaches $m^2$ outside, resulting in a strong confining cavity effect.
In  quartic$+$plateau potentials such as \eqref{Vmassless}, 
the effective mass vanishes away from the lump, but a localized effective mass term is induced on the surface. This still results in a confining cavity for large radii $R_o$ and a more stable evolution.
}
\label{fig:picture}
\end{center}
\end{figure}

We focus the discussion on the simplest conceivable theory of a single real scalar field with only self-interactions in $3+1$ dimensions. (Gravity is unimportant for oscillons, so we `set' $G_N=0$.)
The key property for a theory to contain soft oscillons  is that the scalar potential $V(\phi)$ flattens out to a plateau at large field. 
A simple example  is
\begin{equation}\label{Vmassive}
V(\phi) = \frac{1}{2} \frac{m^2\,\phi^2}{1+\phi^2/\Lambda^2}~.
\end{equation}
This potential is basically flat for $\phi\gg\Lambda$ and massive for $\phi\lesssim\Lambda$. This clearly leads to a confining effect: a localized configuration (or `lump') with large field displacement $\phi_0\gg \Lambda$ in this theory inevitably becomes equivalent to a cavity, the field being effectively massless inside and massive outside, see Fig.~\ref{fig:picture}. The radius $R$ of the cavity, corresponding to the region where $\phi\sim \Lambda$, is naturally expected to increase with the field amplitude $\phi_0$. Balancing the (constant) potential and gradient terms one quickly sees that the amplitude and radius are linked as  $\phi_0 \sim \Lambda m R$, so indeed large radii $R\gg 1/m$ can be obtained with large amplitudes $\phi_0\gg\Lambda$.\\

At such high field amplitudes the oscillons  must be dramatically different than the conventional oscillons. The characteristic oscillation frequency can only be set by the object size
$$
\omega \sim \frac{1}{R_o} \ll m~,
$$
which can be much smaller than the mass threshold. This is in principle the perfect condition for tiny evaporation and so for a long-lived oscillon-like state.\\ 

The `bad news' for massive models like \eqref{Vmassive} is that the confining effect of the mass term is actually too strong. For the cavity effect to lead to oscillons, maintaining the coherence of the field is crucial. An excessive confinement of modes tends to spoil coherence and these oscillons end up having short lifetimes in massive theories. \\

Our main finding is that the cavity effect is instead viable for oscillons in gapless models, by which we mean that the field has a potential but the effective mass vanishes at the minimum. 
Generically the form of these potentials reduces to a quartic coupling (near $\phi=0$) that continues to a plateau at large field. A prototypical example is
\begin{equation}\label{Vmassless}
V(\phi) = \frac{M^2}{\Lambda^2} \frac{\phi^4}{(1+\phi^2/\Lambda^2)^2 }~,
\end{equation}
but the precise form is not important. 
At large field $\phi\gg\Lambda$ the potential plateaus as before but now the field is massless near $\phi=0$, that is outside the cavity. The mass parameter $M$ does not correspond to the mass of asymptotic states. Instead it sets the maximal effective mass that occurs at the cavity surface, where  $\phi\sim \Lambda$, see Fig.~\ref{fig:picture}.\\

Thus, a large amplitude oscillon  also dynamically realizes a cavity, now with the field massless both inside and outside. 
The field being massless outside facilitates the emission of radiation -- a welcome feature because the cavity can possibly shed the coherence-spoiling part of the field. As we will see, in contrast with the gapped case \eqref{Vmassive}, a clear attractor with self-similar evolution indeed emerges in gapless models.\\

Our soft oscillons are different from the oscillons that have recently been found in gapless theories \cite{Dorey:2023sjh}, see also \cite{Lozanov:2017hjm,Piani:2025dpy}. Those oscillons have a fixed size (set by the equivalent of $1/M$). This might have  lead to the conclusion that a generic relation between the size and the mass scale of the theory  would also be present for gapless models. 
Our soft oscillons show that this is not necessarily the case. Their radius is parametrically larger than $1/M$, in fact almost unboundedly so.\\

As we will show in detail, the soft oscillon structure is quite rich: they have a large number of harmonics (up to $n_{max}\sim MR_o/\pi$), with a precise relative harmonic content, that corresponds to a neat space-time evolution in the core. The attractor solution develops spherical front waves (one outgoing and one ingoing) traveling through the interior of the cavity. This configuration has the virtue that near the cavity surface the field is static most of the time, which of course implies that the emission of radiation is minimized. \\

Many of the properties of this new soft oscillon attractor can be understood with a bag model that exploits the cavity effect. In fact, our bag model is remarkably close to the MIT bag model of hadrons \cite{Chodos:1974je}, so much so that it seems appropriate to say that oscillons in  plateau potentials {\em realize}  MIT bags. \\ 

The evaporation of soft oscillons is also quite interesting. Two clearly separated regimes can be identified depending on how the oscillon size compares with a certain critical radius $R_c$. Below $R_c$ the dominant attractor is {\em clean}, it is maintained with fewer modes. Correspondingly the  evaporation rate is low and characterized with the total emitted power scaling linearly with $R_o$. For $R_o>R_c$ instead, a {\em dirty} regime appears, with many more modes excited and emitted power scaling like $R_o^2$. We offer both numerical and analytic evidence of this below.\\

The rest of this article is structured as follows. In Section \ref{sec:potential} we discuss general aspects of the theories under study. We also show the oscillons that are obtained using `random' initial conditions, far from the actual attractor, but which delivers sufficient hints to reveal the attractor.
In Sec.~\ref{sec:bagmodel} we give a simple bag model description of the attractor exploiting the cavity picture. 
In Sec.~\ref{sec:softosc} we show the soft oscillons found numerically.
In Sec.~\ref{sec:evaporation} we discuss evaporation, including both a theoretical estimate and the numerical results in the two regimes, above and below $R_c$.
In Sec.~\ref{sec:anisotropies} we show the effect of anisotropic perturbations using 3D lattice simulation, and we conclude in Sec.~\ref{sec:concl}.

\section{Minimal models}
\label{sec:potential}

We start by specifying the requirements on a model to harbor soft oscillons. We shall restrict the discussion to the simplest model consisting in a single real scalar field $\phi$, and with a standard kinetic term. The theory is then specified by the form of the scalar potential $V(\phi)$. In full generality, this can be written in terms of a generic function $\VV$ as
\begin{equation}
\label{generalV}
V(\phi) = M^2 \, \Lambda^2 \; \VV\left(\frac{\phi}{\Lambda}\right) 
\end{equation}
with $M$ and $\Lambda$ two mass scales. The effective theory can also contain high dimension derivative operators suppressed by $\partial/\Lambda$, which we neglect because the solutions involve tiny gradients compared to $\Lambda$. To ensure that the theory is weakly coupled, we assume that they are well separated as
\begin{equation}\label{ML}
    M\ll \Lambda~.
\end{equation}
This regime is a small deformation of the shift invariant theory ($V=0$). The choice $\VV(x)=1-\cos{x}$ corresponds to an axion (or sine-Gordon), $M$ plays the role of the axion mass and $\Lambda$ the decay constant. \\

We write the potential in more general form because we are interested in the important case of massless $\phi$. This is of course not incompatible with a nonzero potential $V(\phi)$. Rather it only requires a vanishing mass, that is, no quadratic term in the expansion near the vacuum. Assuming a $Z_2$ symmetry for simplicity $V(-\phi)=V(\phi)$, the gapless case is obtained by a generic function $\VV$ of the form
\begin{equation}
    \VV(x)=\sum_{n=2}^{\infty} g_{2n} \, x^{2n}
\end{equation}
where the coefficients $g_{2n}$ being generically $O(1)$. This leads to
\begin{equation}
V(\phi)= \lambda \, \phi^4\;\left( 1 + c_6 \, \frac{\phi^2}{\Lambda^2} + \cdots \right)
\end{equation}
with the quartic coupling identified as 
\begin{equation}\label{quartic}
    \lambda=g_4 \frac{M^2}{\Lambda^2} \, \ll1~.
\end{equation}
At the same time, all of the  coefficients $c_{2n} = g_{2n}/g_4$ are $O(1)$ pure numbers. Of course, we restrict to $\lambda>0$ (otherwise $\phi=0$ is not even a minimum). Notice that this implies that the lowest term, the quartic coupling, is of repulsive nature. Hence, the terms responsible for sustaining  oscillons must be $\phi^6$ and/or higher. That is, the coefficients $c_6$ and higher must be sufficiently negative, something that indeed happens in plateau potentials. We will refer to these potentials as having a {\em quartic}$+${\em plateau} form.\\

The absence of a $\phi^2$ term is certainly a fine-tuning, however this is a secondary concern since our main goal here is to show the new phenomenon that arises in gapless theories. Moreover, it is easy to envisage simple extensions of the minimal theory where masslessness is natural (see below). Importantly, the new degrees of freedom in these extensions do not affect the soft oscillons. For these reasons, we will proceed with the analysis of the gapless theory as the case for which the potential `starts' with the  quartic term.\\

The general form \eqref{generalV}  makes manifest that we are dealing with an effective field theory (EFT) for the possibly massless scalar $\phi$ with an EFT cutoff determined by $\Lambda$ (barring factors of $4\pi$). We are going to look for semiclassical configurations in this EFT and we need to make sure that they are within the validity of the EFT. The separation of scales \eqref{ML}, or equivalently the smallness of the quartic coupling $\lambda$ enforces that the solutions are within the EFT regime. Indeed, introducing 
\begin{equation}
\vp \equiv \frac{\phi}{\Lambda}~,    
\end{equation}
the equation of motion reads
\begin{equation}\label{eom}
\Box\vp - M^2 \,\VV'(\vp) = 0    
\end{equation}
so the nonlinear solutions (amplitude of order $\vp\gtrsim 1$) are going to be characterized with  gradients of order $M\ll\Lambda$. Recently \cite{Dorey:2023sjh} has given explicit examples of oscillons in a gapless model. The shape of the potential assumed in \cite{Dorey:2023sjh} is a double well and differs from our main focus, but the size of their oscillons is defined by the (inverse) mass scale analogous to our $M$, which is present in the model in spite of being gapless.\\


The situation is actually qualitatively similar to massive theories: near $\phi=0$ a field `lump' tends to expand, both because of radiation pressure and from repulsion from $\lambda>0$. 
Formally this is  parallel to what happens for oscillons in massive theories, where the mass term induces a repulsive-like effect. 
At large field, $V$ is attractive if (sufficiently negative) self interactions can compensate the repulsion, and a localized object can be form. 
Naively, one expects that the condition on the set of $g_n$ for this to happen is that $V$ is smaller than either $ m^2 \phi^2/2$ or 
$ \lambda \phi^4$ at $\phi\gtrsim\Lambda$ for the massive and massless theory respectively. \\

In this work we shall mainly discuss theories with plateau potentials
\begin{equation}\label{plateau}
    V(\phi) \simeq {\rm const} \quad {\rm for } \quad  \phi\gtrsim\Lambda.
\end{equation}
This type of potential admits especially interesting solutions with parametrically large size $R_o\gg 1/M$, to which we will refer as `soft' oscillons. 
We anticipate that these solutions contain large field `excursions' $\Delta\phi\gg\Lambda$, significantly larger than the scale $\Lambda$ where the plateau starts.
From the EFT perspective this is not {\em per se}  an obstruction. In the case of axionic models, large field excursions have been studied at length \cite{Kaplan:2015fuy}. The presence of large field-variations in the initial state is also quite clearly not a problem. (In axion models one could compare this to the initial state of a large number of domain walls, a configuration that is certainly captured by the low energy EFT.) 
A more non-trivial condition (that we impose on ourselves) is that the time evolution of such initial state remains within the applicability range of the  EFT. As we shall see, this is indeed the case for our soft oscillons. At all points in the evolution the maximum space- and time- gradients everywhere on the solutions are near the soft scale $M$, well below the cutoff $\Lambda$. \\

For the sake of simplicity, until now we have set $G_N=0$, however this is only an approximation and one needs to check a posteriori whether gravitational effects are small. This can be expressed in several different ways: i) small Newtonian potential at the oscillon surface, ii) small expansion rate inside the oscillon
iii) being far from a black hole ($R_o$ much larger than the Schwarzchild radius), and iv) requiring sub-Planckian field excursions. Interestingly enough, all these requirements reduce to a unique condition, setting an upper limit to the oscillon radius $MR_o \gg M_P/\Lambda$, with $M_P$ the Planck mass. Having in mind sub-Planckian values of $\Lambda$ this still gives room for large oscillons as we defined them $MR_o\gg1$.\\

We now return to the issue of naturalness of the gapless potentials. Setting a vanishing mass ($g_2 =0$ in the language of \eqref{generalV}) is of course a fine tuning. This is not a big concern for the purpose of this work because it does not affect the consistency of the theory. Let us add, though, that the model can certainly be augmented with a mechanism that preserves $m_\phi=0$ naturally. Such mechanisms typically require the introduction of additional degrees of freedom coupled to the scalar $\phi$, and this may imply that the properties of soft oscillons are affected. However very massive objects are unaffected. For example, consider a supersymmetric embedding. The minimal SUSY extension of the real scalar consists in a chiral multiplet, that is an additional scalar $\chi$ and chiral fermion $\psi$. The resulting theory has a total of 4 massless degrees of freedom and the full lagrangian can be obtained from a superpotential $\cal W$, which leads to a potential $V=|d{\cal W}/d\Phi|^2$ with $\Phi = \phi+i \chi$ the complex scalar. To fix ideas, one can think of  ${\cal W} = \Phi^3/(1+\Phi^2)$, which does not correspond precisely to \eqref{Vmassless} but is structurally of quartic$+$plateau form. It is easy to check that a nontrivial background for the real part $\phi$ is consistent classically with all the other fields set to $0$, hence the classical analysis is unaffected. At quantum mechanical level, radiation of superpartners is expected, of course. Quantum radiation is also emitted by the real part $\phi$, that provides the oscillon background. Since the oscillon emits classical radiation, though, one expects that the quantum evaporation is subleading. (A nontrivial effect from quantum radiation is the appearance of a quantum break time \cite{Dvali:2017eba,Dvali:2017ruz}, but we leave this for future work \cite{inprep}.)
In practice, then, we can ignore the effects from any partners that may be responsible for keeping $m_\phi=0$ naturally.\\

Let us point out an additional interesting property of the quartic$+$plateau potential $V(\phi)$. At large field, $V(\phi)$ is approximately shift-invariant.  On the other hand, at small $\phi$, the theory is invariant under dilatations. The quartic$+$plateau potential then represents a renormalization group flow from a shift symmetric to a conformal theory in the infrared. This suggests an embedding in a 5D model with a geometry that interpolates from 5D Anti-de Sitter to 5D Minkowski spacetime. Notice also that the quartic$+$plateau potential arises (in the Enstein frame) for a scale-invariant field nonminimally coupled to gravity \cite{Bezrukov:2007ep,DeSimone:2008ei}.

%

\begin{figure}[t]
    \centering
    \includegraphics[width=0.8\linewidth]{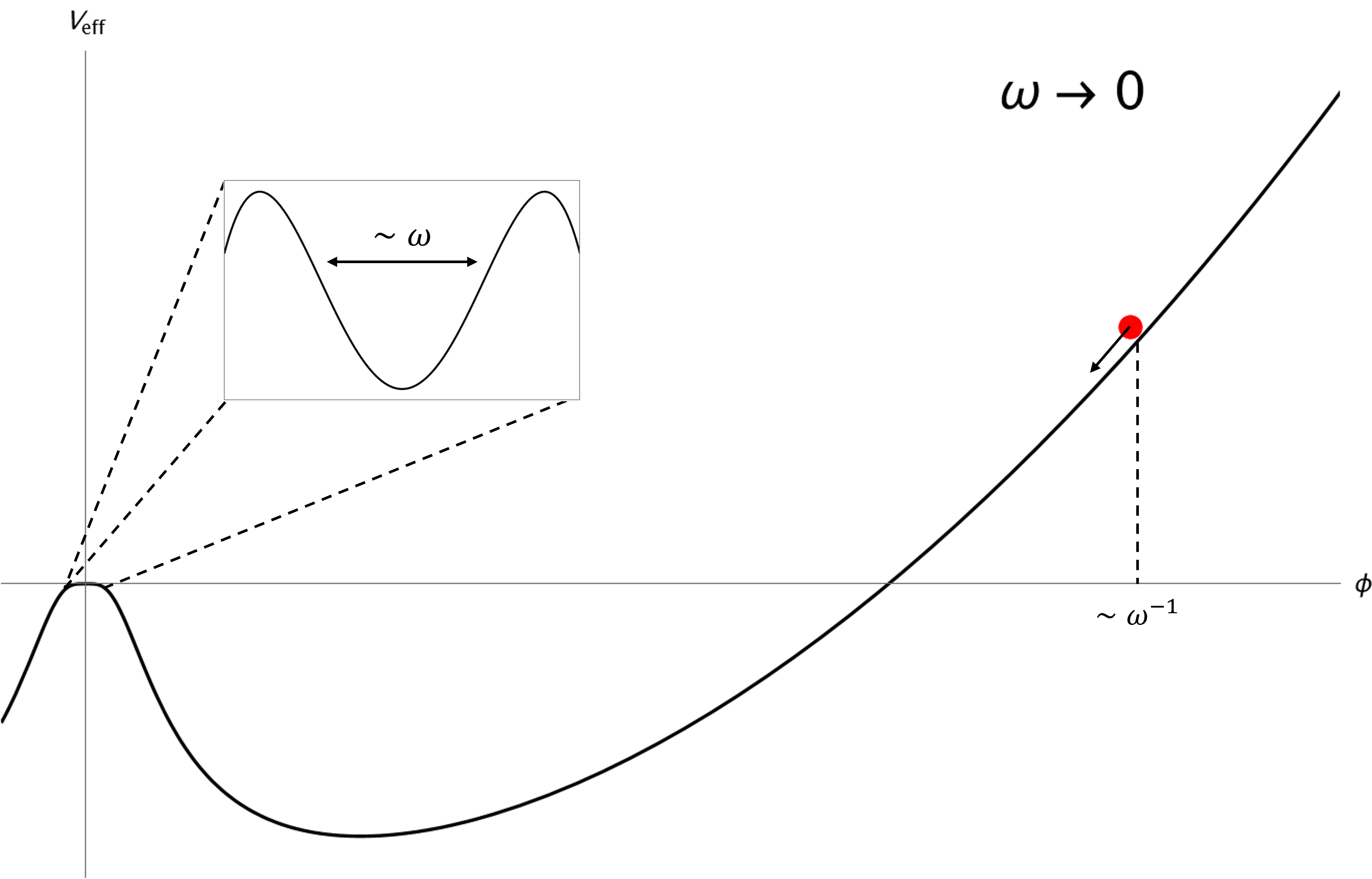}
    \caption{Although there are strictly speaking no bound breather modes in a gapless theory, there is a well-defined shooting problem to try and construct large amplitude and large radius oscillons. There are large, breather-like objects, whose radiation-tail can be asymptotically small.}
    \label{fig:radiationeffective}
\end{figure}

\subsection{Shooting around}
 In massive theories, the scalar potential provides a natural frequency below which one can search for single-frequency breather modes that are bounded at spatial infinity, namely the mass $m$. In standard oscillon literature one then identifies each breather profile of frequency $\omega$ with the fundamental mode of the oscillon that exists at that frequency. To find localized breathers one looks for the unique solution of the ODE

    \beq
        \nabla^2 \phi_{br} + (\omega^2 - m^2) \phi_{br} - V_{eff}'(\phi_{br}) = 0,
        \label{eq:effectivebreathermassive}
    \eeq
    that satisfies the boundary condition $\phi_{br} \rightarrow e^{-\sqrt{m^2 - \omega^2}\, r}$ as $r\rightarrow 0$. The effective potential is given by
    \beq
        V_{eff}(\phi) = \frac{1}{\pi} \int_0^{2 \pi} dt' \, V_{nl}(\phi \cos(t')),
        \label{eq:effectivepotential}
    \eeq
    where $V_{nl}(\phi)$ is the nonlinear part of the scalar potential of the theory under consideration. This framework for understanding oscillons is well established in the literature. However, in gapless theories, where $m^2 = 0$, a similar procedure leads to the conclusion that one can only find breather modes that have a radiative boundary condition $\phi_{br} \rightarrow e^{i \omega r}$ at spatial infinity. Furthermore, there is no unique solution that one can tentatively identify with an oscillon state. This situation seems somewhat dire to try and apply the standard toolkit to search for oscillons in massless theories. For certain potentials, in particular plateau potentials, something interesting happens to the breather equation ~\eqref{eq:effectivebreathermassive} when $m^2 = 0$. Although there is no exponentially decaying solution that asymptotes to 0 (the true vacuum), there is now a unique solution that obeys $\phi_{br} \rightarrow e^{-r} + A \omega$ as $r\rightarrow \infty$. Here $A$ should be interpreted as some $O(1)$ number. In the limit $\omega \rightarrow 0$, there exists a breather solution that is exponentially trapped and asymptotes to the true vacuum of the theory. Note that this solution also has infinite size $R_o$.\\

    We can look for the approximately localized solution of Eq.~\eqref{eq:effectivebreathermassive} through a numerical shooting problem, common in the oscillon literature. One solves the ODE with boundary conditions $\phi_{br}(0) = \phi_0$ and $\phi_{br}'(0) = 0$, and adjusts $\phi_0$ until the solution obeys the appropriate boundary condition at $r\rightarrow\infty$. The procedure is analogous to the classical mechanics problem of releasing a ball in a potential landscape and searching for the correct initial release point so that the ball halts exactly at a local maximum of the potential (a hilltop). A schematic representation of the problem is shown in Fig.~\ref{fig:radiationeffective}. \\

\begin{figure}[t]
\begin{center}
\includegraphics[width=0.8\textwidth]{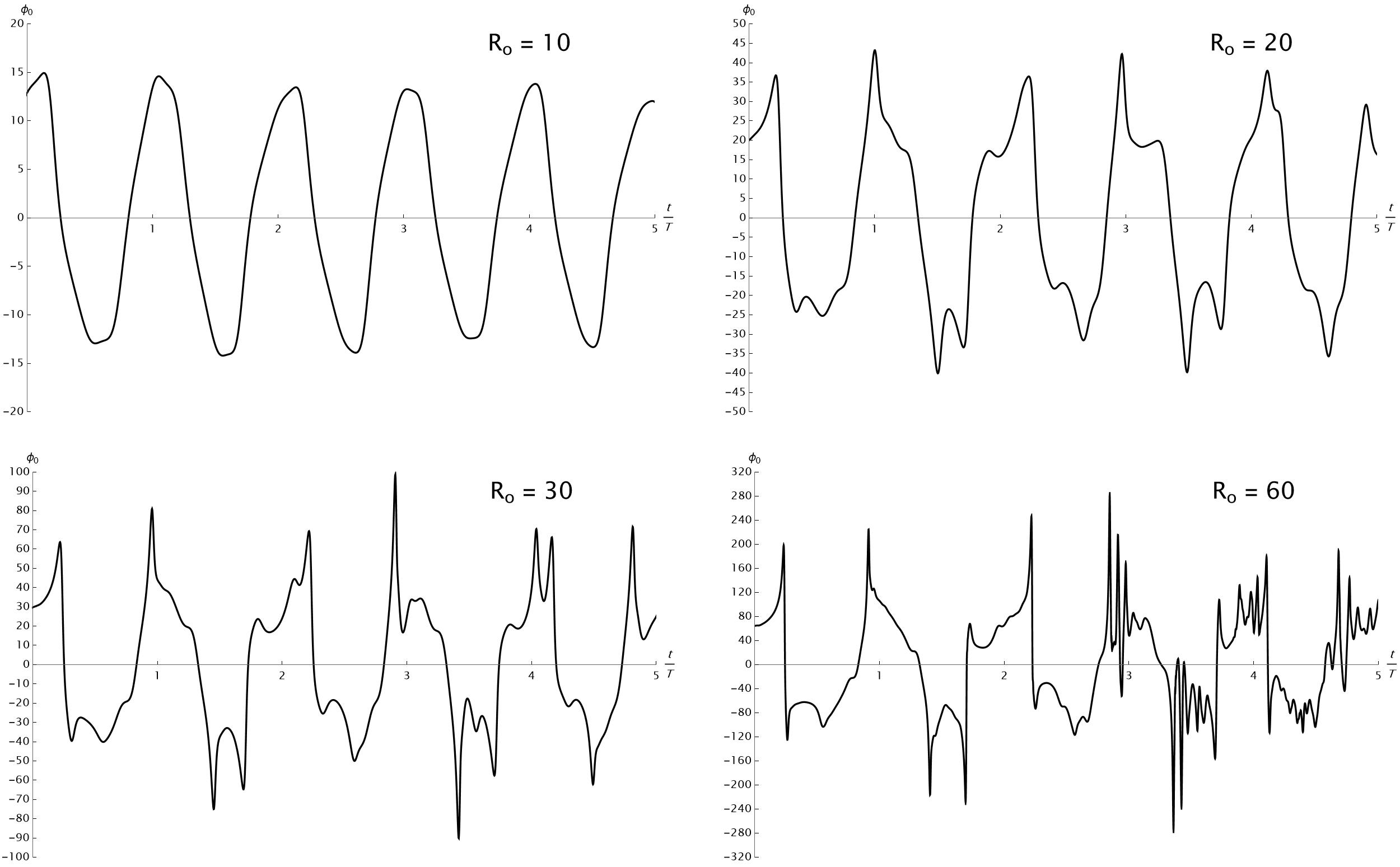}
\caption{The amplitudes of the solutions at the origin $r = 0$ vs time in units of the period of the background. There is an approximate relation between the principal frequency and radius as $R_0 = \pi/\omega$. It is clear that as the solution becomes larger, the frequency content of the solution becomes larger as well. The classic oscillon described by a single harmonic is no longer a sufficient description as $R_0 \rightarrow \infty$.}
\label{fig:amplitudes}
\end{center}
\end{figure}

\begin{figure}[t]
\begin{center}
\includegraphics[width=0.8\textwidth]{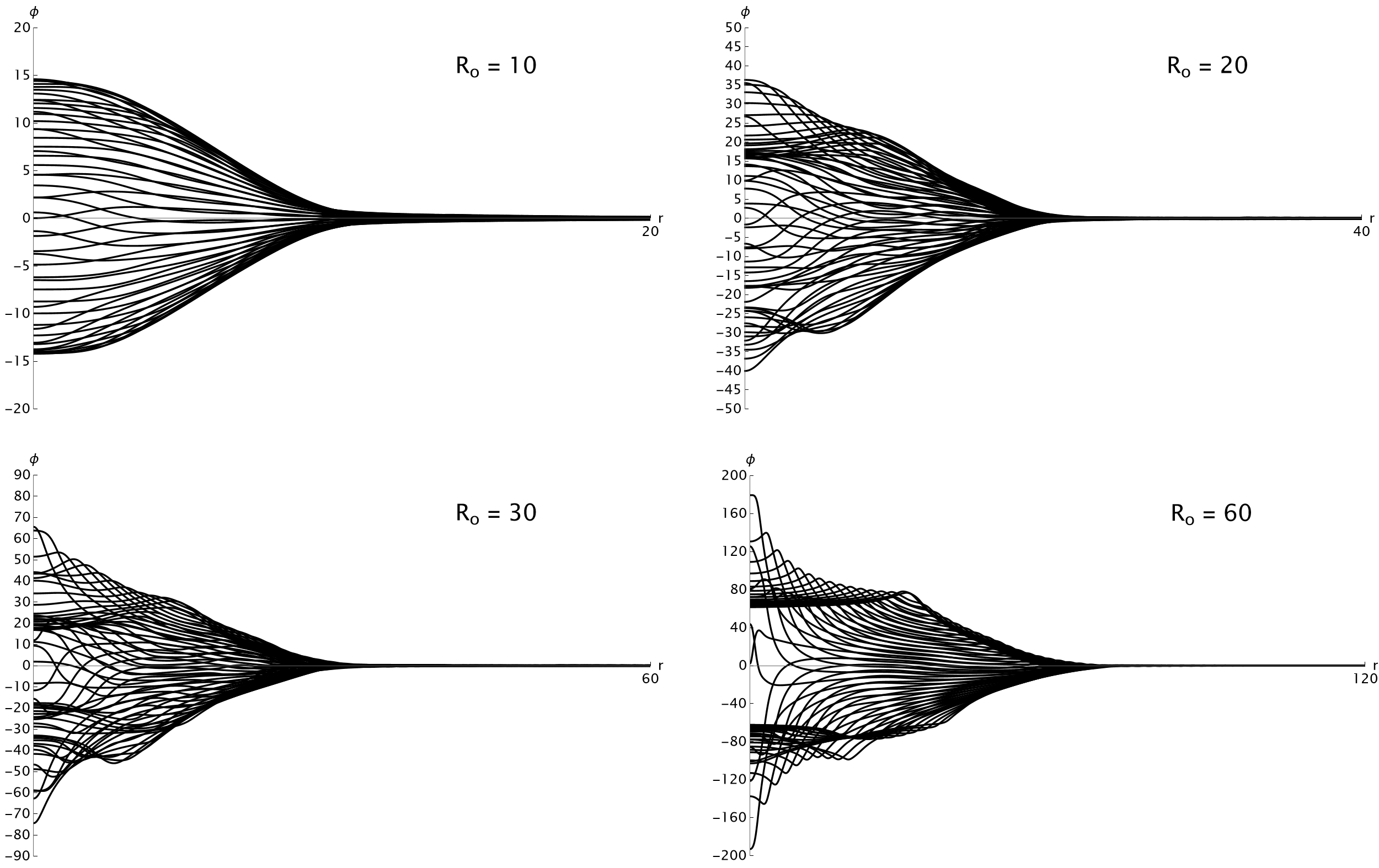}
\caption{The spatial profiles of the solutions, where each line corresponds to a different time. The suggestion of a different type of ansatz emerges.}
\label{fig:spatialprofiles}
\end{center}
\end{figure}

Since we are only interested in breather solutions that asymptote to something that is close to the true vacuum of the theory, we are necessarily interested in small frequencies $\omega \rightarrow 0$. Unlike standard oscillons where the size of the object is set by the mass, $R_o \sim m^{-1}$, these breather states can be arbitrarily large. In particular one has the relation $R_o \approx \pi/\omega$, so that $R_o \rightarrow \infty$ as $\omega \rightarrow 0$. It is interesting to see whether these ansatzes obtained through the shooting problem, remain localized when evolved with the full equations of motion. In Figs.~\ref{fig:amplitudes} and \ref{fig:spatialprofiles}, we show the result of numerically evolving a field configuration in a typical massless plateau potential, starting from the initial conditions $\phi(t = 0, r) = \phi_{br}(r)$ and $\dot{\phi}(t = 0, r) = 0$. We are intentionally vague concerning the exact shape of the potential, as this work will show that the type of soft oscillons that we describe can exist in any plateau potential.\\

Although the spatial-temporal profiles shown in Figs.~\ref{fig:amplitudes} and \ref{fig:spatialprofiles} still show an oscillatory envelope, typical for oscillons, it is abundantly clear that the standard picture in which the oscillon is described by a bounded breather mode sourcing a small radiative tail, is no longer a sufficient description of the oscillatory solutions we find. In particular, higher harmonics can become of the same order as the fundamental mode as $R_o \to \infty$. To make progress in understanding these solutions, we will have to find a better ansatz that describes these oscillons with large radius $R_o$ and amplitude $\phi_0$, and small frequency $\omega$. We will proceed to construct such an ansatz in Sec.~\ref{sec:bagmodel}, inspired by the results of Figs.~\ref{fig:amplitudes} and \ref{fig:spatialprofiles}. Note in particular that in Fig.~\ref{fig:spatialprofiles}, that as $R_o \to \infty$, the breather solution dynamically settles into a constant interior, connected to the vacuum at $r \to \infty$ by a static envelope.

\section{Bag model}
\label{sec:bagmodel}

From the arguments of Sec.~\ref{sec:potential} and the numerical time evolution of a `random' initial condition, it is clear that the oscillons for plateau potentials in the soft limit $M R_o \gg1$ have completely  different characteristics compared to any oscillon discussed previously in the literature. The purpose of this Section is to get some first glimpse of this new regime. Inspired by the {\em cavity effect} inherent to plateau potentials, the first toy model consists in ignoring the radiation from  the oscillon and view it simply as a perfect cavity populated with a massless field. We shall refer to this as a {\em bag model}. In fact, this approximation is very reminiscent of the MIT bag model for hadrons \cite{Chodos:1974je}. \\

As we will see, even at zero-th order in the leakage of radiation a special configuration emerges in this bag model. 
Moreover, this sets the language to compute the evaporation rate. To that end, a closer look at the cavity surface (which we also refer to as {\em skin}) will be needed. We do so in Sec.~\ref{sec:skin}.

\subsection{Spherical cavity}
\label{sec:cavity}

Let us now bring some light into why the peculiar oscillon announced in the Introduction should exist. 
The key observation is that a configuration $\phi(t,r)$ with a large field displacement living in a plateau potential, has an effective mass that is a localized surface mass term
$$
V''(\phi(r,t)) \simeq v(t) \delta(r-R_o)
$$
In the $M R_o \gg1 $ limit, $v(t)$ stays constant during most of the oscillation period, so it can be approximated by a constant. Its value  is set by the characteristic mass in the theory, $M$. 
Therefore, the first order approximation (accurate up to $1/M R_o$ corrections) reduces to that of a massless free field in a cavity\footnote{Ref.~\cite{Alexeeva:2023rfi} has recently discussed a somewhat similar setup with the nonlinear field equation restricted to a `ball' or cavity of fixed radius, intended to capture the small amplitude regime. The main difference with our case is that we consider large amplitude and, more importantly, the cavity arises dynamically in our setup.} with a large surface mass term (compared to the cavity characteristic frequencies, of order $1/R_o $). This motivates the approximation of simply  ignoring the exterior region and switching to the problem of a massless field inside a perfect cavity with a Dirichlet boundary condition $\phi(r=R_o)=0$.  This is of course the equivalent of working at zero-th order in the emitted radiation. Notice that this picture is the same as \cite{Chodos:1974je} only with the field satisfying Dirichlet instead of Neumann boundary condition on the bag.\\

In this approximation an accidental superposition principle emerges even though the theory is nonlinear. Among the infinitely many available solutions, we must seek a special configuration that minimizes the energy loss of the oscillon. The existence and longevity of oscillons is always connected to a reduced time-dependence in the oscillon core. The most special solution in the cavity then is one where the field is actually {\em constant} near the cavity surface during most of the oscillation period. Clearly the only static near-surface solution that vanishes at the surface must be proportional to $1-R_o/r$. This must be matched to some other exact solution of $\Box\phi=0$ to avoid a singularity at $r=0$. This is certainly possible thanks to the following property of the massless wave equation. Any pair of solutions of $\Box\phi=0$, say $\phi_{1,2}$ (say, with different values of $X\equiv\partial_\mu\phi\partial^\mu\phi=\dot\phi^2-\phi'^2$) that match continuously on a light-like region (a radiation front) delivers a new piece-wise defined solution that coincides with $\phi_1$ or $\phi_2$ on either side of the front. This allows to extend the near-surface static solution inwards easily by looking for solutions that overlap well on radiation fronts. It is easy to see that in each quarter of a cycle the 3 solutions involved are: $1-R_o/r$ (with $X<0$),  $t/r$ ($X>0$) and a constant ($X=0$), and we need two spherical fronts, one out-going and the other in-going.\\

\begin{figure}[t]
\begin{center}
\includegraphics[width=\textwidth]{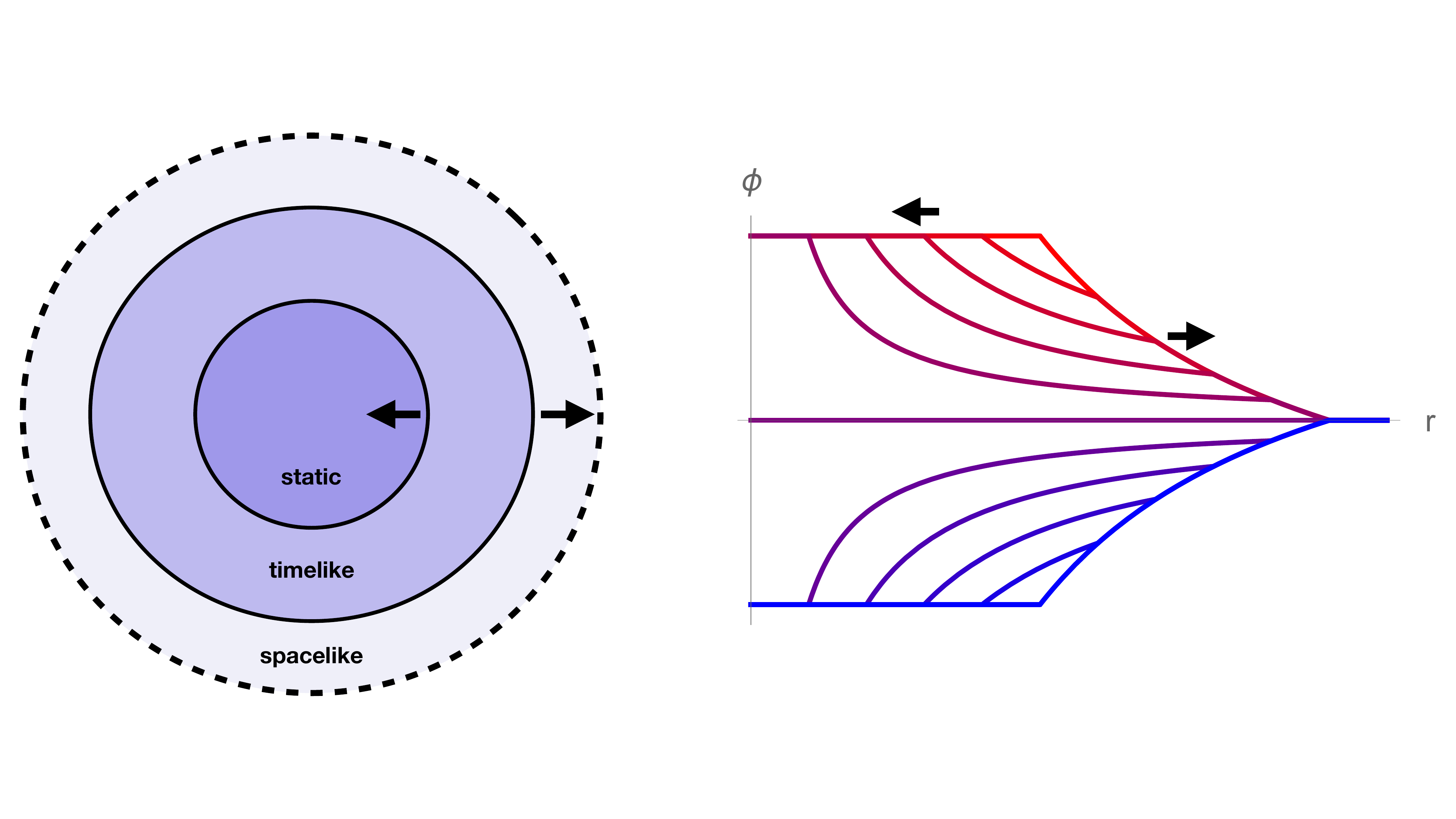}
\caption{Sketch of the bag configuration \eqref{phi_bag}, consisting in two concentric spherical fronts that bounce off the cavity surface. Snapshots of the field profile separated by equal times and for half a cycle are shown on the right. The fronts correspond to the points with discontinuous first derivative.}
\label{fig:bag}
\end{center}
\end{figure}

The full solution can be written piece-wise\footnote{We thank A. Wereszczyński for pointing out that potentials of the form $V(\phi)=|\phi|$ admit exact oscillon solutions in piece-wise form \cite{Arodz:2007jh}.} in spacetime regions separated by the exterior and interior concentrical  fronts $r_{ext,\,int}(t)$, 
\begin{equation}\label{phi_bag}
\phi_{bag}(t,r)= \phi_0 \,
\begin{cases} \displaystyle
    {\rm sq}\left(\frac{t}{2R_o}\right) &\quad r< r_{int}(t) \\[3mm]
    \displaystyle
    {\rm tr}\left(\frac{t}{2R_o}\right)\frac{R_o}{2r} &\quad r_{int}(t)< r< r_{ext}(t) \\[3mm]
    \displaystyle
    {\rm sq}\left(\frac{t}{2R_o}\right)\;\left(\frac{R_o}{r}-1\right) &\quad r_{ext}(t)<r< R_o \\[1mm]
    \displaystyle
    0 &\quad r\geq R_o ~,
   \end{cases}
\end{equation}
with ${\rm sq}(x)={\rm sign}(\cos{x})$  and ${\rm tr}(x)=\frac2\pi \arcsin{\cos{\pi\,x}}$ the unit period square and  triangular waves, and 
\begin{align}
    r_{int}(t) & = \frac{R_o}{2} \,\frac{{\rm tr}(t/2R_o)}{{\rm sq}(t/2R_o)}= \frac{R_o}{2}\, \frac{{\rm tr}(t/R_o)+1}{2} \nonumber \\
    r_{ext}(t) &= R_o- r_{int}(t)~. 
\end{align}
A sketch of the profile and motion of this field configuration is shown in Fig.~\ref{fig:bag}. 
Notice the peculiar structure of  \eqref{phi_bag}, consisting in  3 layers:  constant field profile near the center, time-like gradient in between the fronts and the static outer profile. These layers are separated by an inward and an outward front traveling at the speed of light, that cross each other at $r=R_o/2$.\footnote{A natural question is whether solutions of this form exist in dimensions different than $3 + 1$. Leaving a comprehensive study of soft oscillons in other dimensions for the future, we have checked that similar behavior does occur though details like the shape of the profile and the lifetime vary significantly. Note in particular that the functional form of the outer region ($r > r_{ext}$) in \ref{phi_bag} changes in other dimensions.}\\

It is clear from \eqref{phi_bag} and Fig.~\ref{fig:bag} that the field evolution is coherent in the cavity in the sense that it changes sign and reaches maximum values globally. It is also clear that the configuration has a well defined and constant radius -- where $\phi=0$ in this approximation. (In the full numerical solutions this is replaced by where $\phi$ is  of order $\Lambda$, and it coincides with where most of the energy is concentrated.) Interestingly, this constant overall radius coexists with the bouncing motion within the cavity due to the spherical fronts. Notice also that twice per cycle the inner front focuses at the origin, at which event also the field at the center abruptly changes sign. The large focusing is not a problem in the bag approximation because the field is effectively free while the amplitude is large. The sign flip (taking place in all the cavity) instead is not under control in this approximation since it implies probing small values of $\phi$ where the potential is noticeable. This leads to a physical effect, the `dirty' regime of evaporation discussed in Sec.~\ref{sec:evaporation}. However, since the sign flip is fast this is still a small effect and the idealized bag picture is still useful. \\

It is straightforward to compute the energy of the bag configuration \eqref{phi_bag},
\beq\label{bagE1}
E_{bag} = 4\pi \int r^2 \left(\frac{\phi'^2}{2}+V_0\right)\Big|_{\rm bounce}
=8\pi \, \phi_0^2 \,R + \frac{4\pi}{3} V_0\,R^3
\eeq
where $V_0=\Lambda^2\,M^2$ is the value of the potential (everywhere in the cavity interior we are on the plateau) and the moment of the bounce refers to when the fronts meet, which is also when $\dot\phi=0$ everywhere in the bag approximation.
For reasons that will become clear in Sec.~\ref{sec:skin},  the least radiating field configuration satisfies that the field gradient at the cavity surface relates to the value of the potential $V_0$ as
\begin{equation}\label{phiprime}
    \frac{\phi'^2(R_o)}{2}=V_0.
\end{equation}
(Notice that this is recognized as one of the two boundary conditions in the MIT bag model \cite{Chodos:1974je} for a static bag.)
One reads from \eqref{phi_bag} that $\phi'(R) = \phi_0/R_0$ which leads to
\begin{equation}
    \phi_0 = \sqrt{2}\,\Lambda\, M\, R_o 
    \label{eq:bagampcent}
\end{equation}
and
\beq\label{bagE}
E_{bag} =\frac{16\pi}{3}\,M^2 \Lambda^2 R_o^3 ~.
\eeq
~\\

\subsection{Quantum fingerprint}
With this clear notion of the soft oscillon, we can move to a quantum mechanical language in the spirit of \cite{Dvali:2011aa, Dvali:2012en,Dvali:2012gb,Dvali:2017eba,Dvali:2018tqi,Dvali:2020wqi,Dvali:2021tez,Dvali:2021rlf,Dvali:2023qlk}.
Thanks to the emergence of the cavity/bag the obvious natural procedure is to express the general state of the quantum field in a spherical cavity of radius $R_o$ by the occupation numbers of each cavity mode
\beq\label{Nnlm}
N_{nlm}  ~.
\eeq
 
The total energy is
\beq\label{Esum}
E = \sum_{nlm} N_{nlm}  w_{nlm}
\eeq
For a perfect (non-leaking) cavity,
\beq\label{w_n}
w_{nlm} = \frac{u_{n\,l}}{R_o}
\eeq
with $u_{n\,l}$ the zeros of the spherical Bessel function $j_l(u)$. For large $n,\,l$, $u_{n\,l}\simeq \pi n + \pi l/2$. 
In a leaking cavity approximation where the field propagates outside, these modes would acquire a small imaginary part. Implicitly we are neglegting this, but it is clear that these are really resonant quasi-normal modes.
For $l=0$, the eigenfrequencies are 
\begin{equation}
    w_n=\frac{n\,\pi}{R_o}
\end{equation}
and the $l=0$ modes take the form
\begin{equation}\label{phin}
    \phi_n = \frac{A}{\sqrt n}\frac{\sin{\frac{n\pi r}{R_o}}}{r}
\end{equation}
with $A$ an $n-$independent normalization constant. \\

In the bag approximation \eqref{phi_bag}, the field at the origin $\phi(r=0)$ evolves in time as a square wave, whose Fourier transform  goes like $1/n$ (odd $n$'s only). Thus the field is a sum of eigenmodes of the form $\phi(0)\sim \sum \phi_n(0)/n^{3/2}$, and we can read off the occupation numbers (recalling that we are dealing with a free field so the occupation numbers are quadratic in $\phi_n$). Putting it all together, the bag model \eqref{phi_bag}  corresponds to a large occupation number of s-wave modes with a power law distribution 
\begin{equation}\label{Nbar}
N_{n00} = \frac{128}{3\pi^2}\;\frac{\Lambda^2 M^2\, R_o^4 }{ n^3 } \; \delta_{n,\,{\rm odd}} ~.
\end{equation}
The overall amplitude follows by matched by requiring that the total energy matches \eqref{bagE} and using $\sum_{n\,odd}n^{-2}=\pi^2/8$. \\

The occupation numbers \eqref{Nbar} exhibit some interesting properties: they follow a power-law distribution as expected by the absence of scales in the idealized bag model. The spectrum is very `red', strongly dominated by the fundamental mode. For comparison, having the same amount of energy stored only in the fundamental mode leads to $N_{100} = \frac{16\pi^2}{3}\;{\Lambda^2 M^2\, R_o^4 }$, which is larger than $N_{100}$ in \eqref{Nbar} by a factor $\simeq12$. 
Notice also  that \eqref{Nbar} scales inversely with the coupling constant $\frac{1}{\lambda} \sim \frac{\Lambda^2}{M^2}$ as is common for oscillons in massive theories. In the soft limit $MR_o\gg1$, it also scales with $R_o$, and with quite a strong dependence. \\

Recall that in the gapless theory the cavity effect arises from the large surface mass term, of order $M \delta(r-R_o)$, and the width of the delta function also of order $1/M$. Therefore the expressions are approximate, with corrections suppressed by $1/(M R_o)$ in the soft end of the spectrum. A finite skin mass term also implies that the total number of modes trapped in the cavity is finite. One expects that modes with frequencies 
$$
w_n \lesssim M
$$
are `confined’ efficiently because $M$ is the mass scale that characterizes the effective mass from the skin. Therefore one expects the highest harmonic have
\begin{equation}\label{nmax}
    n_{max} \simeq \frac{M R_o}{\pi}~.
\end{equation}

For this reason, a better picture of the background mode composition of the background is to cut it off beyond $n>\nmax$,
\begin{equation}\label{Nbar2}
N^{(bag)}_{n00} = 
\begin{cases}\displaystyle
    \frac{128\,\pi^2}{3}\frac{\Lambda^2}{M^2}\;\frac{ \nmax^4 }{ n^3 } \;  & n<\nmax \;\; {\rm and \; odd}\\
    0 & n>\nmax 
\end{cases}  
\end{equation}

An adiabatic evolution where the soft oscillon slowly evaporates and decreases its radius $R_o(t)$ translates in  occupation number language in that the modes become depleated adiabatically by the time dependence implicit in $\nmax = M R_o(t) /\pi$.

\subsection{Cavity skin profile}
\label{sec:skin}

Let us now move one step beyond the idealized bag model and look at the structure of the field near the edge of the effective cavity in the nonlinear theory. Namely,  we are now aiming at solving the full equation of motion \eqref{eom}, and finding the static near-surface profile.  
In the $M R_o \gg1$ limit, zooming in on a narrow angular region the field profile near $r\simeq R_o$ is well approximated by a planar form where $\phi$ depends only on one cartesian coordinate, which we call $z$. 
Plugging the static ansatz $\phi(z)$ into the field equation leads to
\beq\label{HDWeq}
\frac{\phi'(z)^2}{2}=V(\phi(z))~.
\eeq
One immediately notices that in the interior region where $\phi$ has climbed the plateau $V(\phi)=V_0= $const the field gradient is fixed. This justifies \eqref{phiprime}. In hindsight, it follows from staticity  near  the surface.\\

The solution to \eqref{HDWeq} can be viewed in a sense as `half' a domain wall (DW), since it interpolates from the vacuum  $\phi=0$ at infinity to a large field value (analogous to the maximum in the potential for a theory with domain walls). This solution plays an important role, we will refer to it as the {\em half-DW} and will denote its profile as $\phi_{hDW}$.\\

\begin{figure}[t]
\begin{center}
\includegraphics[width=.7\textwidth]{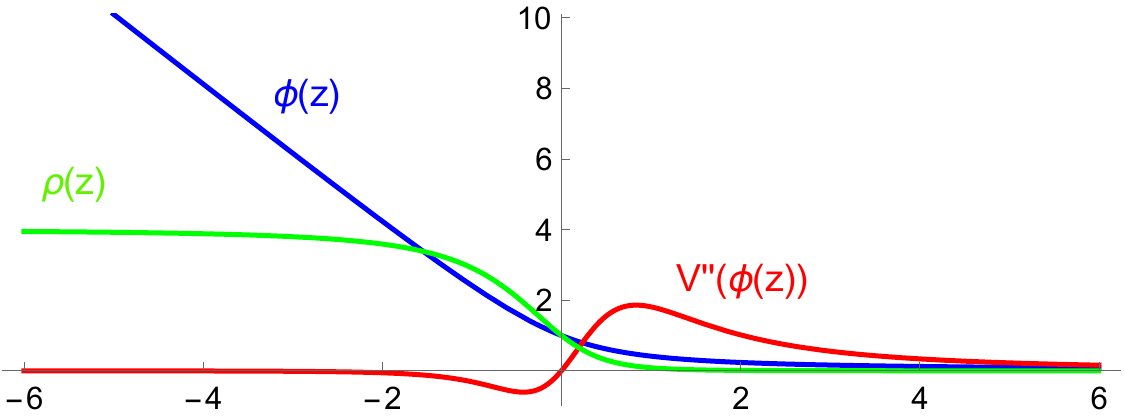}
\caption{Field profile in units of $\Lambda$ (blue), energy density $\rho(z)=\phi'^2/2+V(\phi)$ in units of $M^2\Lambda^2$ (green) and effective mass-squared in units of $M^2$, for the half-DW $\phi_{hDW}(z)$ \eqref{phi_hDW_ex} in the example theory \eqref{hDW_ex}. This captures  behaviour near the surface of the cavity/oscillon. The horizontal axis is $z\, M$. }
\label{fig:halfDW}
\end{center}
\end{figure}

For quartic$+$plateau potentials,  $\phi_{hDW}(z)$ generically approaches the vacuum asymptotically like $1/z$ for $z\to+\infty$, and grows linearly in $z$ towards $z\to-\infty$. In between the two regimes $\phi_{hDW}\sim\Lambda$, which corresponds to the cavity skin. The precise form of the potential is not important. For the sake of illustration it is convenient to consider the particular choice
\begin{equation}\label{hDW_ex}
   V(\phi) = 2 \frac{M^2}{\Lambda^2} \frac{\phi^4}{(1+\phi^2/\Lambda^2)^2 }  ~, 
\end{equation}
for which the half-DW solution admits a  simple closed form expression, 
\begin{equation}\label{phi_hDW_ex}
    \phi_{hDW}(z) = \Lambda\left( \sqrt{1+(M\,z)^2}-M\,z\;\right)~,
\end{equation}
which we show in Fig.~\ref{fig:halfDW}.
As for regular DWs, the stress tensor has the form
\beq
T_{\mu\nu}^{hDW} = \rho(z) \; {\rm diag} \left( 1 ,\, 0 ,\, -1,\,-1\, \right)
\eeq
implying  equation of state parameter $w=-1$ and $0$ in the transverse  and longitudinal directions respectively.  We denote  by $\rho(z)=(\phi'^2/2+V)|_{{hDW}}$ the energy density profile, which interpolates from $0$ at $z\to+\infty$ to a constant in the interior region, see Fig.~\ref{fig:halfDW}. (These equations of state are realized only in the near-surface region. The innermost region of \eqref{phi_bag} has constant field, so $w=-1$ in all directions.)\\

The half-DW profile is also useful to have a finer view of the confining mechanism. Having such large occupation numbers, the oscillon can be treated as a large semiclassical background. As usual, it is convenient to split the field $\phi$ into background and perturbations $\delta\phi$. Near the surface, the oscillon is well approximated by the half-DW so we write
\begin{equation}
    \phi=\phi_{hDW}(z)+\delta\phi(t,z)
\end{equation}
and the equation of motion leads to the Schrödinger-like equation for $\delta\phi$,
\begin{equation}\label{schrodinger}
    \left (\Box + V''(\phi_{hDW}(z)) \right) \delta\phi =0
\end{equation}
with the Schrödinger potential equal to the effective mass-squared evaluated on the half-DW profile, which is shown in Fig.~\ref{fig:halfDW} for the potential \eqref{hDW_ex}. It is clear from the Figure that the Schrödinger potential is more positive than negative, and should lead to bound states in the volume of the cavity.  \\

With Eq.~\eqref{schrodinger} we compute the  transmission and reflection coefficients for an incident wave that approaches the cavity surface from within and which of course is related to the oscillon evaporation. We analyse this in more detail in Sec.~\ref{sec:evaporation}. It may suffice to say now that from the form of the Schrödinger potential one expects that low frequency waves ($w\ll M$) are reflected  while high frequency modes ($w\gtrsim M$) are transmitted through the skin.

\section{The soft oscillon branch}
\label{sec:softosc}
\begin{figure}[h]
    \centering
    \includegraphics[width=0.8\linewidth]{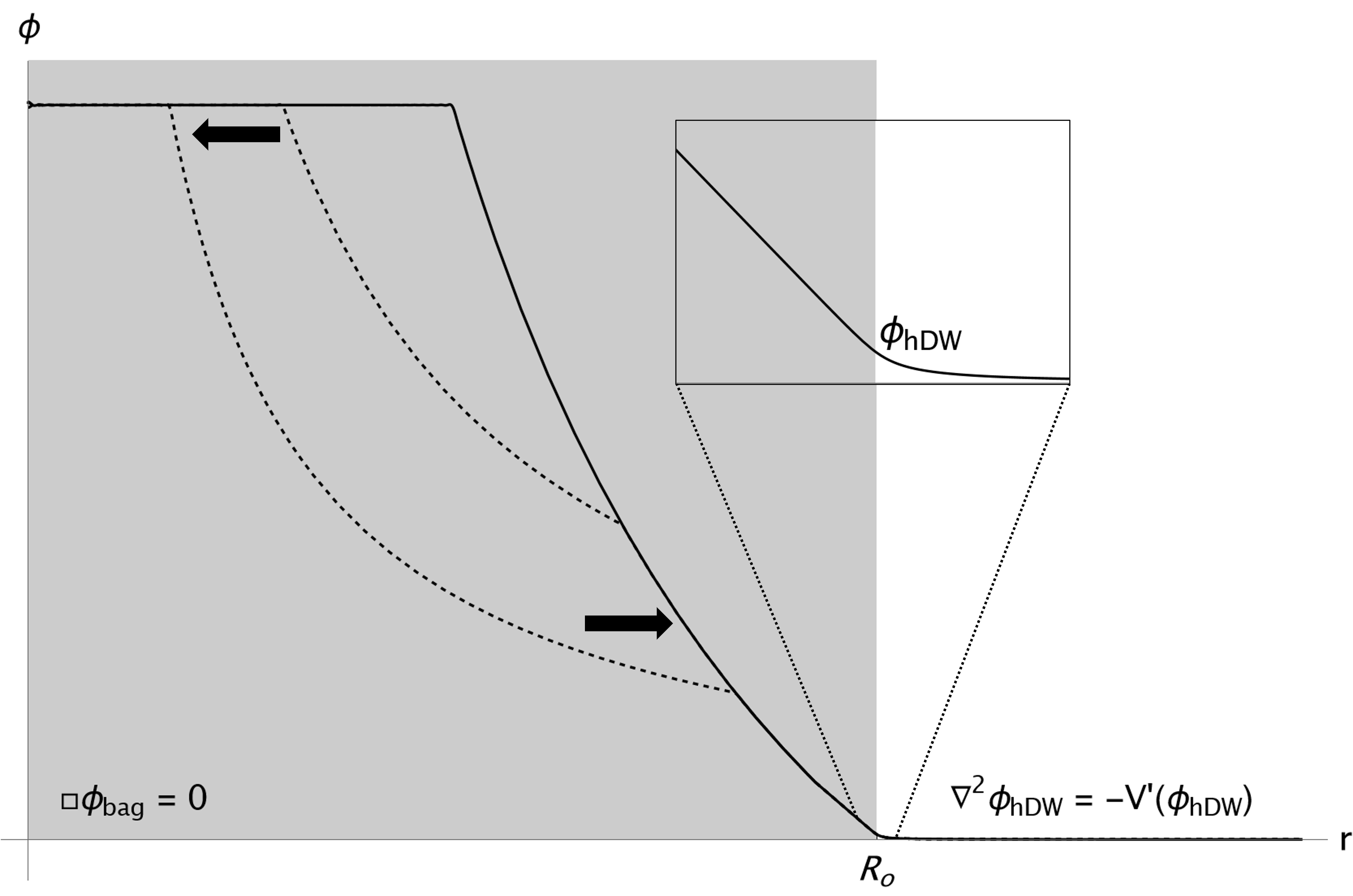}
    \caption{The soft oscillons can be thought of as a matching of two solutions of the equations of motion. At large field amplitudes $\phi \gg \Lambda$, the field is approximately free and the bag-model solution is exact (grey region). The bag-model configurations asymptotes to the half-DW solution that acts as a confining static 'skin' that matches the $\phi \gg \Lambda$ domain (grey) to the vacuum $\phi \approx 0$ domain (white). What's unique about the bag-model is the fact that the boundary region remains static as the outgoing front travels to $R_o$ (dashed lines). This ansatz thus solves the equations of motion exactly until $t < R_o/2$.}
    \label{fig:softoscsketch}
\end{figure}

Based on the considerations outlined in the previous sections, we have a general recipe to construct potentially long-lived oscillons of arbitrary size $R_o$ (and equivalently large amplitude $\phi_0$) in real scalar field theories with a plateau potential. Schematically, this novel ansatz can be understood and constructed based on the following steps:
\begin{enumerate}
    \item Construct the half-DW of the theory explicitly by looking for static solutions of the equations of motion (this can be done using the Bogomolny equations typically used when looking for Domain Wall solutions). The half-DW asymptotes to $\phi_{hDW}(z) \propto z$ as $z \rightarrow \infty$. The half-DW will serve as the confining skin of the bag model, cavity solution.
    \item Match the bag model solution, given in Eq.~\ref{phi_bag}, to the half-DW, shifted so that $\phi_{hDW}(R_o) \approx \mathcal{O}(1)$, where $r = R_o$ is the size of the oscillon. This can be done by varying the amplitude $\phi_0$ in Eq.~\ref{phi_bag}, which is a free parameter of the bag model ansatz.
    \item By performing this matching we obtain a piecewise ansatz for the field $\phi(t,r)$
    \begin{equation}
\phi(t, r) = 
\begin{cases} 
      \phi_{bag}(t,r) & r< R_o \\
      \phi_{hDW}\left(-(r- R_o)\right) &r\geq R_o \\
   \end{cases}
   \label{eq:softoscansatz}
\end{equation}
\end{enumerate}
Since close to $r = R_o$ the bag model solution stays static (like the half-DW solution), this ansatz is almost exact until $t \sim R_o/2$, at which point the outgoing front of the bag model reaches the static skin. A sketch of the setup is shown in Fig.~\ref{fig:softoscsketch}.
A short remark concerning nomenclature is in order. In this work we use the following terms to refer to various concepts
\begin{enumerate}
    \item \textit{Cavity} or \textit{bulk}: the inner region of the field, with $r < R_o$. This is the region where the field is effectively free for most of its evolution.
    \item \textit{Skin} or \textit{half-DW}: the outer region, with $r \geq R_o$. The field in this region acts as a confining potential for the free inner region.
    \item \textit{Cycle}: A cycle is defined by the time in which the ingoing and outgoing fronts of the bag model travel a distance of $\Delta r =  R_o$, flipping the overall sign of the field in the process. An entire period thus contains two cycles.
\end{enumerate}

For notational convenience, we also define:
\begin{equation}
    \mu \equiv \sqrt{V''_{max}(\phi)}
\end{equation}
Where $V''_{max}(\phi)$ is the maximum value of the effective mass-squared for the potential under consideration. This sets the scale of the highest frequency mode trapped in the cavity. The cavity and skin correspond to the gray and white regions in Fig.~\ref{fig:softoscsketch} respectively. As we'll see this way of thinking about soft oscillons is an adequate representation of the physics at play. \\

Whenever $t\mod{R_o/2} = odd$, the field value of this ansatz is $O(\Lambda)$ everywhere. The free cavity and hDW skin picture are no longer adequate descriptions of the configuration as nonlinearities of the potential are felt in the cavity. To understand what the effects of those nonlinearities are, one needs to understand how the outgoing front scatters off the static boundary, and the whole configuration flips sign. This process is best studied using numerical simulation. We will offer an analytic understanding of the decay processes of these soft oscillons in Sec.~\ref{sec:evaporation}. In this section we will show what happens when we dynamically evolve our soft oscillon ansatz in a typical plateau potential.

\subsection{Numerical solutions}
\begin{figure}[t]
    \centering
    \includegraphics[width=0.9\linewidth]{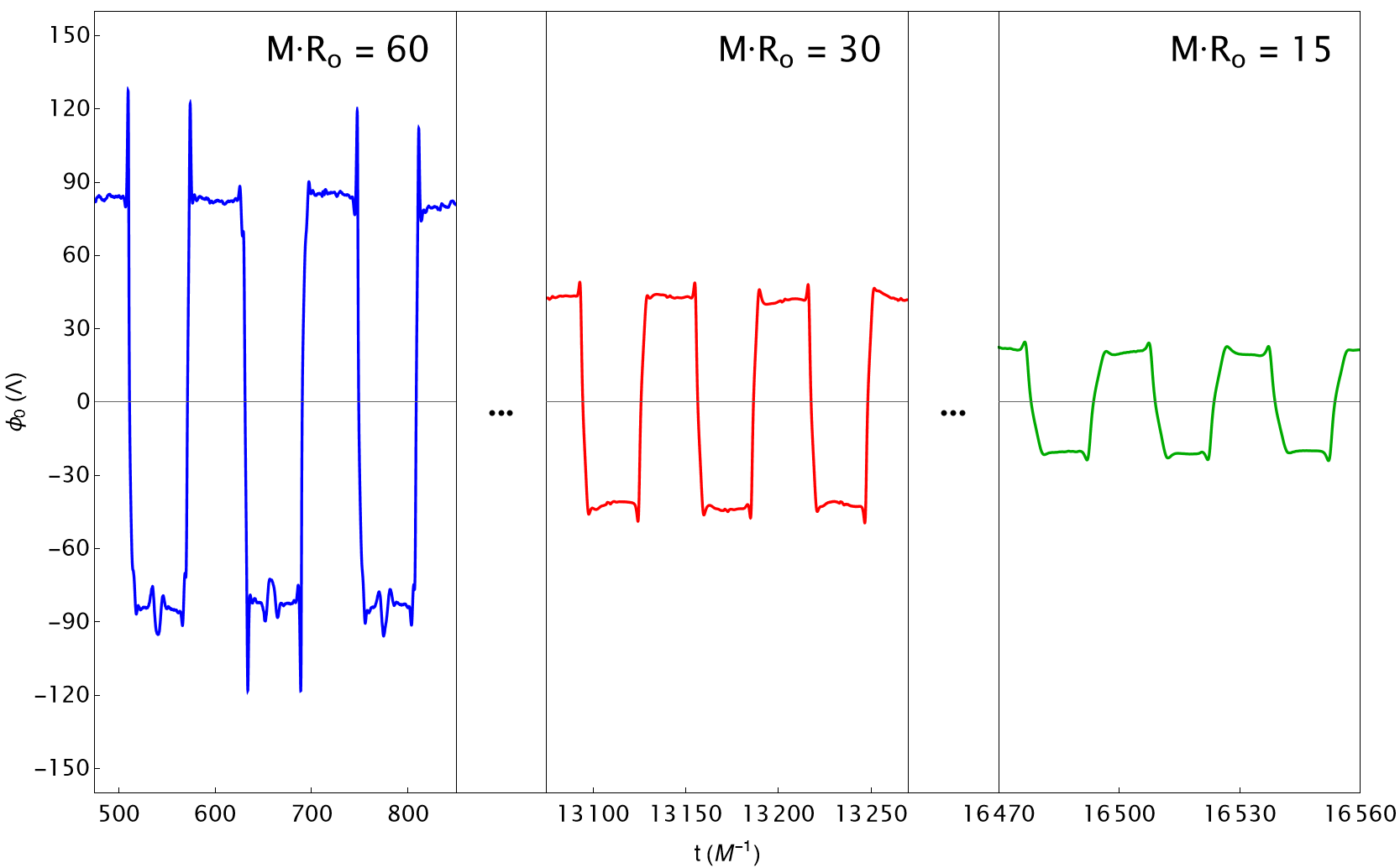}
    \caption{The evolution of the field at the origin $r = 0$, written as $\phi_0 = \phi(t, 0)$. The field shows a clear oscillatory pattern, similar to standard oscillons, but with a much higher harmonic content. The pattern is shown for a configuration of size $M R_o = 60$ (blue), $M R_o = 30$ (red) and $M R_o = 15$ (green). The triple dots are show to indicate that these configuration adiabatically transitions through different sizes. As the oscillon shrinks, its amplitude and period of oscillation shrink as well.}
    \label{fig:ampnumerics}
\end{figure}

\begin{figure}[h]
    \centering
    \includegraphics[width=0.9\linewidth]{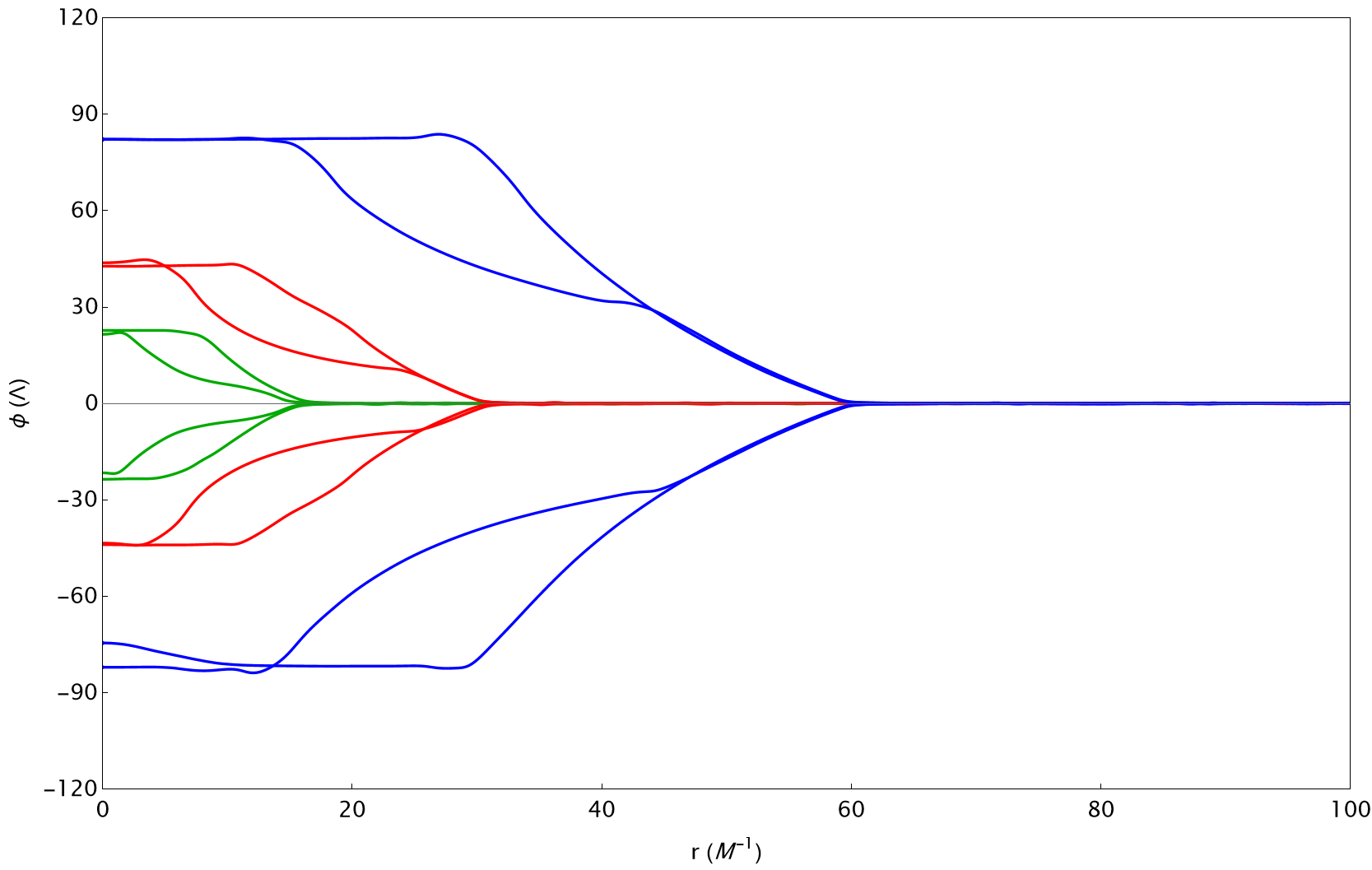}
    \caption{The spatial profiles of soft oscillons of three distinct sizes, (the same color coding as in Fig.~\ref{fig:ampnumerics}), taken at different fraction of each respective period $T = 2 R_o$; namely, $t \approx 0$, $t \approx R_o/4$, $t \approx 3 R_o/4$ and $t\approx R_o$. Even as the soft oscillon shrinks in size, it's oscillation retains remarkable regularity.}
    \label{fig:spatialnumerics}
\end{figure}

We will focus the study of the dynamical evolution of the soft oscillon ansatz on a massless scalar field with plateau potential of the explicit form:
\beq
V(\phi) = M^2 \Lambda^2 \frac{\left(\phi/\Lambda\right)^4}{1 + \left(\phi/\Lambda\right)^4},
\eeq
leaving discussion concerning other potentials to Sec.~\ref{sec:potcomparisons}. 
To obtain our results we solve the equation of motion of the scalar field
\beq
\Ddot{\phi} - \nabla^2 \phi + V'(\phi) = 0,
\label{eq:scalareom}
\eeq
using finite difference methods. Additionally we impose spherical symmetry, so that $\nabla^2 \rightarrow \frac{1}{r^2}\partial_r r^2 \partial_r$. The importance of aspherical modes will be discussed in Sec.~\ref{sec:anisotropies}. Finally, we impose a Neumann boundary condition at $r = 0$, namely $\partial_r \phi(t, 0) = 0$, and a radiative boundary condition at $r \rightarrow \infty$, namely $\dot{\phi}(t, r) = \frac{1}{r} \phi(t,r) + \partial_r \phi(t,r)$. We initialize the field as in eq.~\eqref{eq:softoscansatz} at $t = 0$ (so at maximum amplitude with zero velocity).\footnote{Note that initializing with just the bag model ansatz, setting $\phi(0,r) = 0$ for $r \geq R_o$ does not make a big difference for our results.}
In Figs.~\ref{fig:ampnumerics} and \ref{fig:spatialnumerics}, we show the spatio-temporal evolution of a configuration of initial size $MR_o = 60$.  Initializing at this size, the oscillon is able to adiabatically transition through \textit{soft} configurations of ever smaller sizes. The profiles of different colors in Figs.~\ref{fig:ampnumerics} and \ref{fig:spatialnumerics} are thus obtained from the same initialization.\\

\begin{figure}[h]
    \centering
    \includegraphics[width=1\linewidth]{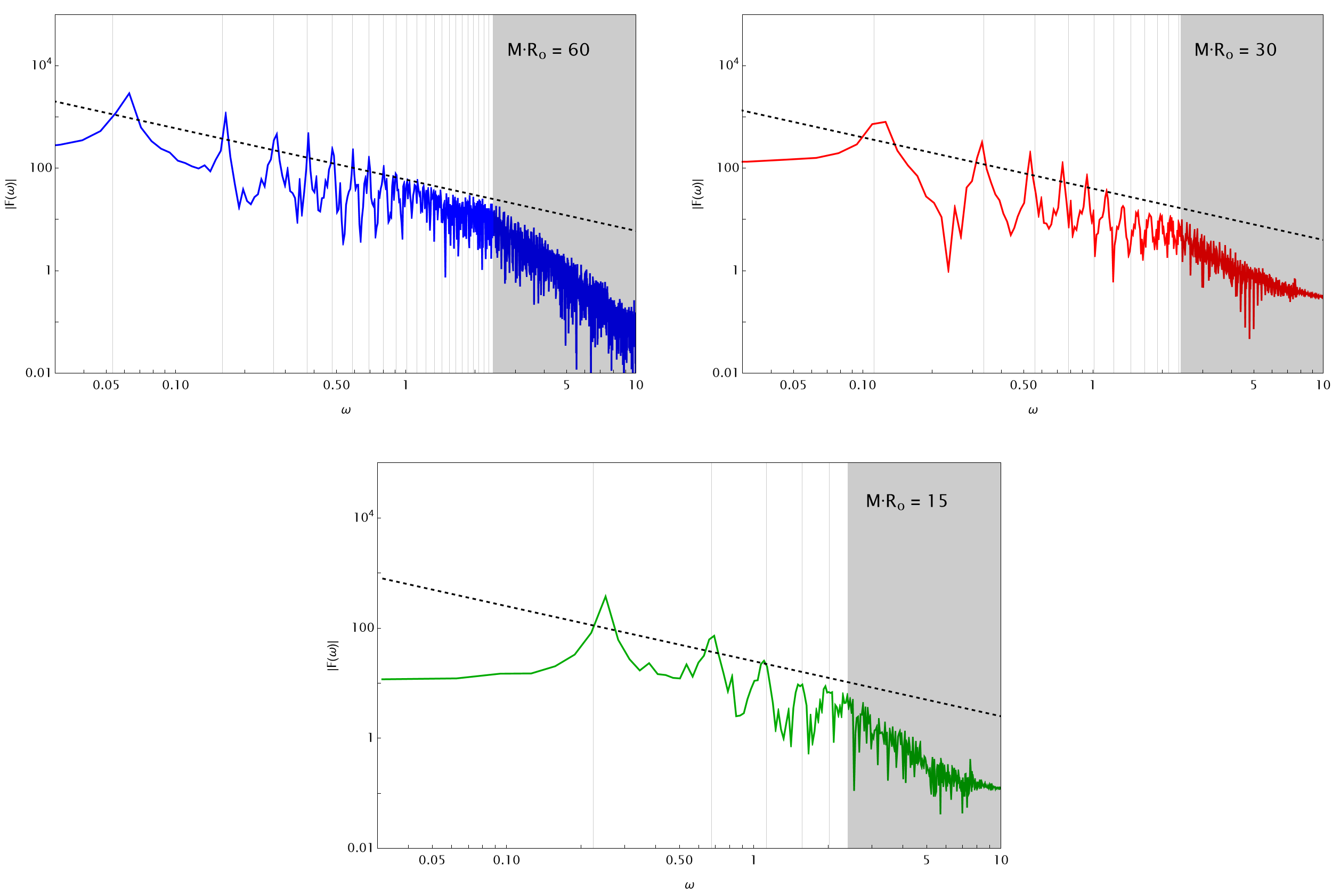}
    \caption{The Fourier decomposition of the time series $\phi_0$, of the field oscillation at the origin, for an oscillon of different sizes (color coding as in Fig.~\ref{fig:ampnumerics}). The peaks corresponding to the fundamental cavity mode $\omega_0 = \pi/R_0$ and its overtones $\omega_n = n \pi/R_0$ are clearly visible and shown by the vertical gridlines. They follow the power law $\propto1/\omega$ predicted by the bag model (dashed lines). When $\omega \gtrsim  \mu$, the half-DW `skin' of the soft oscillon, no longer traps any of the modes, resulting in the broken power law that is observed.}
    \label{fig:spectrumcenter}
\end{figure}

It is interesting to investigate whether the spectrum of the signal at $r = 0$, namely $\phi_0 \equiv \phi(t,0)$ in Fig.~\ref{fig:ampnumerics}, follows the power law scaling of the bag model. Performing a Fourier transform of the signals shown in Fig.~\ref{fig:ampnumerics}, one obtains the results shown in Fig.~\ref{fig:spectrumcenter}.

Fig.~\ref{fig:spectrumcenter} shows that the frequency modes at the origin follow a broken power law. This type of spectrum can be understood within the cavity and skin picture of Sec.~\ref{sec:bagmodel}. Namely, modes that have frequency below the height of the cavity $\mu$, with here $\mu \approx 1.5 M$, are trapped. Their amplitude follows a power law $\phi_\omega \propto 1/\omega$, shown by the black dashed line in Fig.~\ref{fig:spectrumcenter}. This agrees with the scaling of the bag model. Moreover, the spectrum should be peaked at the cavity eigenmodes, given by $\omega_n = n \pi/R_o$, with $n = 1, 3, 5, 7...$. These eigen frequencies are clearly visible in the spectrum, shown by vertical gridlines. High frequency modes, with $\omega >\mu$, are able to tunnel through the hDW skin. The region with $\omega > \mu$ is shown in gray in Fig.~\ref{fig:spectrumcenter}. These modes are certainly transmitted through the cavity skin, and the relatively fast decay beyond $\omega>\mu$ confirms (and refines) the expectations drawn in Sec.~\ref{sec:bagmodel}.\\

Our numerics confirm that the ansatz of Eq.~\ref{eq:softoscansatz} has considerable longevity. In particular, the solution survives past the nonlinear scattering event that takes place at $t = R_o/2$. Also, our understanding of these soft oscillons as a collection of spherical modes within a free cavity, trapped by the hDW skin is adequate. The field transitions through soft oscillons of different sizes as it loses energy. Equation \eqref{eq:softoscansatz} does not explain how the configurations loses energy through nonlinear interactions, nor how the global evolution of the parameters (such as its size $R_o$) proceeds. We focus on these questions next. 

\subsection{Radiation and global evolution}
\begin{figure}[h!]
    \centering
    \includegraphics[width=\textwidth]{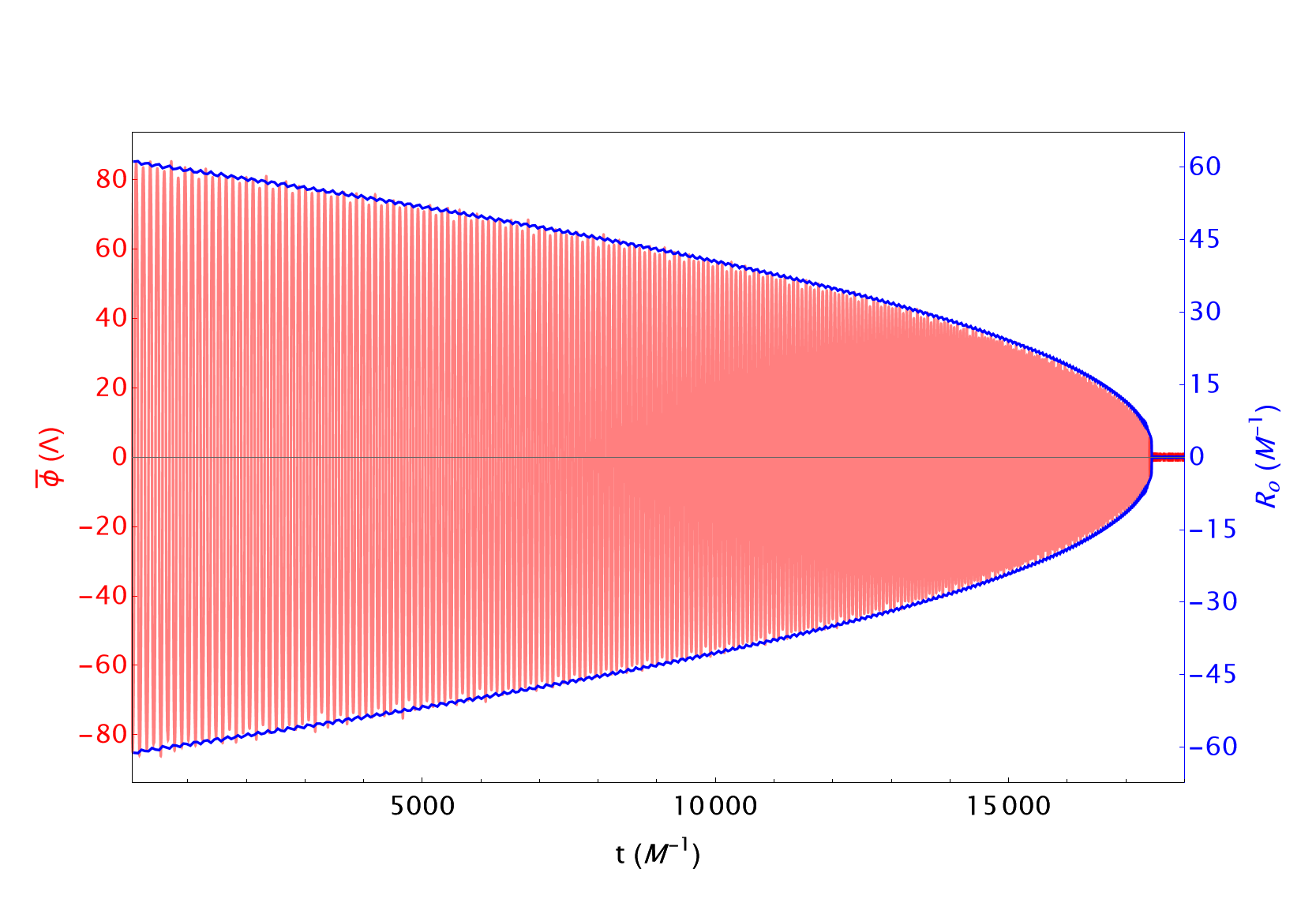}
    \caption{The evolution of the average bulk amplitude of the field $\bar{\phi}$ (red, left axis) and the size of the configuration (blue, right axis) over time. $\bar{\phi}$ oscillates rapidly and becomes almost indiscernible as the soft oscillon (and its period) shrinks. The size of the oscillon acts as a perfect envelope of the amplitude, as the bag model predicts $\phi_0 \propto R_o$. The proportionality factor here is $\approx \sqrt{2}$. Note that the $R_o < 0$ line here is just a visual guide and is simply obtained by mirroring the physical $R_o > 0$ measurements.}
    \label{fig:Roevolve}
\end{figure}

As discussed, the cavity and skin picture breaks down when $t \mod R_o/2 \approx m$, where $m$ is an odd integer. On the one hand because $V'(\phi) \neq 0$ within the bulk, resulting in an extra source term within the cavity. Moreover, the boundary of the bag model solution (see Fig.~\ref{fig:softoscsketch}) is no longer static at this point in time, as the skin flips sign. We call this the nonlinear scattering event, where part of the outgoing wave is transmitted through the hDW, losing a percentage of its energy and flipping the overall sign of the solution. The soft oscillons thus lose energy in bursts of radiation, separated by $\Delta t = R_o$. These bursts are clearly visible at large distances $r > R_o$. Notice that oscillons emitting radiation bursts have also been found in $1+1$ dimensions \cite{Adam:2019prh,GarciaMartin-Caro:2025zkc}. We will provide a better understanding of this evaporation mechanism in Sec.~\ref{sec:evaporation}. Since the solution pulsates energy at a rate proportional to $R_o^{-1}$, and the energy of the soft oscillon scales with $E\propto R_o^3$, the configuration necessarily has to readjust its size and therefore its amplitude in order to conserve energy. In Fig.~\ref{fig:Roevolve}, we plot the temporal evolution of the size $R_o$ and bulk average amplitude $\bar{\phi}$, the global parameters that characterize the soft oscillon. We define the size $R_o$ of the oscillon as the radius $r$ where the energy density $\rho(t, r) = \frac{1}{2} \dot{\phi}^2  + \frac{1}{2}(\nabla\phi)^2 + V(\phi)$ is equal to $\rho(t, R_o) \equiv 0.1 \Lambda^2 M^2$. The bulk average amplitude $\bar{\phi}$ is then computed by averaging the field amplitude within the region $r \leq R_o/2$.
Fig.~\ref{fig:Roevolve} confirms that the oscillon is able to adiabatically change its size and amplitude as it pulsates away energy. The configuration transitions through a set of soft oscillons of different sizes and amplitude. The size of the solution is a perfect envelope of the oscillating bulk amplitude, confirming the understanding that there is a linear relation between $R_o$ and the field amplitude $\phi_0$ in the bag model picture. We measure $\bar{\phi} \approx \sqrt{2} (M R_o) \Lambda$, which agrees with our analytic expectation of Eq. ~\eqref{eq:bagampcent}. It is remarkable that an initial configuration of size $R_o = 60 M^{-1}$ is able to survive for times that are much larger than its light crossing time $t_c = 2 R_o$. In fact, the decay around $t = 1.7 \cdot 10^4 M^{-1}$ happens after about $140 t_c$. Note that the evolution of the size of the oscillon $R_o(t)$ is approximately given by $R_o(t) = \sqrt{R_o(0)^2 - C t}$, with $C \approx 0.22$.\\

\begin{figure}[h]
    \centering
    \includegraphics[width=0.7\linewidth]{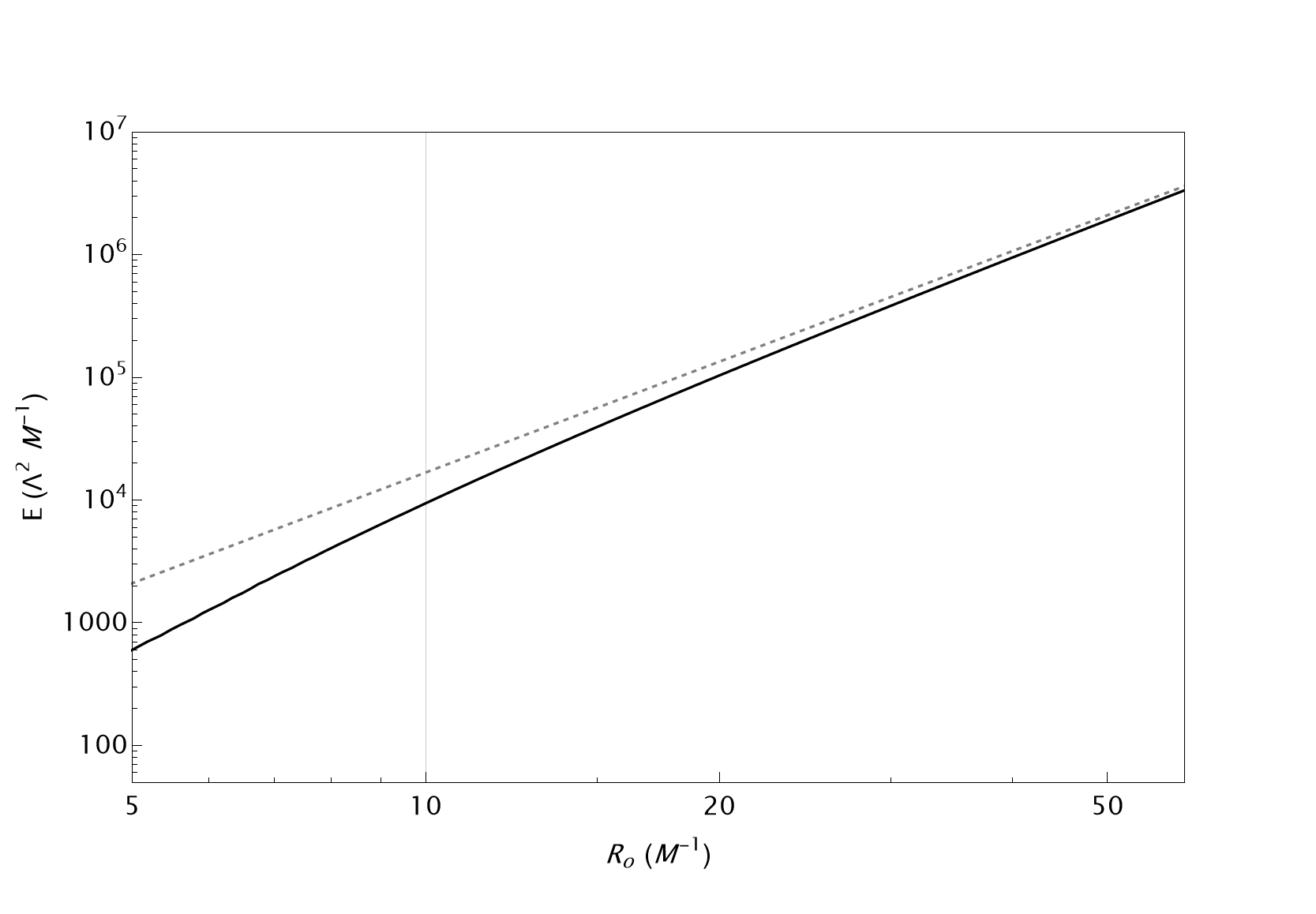}
    \caption{The energy contained within the soft oscillon configuration versus its size. The scaling of Eq.~\ref{bagE} (shown with the dashed line) is accurate until only a few modes are trapped by the half-DW skin, when $(MR_o)/\pi \sim \mathcal{O}(1)$, indicated by the vertical gridline. Here the bag model approximation is expected to break down.}
    \label{fig:Eevolve}
\end{figure}

We can obtain an estimate for the decay rate of these objects by simultaneously measuring how the energy of the object depends on $R_o$. Note that eq.~\ref{bagE}, suggests a volume scaling between the two. In Fig.~\ref{fig:Eevolve} we confirm this scaling. Only when the amount of modes trapped within the cavity becomes $\mathcal{O}(1)$, the approximation breaks down.
As expected, the energy scales with the volume of the configuration, $E\propto R_o^3$, until the number of modes in the cavity becomes $\mathcal{O}(1)$.\\

\begin{figure}[h!]
    \centering
    \includegraphics[width=1.1\linewidth]{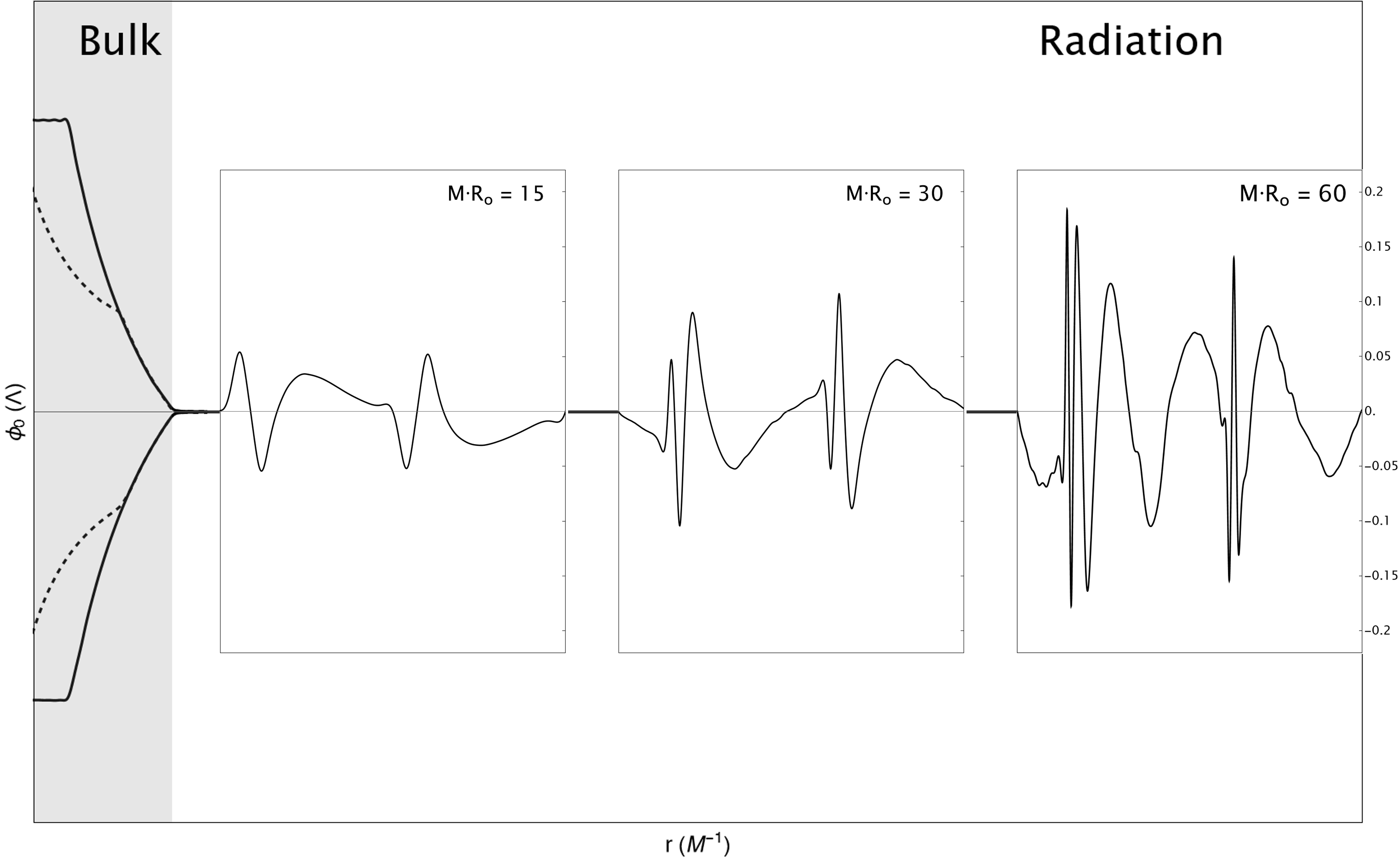}
    \caption{A schematic representation of how the soft oscillon loses energy. The oscillating bulk (right) emits pulses at regular intervals depending on the size of the configuration, $\Delta t = R_o$. The wavepackets travel to spatial infinity $r\to\infty$, where they can be measured by an observer. As the bulk shrinks the waveforms of the packets change as well, changing their frequency content. We show here the numerical wavepackets emitted by an oscillon of three different sizes. As the oscillon shrinks, the packets originating from the largest configurations have travelled the farthest.}
    \label{fig:waveforms}
\end{figure}

The evolution of the soft oscillon thus proceeds by a pulsating ejection of energy through wavepackets that travel to $r \to \infty$. The shape of the ejected wavepacket depends on the size of the oscillon at the time of ejection. As the configuration adiabatically changes size, different shapes of pulses can be found at various distances from the bulk. A schematic representation of this situation is shown in Fig.~\ref{fig:waveforms}, where the exact waveforms are taken from our numerics. Note that the distance between the packets is not to scale, as much more time is spent in the larger configurations. The separation between each wavepacket is equal to $r = R_o$, as the time between pulses is given by $\Delta t_{pulse} = R_o$, and the wavepackets move at the speed of light which we have set to $c \equiv 1$.\\

\begin{figure}[h!]
    \centering
    \includegraphics[width=1\linewidth]{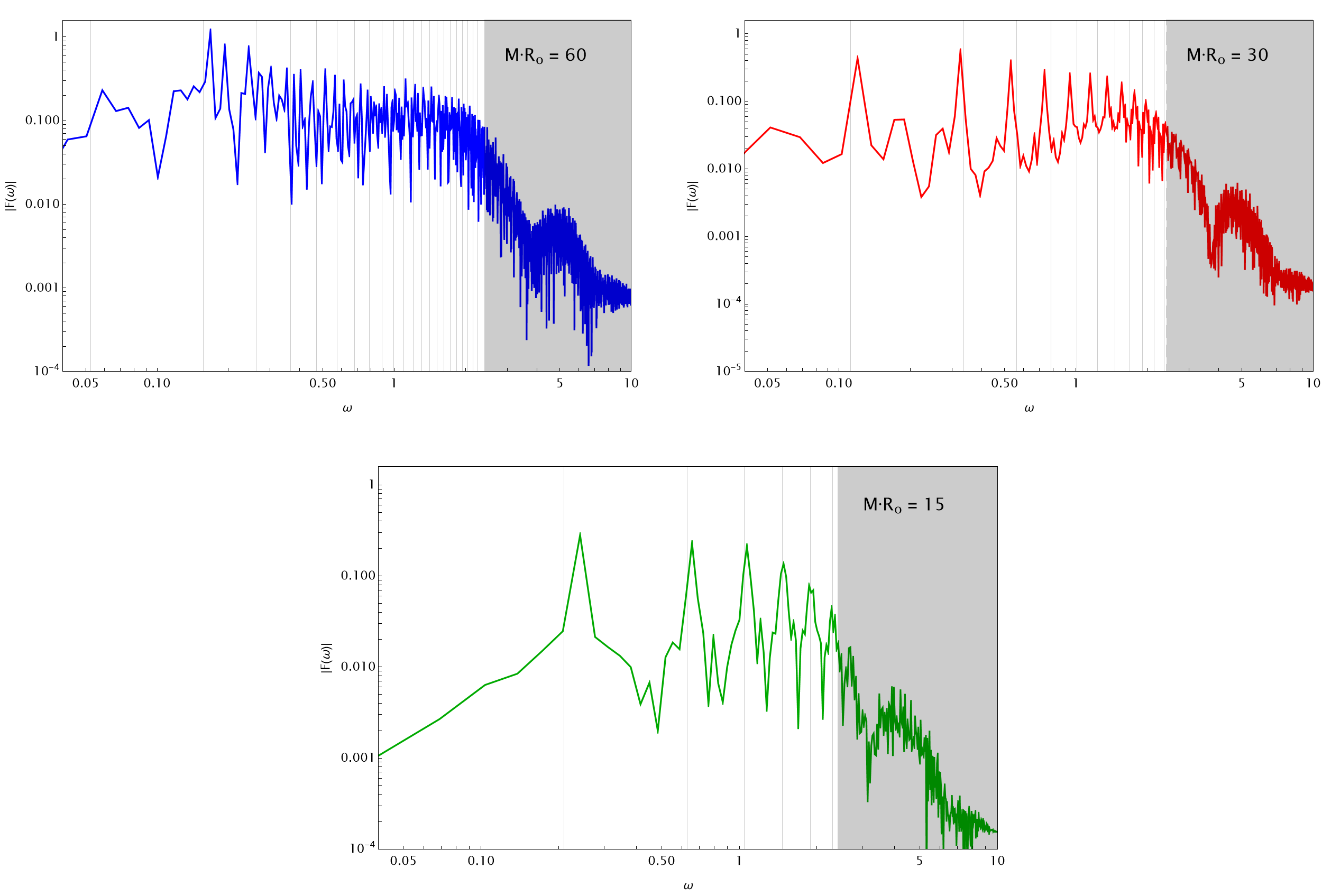}
    \caption{The Fourier transforms of the waveforms shown in Fig.~\ref{fig:waveforms}. As for the bulk, the spectrum clearly shows the spikes associated to the cavity modes. However, the spectrum is now approximately flat. This can be understood intuitively by noting that, although the lower frequency modes have a higher occupancy in the cavity, they are also transmitted to a lesser degree by the `skin'.}
    \label{fig:specwaveforms}
\end{figure}

These waveforms carry information about the 
properties of the soft oscillon to spatial infinity. The time delay between the pulses is a direct measure of the size $R_o$ of the oscillon. Moreover, each emitted packet has a unique frequency composition. The spectrum of the three waveforms, shown in Fig.~\ref{fig:waveforms}, are shown in Fig.~{\ref{fig:specwaveforms}}. To be precise, the spectra are computed by Fourier transforming five subsequent pulses of the type shown in Fig.~\ref{fig:waveforms}. This gives a cleaner result.\\

Contrary to the bulk modes that follow the $1/n$ spectrum of the bag model, the spectrum of emitted pulses is nearly flat. The spikes of the bulk modes are still clearly visible. However, as high frequency modes are transmitted more easily through the skin at $r = R_o$, the result is a flat overall spectrum in the emitted wave. Again, for $\omega \geq \mu$, the spectrum decays, as happens in the bulk. This indicates that the adiabatic evolution of the soft oscillon does not lead to a significant hardening of the cavity modes. \\

As the size $R_o$ increases (see upper left panel of Fig.~\ref{fig:specwaveforms}), the radiation spectrum becomes less flat, and the spikes associated to the bulk modes are less visible. For soft oscillons whose size $R_o$ is larger than some critical value $R_c$, we observe an instability that causes a change in the way the soft oscillon evolves. In sec.~\ref{sec:evaporation} we will argue that this instability is tied to a `dirtying' of the spectrum of bulk modes and therefore the radiated waveforms, which starts to become visible in Fig.~\ref{fig:specwaveforms}. The evolution we have described thus far is tied to a `clean' spectrum in the bulk, associated to the bag model. At large size, the spectrum in the bulk becomes `dirty' and we enter a different regime of evolution.

\subsection{Clean and dirty regimes}
The results of the previous section show that at least some soft oscillons are able to pulsate away energy without significantly changing the relative amplitude of modes contained within the cavity. These oscillons simply transition through bag model solutions of ever decreasing size within the bulk. If we choose to initialize a soft oscillon with a smaller size, the global evolution of parameters as shown in Fig.~\ref{fig:Roevolve} is simply shifted accordingly. We characterize this regime of evolution as the `clean regime'.\\ 

As mentioned previously the clean regime does not persist to arbitrary size $R_o$. In fact, there seems to be a critical radius, which we call $R_c$, above which a soft oscillon is unable to sustain the particular cavity mode composition required for the clean regime. Starting with the ansatz of eq.~\eqref{eq:softoscansatz} with $R_o > R_c$, the global evolution of parameters initially follows that of the clean regime. However, after a number of periods of the configuration, nonlinearities result in a significant deviation in the bulk with respect to the bag model. We colloquially call this moment in time the classical breaktime $t_{br}$ of the soft oscillon ansatz. This deviation from the bag model in the bulk results in a different evolution of the global parameters of the solution. We refer to this altered behavior as the dirty regime. The critical radius $R_c$ can be understood as the maximum radius for which a soft oscillon is able to stay in the clean regime despite the nonlinear scattering event taking place twice every period. In sec.~\ref{sec:nse} we will show why $R_c$ must necessarily exist for any potential although its precise value is model dependent.\\

\begin{figure}[t]
    \centering
    \includegraphics[width=1\textwidth]{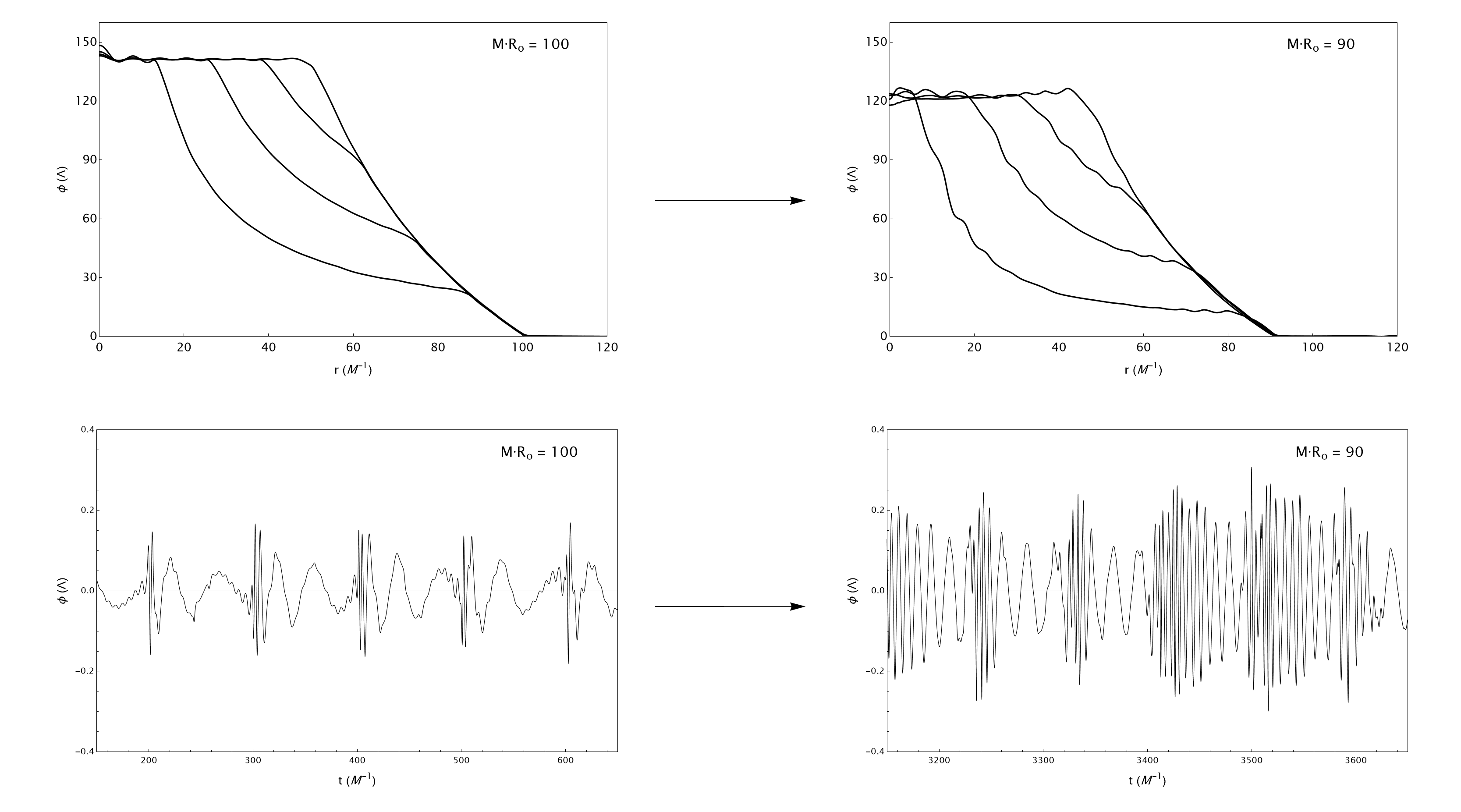}
    \caption{Above the critical radius, a soft oscillon only evolves in an ordered manner for a limited amount of time. This classical `breaking' of the solution is visible at a qualitative level in the spatial profile (top row) and the radiated pulses (bottom row). Schematically, the evolution proceeds in a more perturbed and disorganized manner. The spatial profile becomes more `wobbly', the radiated pulses are less periodic and localized, and the peaks in the spectrum characteristic of the bag model become less visible.}
    \label{fig:criticalgrid}
\end{figure}

The `dirtying' of the configuration is visible at a qualitative level in the field profile. In Fig.~\ref{fig:criticalgrid} we show the difference in the bulk and radiated pulses, before and after the classical breaktime. The field shows clear wobbles with respect to the bag model solution. Moreover, the radiated pulses are less regular and more erratic.\\

\begin{figure}[h!]
    \centering
    \includegraphics[width=1\textwidth]{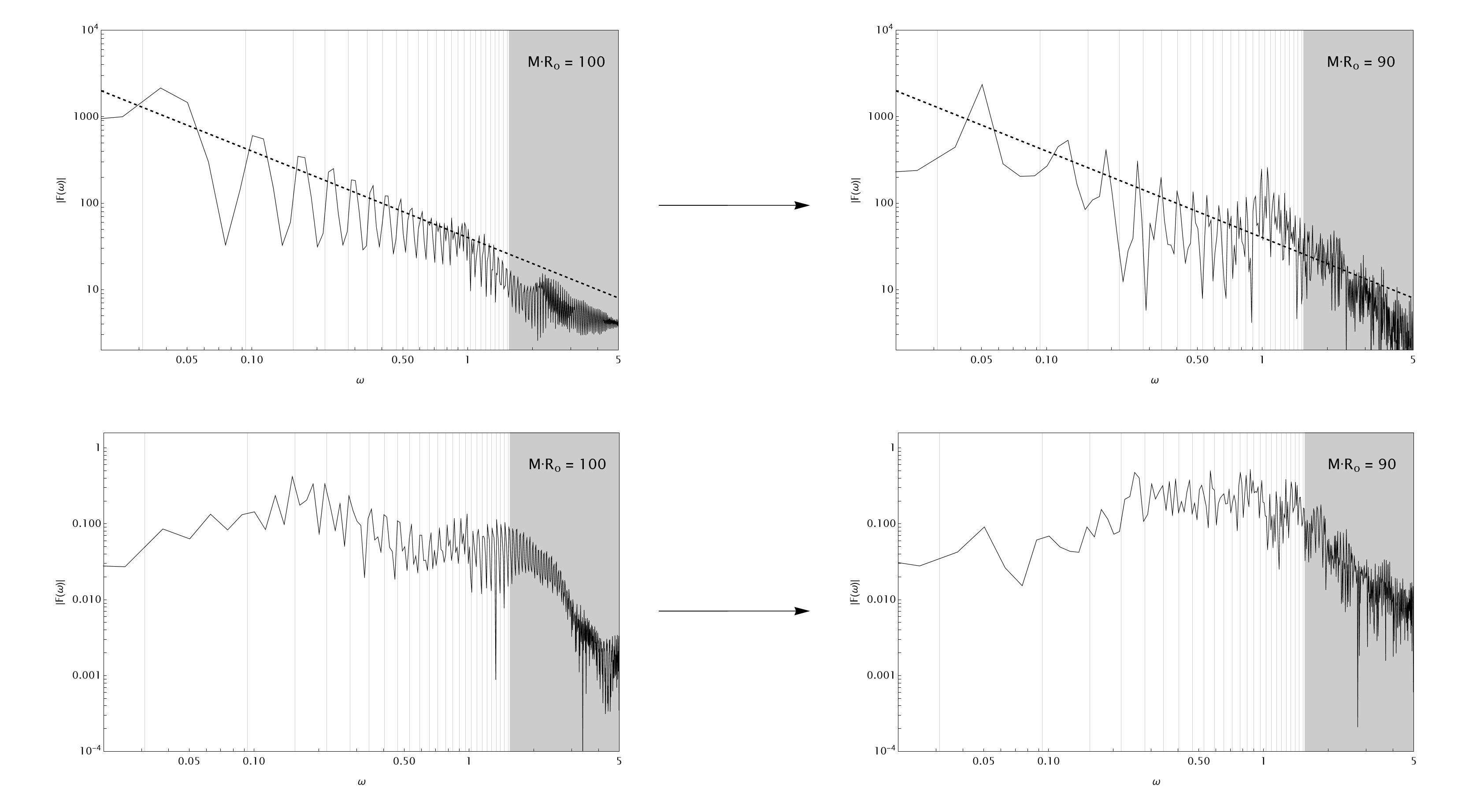}
    \caption{The breaking of the solution is also clearly visible in the spectrum of the field at the origin $\phi_0$ (top row) and the spectrum of outgoing radiation (bottom row). The modes contained in the bulk get blue shifted with respect to the $1/n$ dependence associated to the bag model. Moreover, the spectrum of radiation is less regular.}
    \label{fig:criticalgridsec}
\end{figure}

Beside the qualitative differences in the field profile, it is interesting to check how the spectra of cavity and radiation modes change in the dirty regime. In Fig.~\ref{fig:criticalgridsec} we show how breaking can be interpreted as a deviation from the bag model ansatz in the bulk (top row). In particular, we observe a hardening towards high frequency modes, where the deviation from the $1/n$ dependence is clearly visible. The spectrum of radiation in turn loses the peaks associated to the bag model and shows no clear structure (bottom row). \\

\begin{figure}[t]
    \centering
    \includegraphics[width=0.8\linewidth]{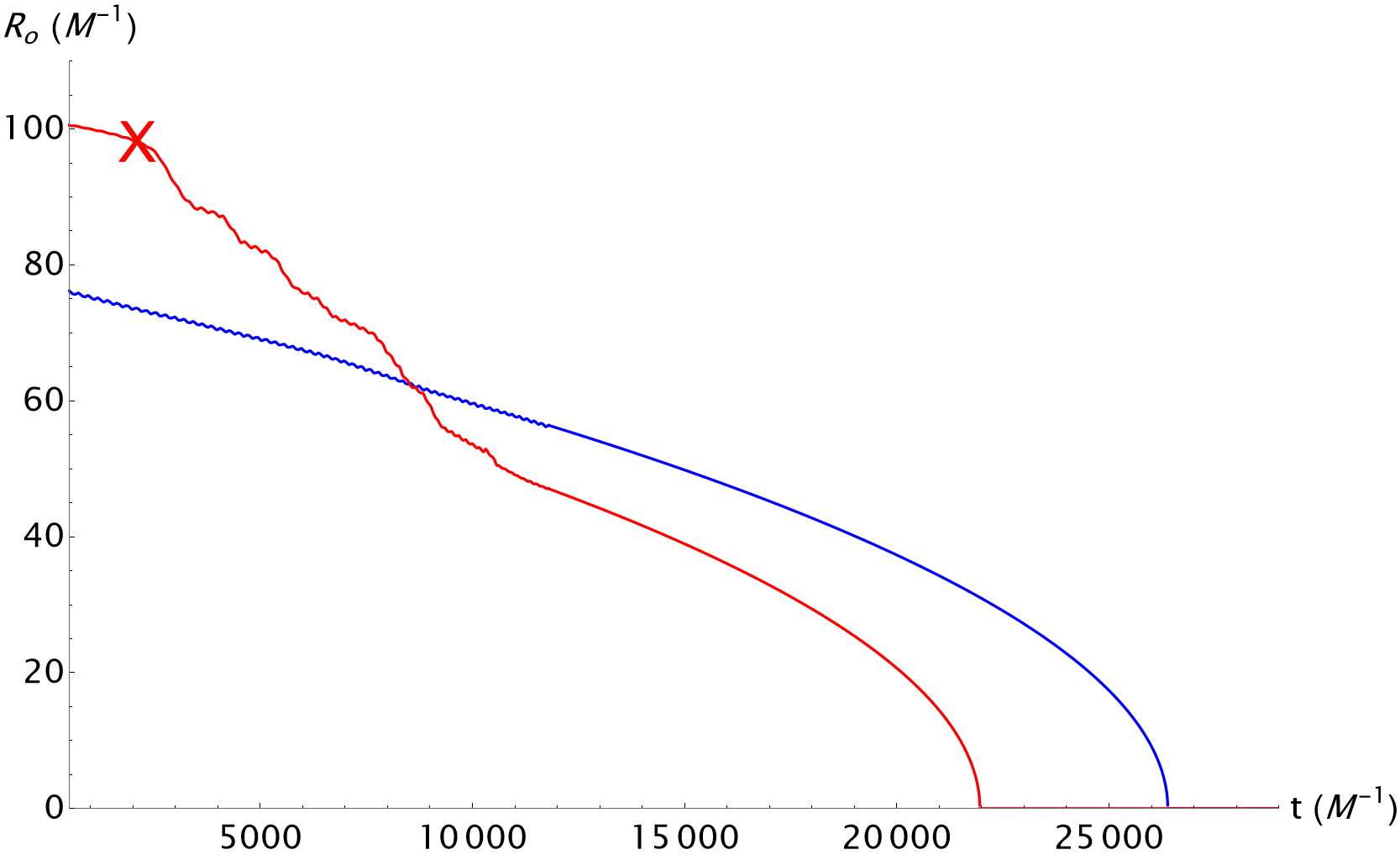}
    \caption{There exists a critical value of the soft oscillon $R_c$, above which the evolution of the size $R_o(t)$, deviates from the previously observed $R_o(t) \approx \sqrt{R_o(0)^2 - C t}$, after a certain amount of cycles. 
    After that the evolution of the size of the oscillon is less predictable, and almost chaotic. This is illustrated by the red curve, representing a soft oscillon with initial size $M R_o(0) = 100$. 
    The classical break time is marked with the red cross and arises after about ten cycles.
    After some time the configuration is attracted to a smaller soft oscillon which isn't sensitive to breaking. If however, the initial oscillon is smaller, no breaking occurs whatsoever (blue curve).}
    \label{fig:criticalradiusbreaking}
\end{figure}

The deviation from the bag model associated to the dirty regime results in a different evolution of the global parameters characterizing the soft oscillon. This is captured by the evolution of its size $R_o(t)$. Although initially $R_o(t) =\sqrt{R_o(0)^2 - Ct}$ the dirtying of the bulk eventually results in an increase of the evaporation rate after the classical breaktime. The size $R_o(t)$ therefore reduces more rapidly after $t_{br}$. In Fig.~\ref{fig:criticalradiusbreaking} we show $R_o(t)$ for oscillons initialized below and above $R_c$ respectively. The blue line that respects $R_o(0) < R_c$ evolves as expected for the clean regime, until its eventual decay. The red line, which has $R_o(0) > R_c$, undergoes classical breaking, indicated by the cross, after which its size decays at an increased rate. When the size of the configuration has reduced below $R_c$, the oscillon is able to reenter the clean regime showing the attractive nature of the soft oscillon ansatz below $R_c$. In sec.~\ref{sec:evaporation} we will provide an analytic understanding of how and why this happens.\\

\begin{figure}[t]
    \centering
    \includegraphics[width=0.8\linewidth]{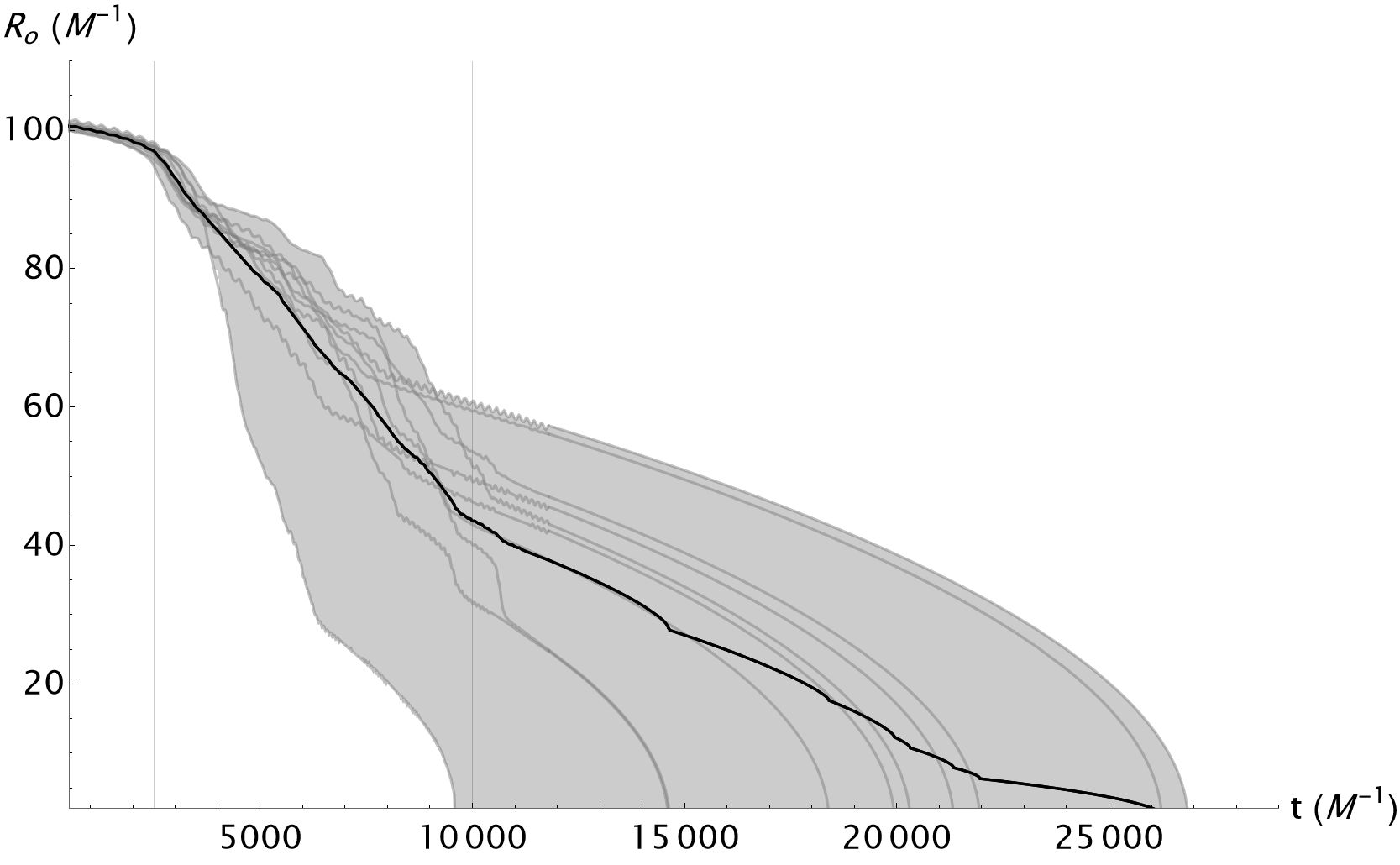}
    \caption{We show the stochastic nature of the classical breaking of the soft oscillons. Each light gray curve represents a different initialization of the soft oscillon with initial size $MR_o(t_0) = 100$, where each initialization is perturbed by a small fluctuation of order $5\%$ of the background soft oscillon ansatz. The black curve represents the mean of each of the individual curves, while the gray region simply demarcates the maximum and minimum values of the simulated curves. After breaking after approximately ten cycles of the soft oscillon, each curve follows its own unique path, until it finds a smaller soft oscillon that is not sensitive to breaking. The mean of all the curves follows an approximate linear decay $R_o(t) \approx R_o(t_{br}) - v t$ for some time after breaking, until most curves are attracted to a `non-breaking' smaller oscillon. The period where the evolution is approximately linear is indicated as the region between the two gridlines.}
    \label{fig:stochasticbreaking}
\end{figure}

Inspecting Fig.~\ref{fig:criticalradiusbreaking}, the evolution of $R_o(t)$ in the dirty regime seems unpredictable, lacking a clear power law description. We confirm that in the dirty regime, the evolution of $R_o(t)$ is highly dependent on the exact distribution of bulk modes. In other words, a small difference in the way in which the cavity deviates from the bag model ansatz, can result in a very different evolution of $R_o(t)$ in the dirty regime. Note that Fig.~\ref{fig:criticalradiusbreaking} is produced by starting from the `prefect' soft oscillon ansatz of \eqref{eq:softoscansatz}. We can study the evolution of $R_o(t)$ by slightly perturbing this initial condition. In Fig.~\ref{fig:stochasticbreaking} we show the stochastic nature of the dirty regime. Each light gray line corresponds to a different initialization of a soft oscillon. We randomly perturb a finite number of the bulk modes by $\mathcal{O}(5\%)$ with respect to the bag model to obtain distinct evolutions of $R_o(t)$. After the classical breaktime, each configuration evolves independently until it is attracted to a soft oscillon smaller than the critical radius. The result is a considerable spread in the evolution of $R_o(t)$ (gray region), showing that the dirty regime is quite sensitive to small differences in the bulk mode population. This large spread is not visible if we perform the same procedure for a soft oscillon below the critical radius, as it never exits the clean regime. \\

Although Fig.~\ref{fig:stochasticbreaking} seems to suggest that it is impossible to predict the evolution of a particular configuration in the dirty regime, the average over all the different initializations (black line) is approximately linear for a long time. In particular, this is true when most of the configurations are in the dirty regime, indicated by the two gridlines. So although we can not make any predictions about a particular initialization, we can say something about the stochastic ensemble. The average global evolution in this regime can be summarized as following a very simple rule, namely that the overall radius decreases at a constant velocity
\begin{equation}\label{Rdot}
    \dot R_o = - v\, t \qquad {\rm with} \quad v\simeq 0.6 \cdot 10^{-2}~.
\end{equation}
Since the energy still scales as $E\propto R_o^3$, we conclude that in the dirty regime the average evaporation rate of the soft oscillon scales with its surface $\dot{E} \propto R_o^2$. We provide an analytic understanding of this rate in sec.~\ref{sec:evaporation}.\\

\begin{figure}[t]
\begin{center}
\includegraphics[width=0.8\textwidth]{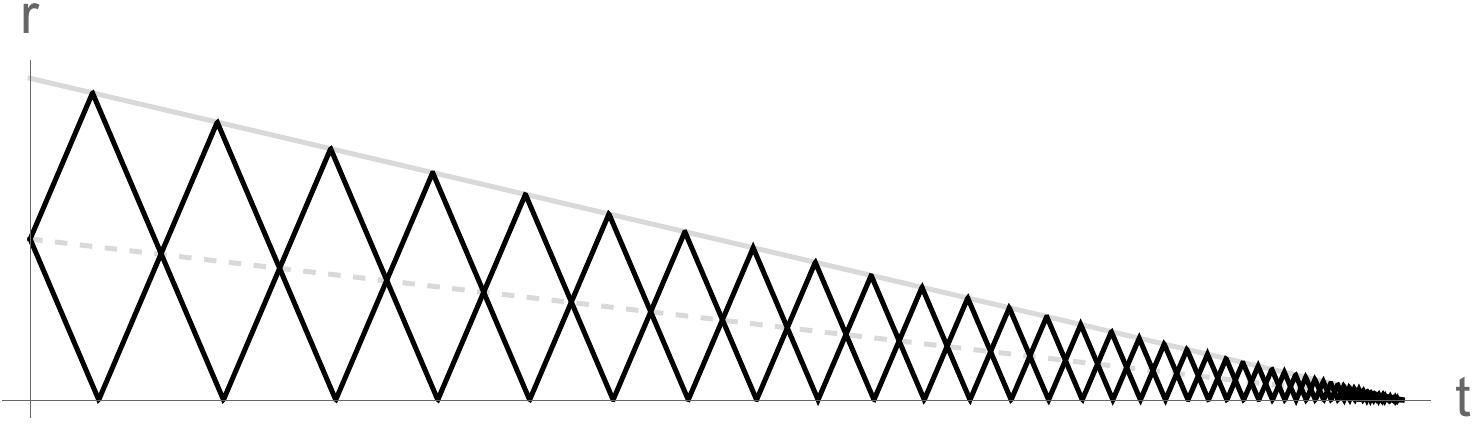}
\caption{Sketch of the evolution, taking place at a constant slow decrease in radius, $\dot R = v\ll1$. Solid black lines represent the in- and out-going fronts in the oscillon core (with fixed slopes). With an evolving oscillon radius $R_o(t)$ the structure of fronts that meet at $R_o(t)/2$ is preserved at all times. }
\label{fig:pingpong}
\end{center}
\end{figure}

A sketch of this evolution \eqref{Rdot} together with the two radiation fronts in the oscillon core is given in Fig.~\ref{fig:pingpong}. Oscillons deep in this regime, that is with initial $R_o\gg R_c$, are expected to have a lifetime
$$
\Delta T \sim \frac{R_o}{v}~,
$$
still significantly larger than the initial light-crossing time.
Moreover, Fig.~\ref{fig:pingpong} is useful to visualize the large number of self-similar bounces that the oscillon goes through in this time. It's easy to see that the overall radius reduces in each cycle by a fixed multiplicative factor $(1-v)/(1+v)$.
The total number of cycles in going from an initial radius $R_i$ to a final radius of order $1/M$ is
\begin{equation}
N_{bounces}\sim \frac{1}{2 v}\,\log\left( M\, R_i \right)~.
\end{equation}
In other words, while the lifetime $\Delta T$ in units of the initial radius $R_i$ is basically a constant, 
the number of cycles over which the evolution is approximately self-similar increases unboundedly with $R_o$.\\

To conclude this section we would like to address questions concerning the general attractiveness of soft oscillons in spherical symmetry. 

\subsection{Attractiveness}
\label{sec:attractiveness}
So far, we have described the properties of soft oscillons in detail. The dynamics of these oscillons can be split into two distinct forms of evolution, which we called the clean and dirty regimes respectively. Our discussion has focused on the properties of configurations initialized with the `perfect' ansatz of eq.~\eqref{eq:softoscansatz}, or a slightly perturbed version of it. This is sufficient to describe the general properties of soft oscillons, which we understand to be the configuration of size $R_o$ that minimizes the evaporation rate of a localized wavepacket of that size. However, it remains unclear to what extent the decay rate of a generic spherical wavepacket of radius $R$ and amplitude $A$ is reduced due to the continuous set of soft oscillons characterized by their size $R_o$. We already showed that a soft oscillon in the dirty regime can reenter the clean regime below the critical radius, hinting at some form of attractor properties below the critical radius.\\

\begin{figure}[h!]
    \centering
    \includegraphics[width=\textwidth]{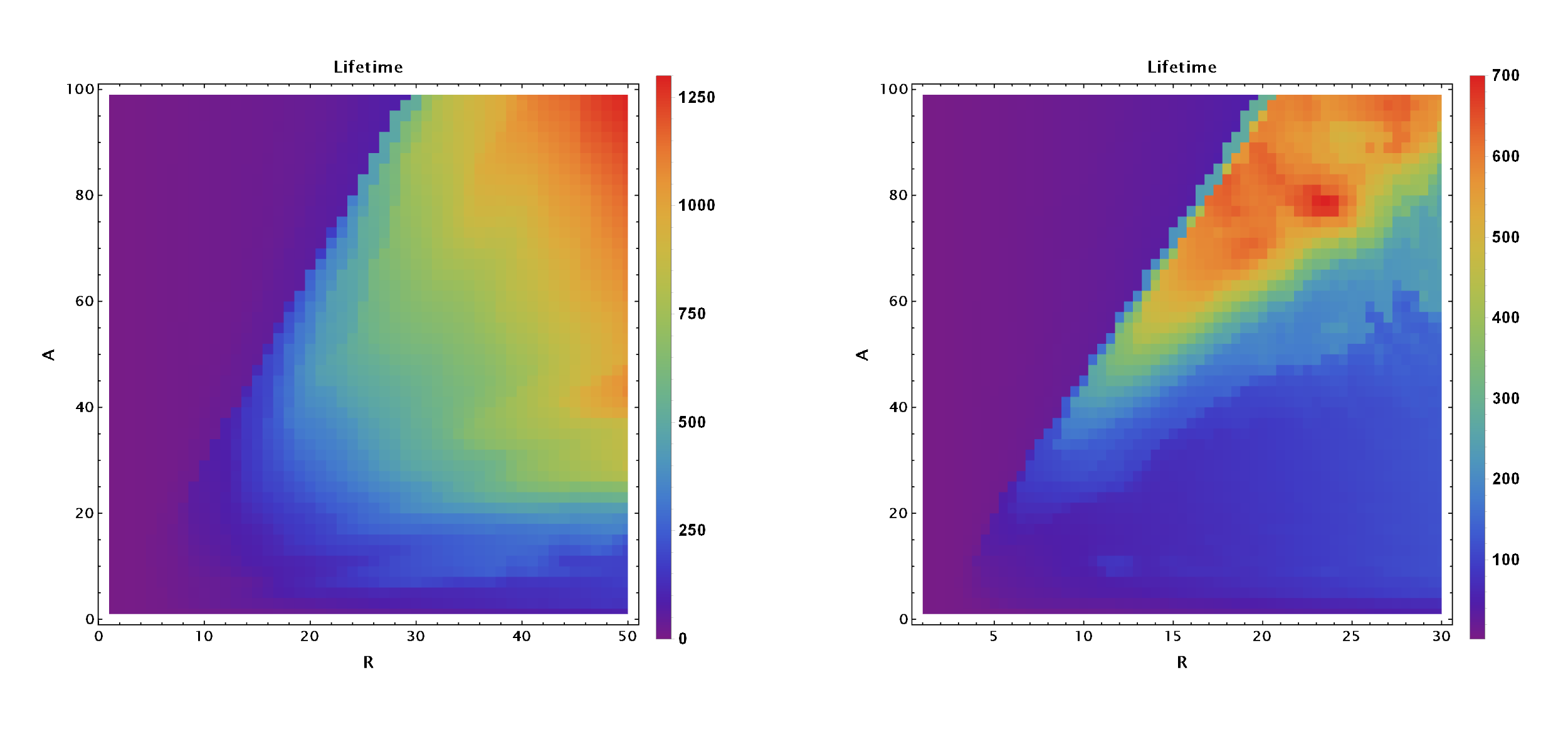}
    \caption{The lifetime defined as the time after which a general wavepacket loses 20 percent of its initial mass. On the left we show the bag model initialization while on the right we use a Gaussian packet, see text for the definitions of $A$ and $R$. The scan clearly shows an island of attractiveness in both cases.}
    \label{fig:attractors}
\end{figure}

To study this question we analyze the decay rate of a localized field configuration of size $R$ and amplitude $A$. We do this for two different parametrizations of the field:
\begin{enumerate}
    \item The bag model initialization of size $R_o = R$ where we now allow $\phi_0 \equiv A$ to be a free parameter so that formally $\phi(0,r) = \frac{A}{\sqrt{2} R}\phi_{bag}(0,r/R) $;
    \item A gaussian packet so that $\phi(0,r) = A e^{-(r/R)^2}$.
\end{enumerate}
For these initializations, we perform a scan over the free parameters $A$ and $R$ and measure the time it takes for the $20\%$ of the initially localized energy to dissipate, a measure of the lifetime of the configuration. We show the results of this scan in Fig.~\ref{fig:attractors}. The red colors correspond to a longer lifetime. \\

\begin{figure}[h!]
    \centering
    \includegraphics[width=0.8\linewidth]{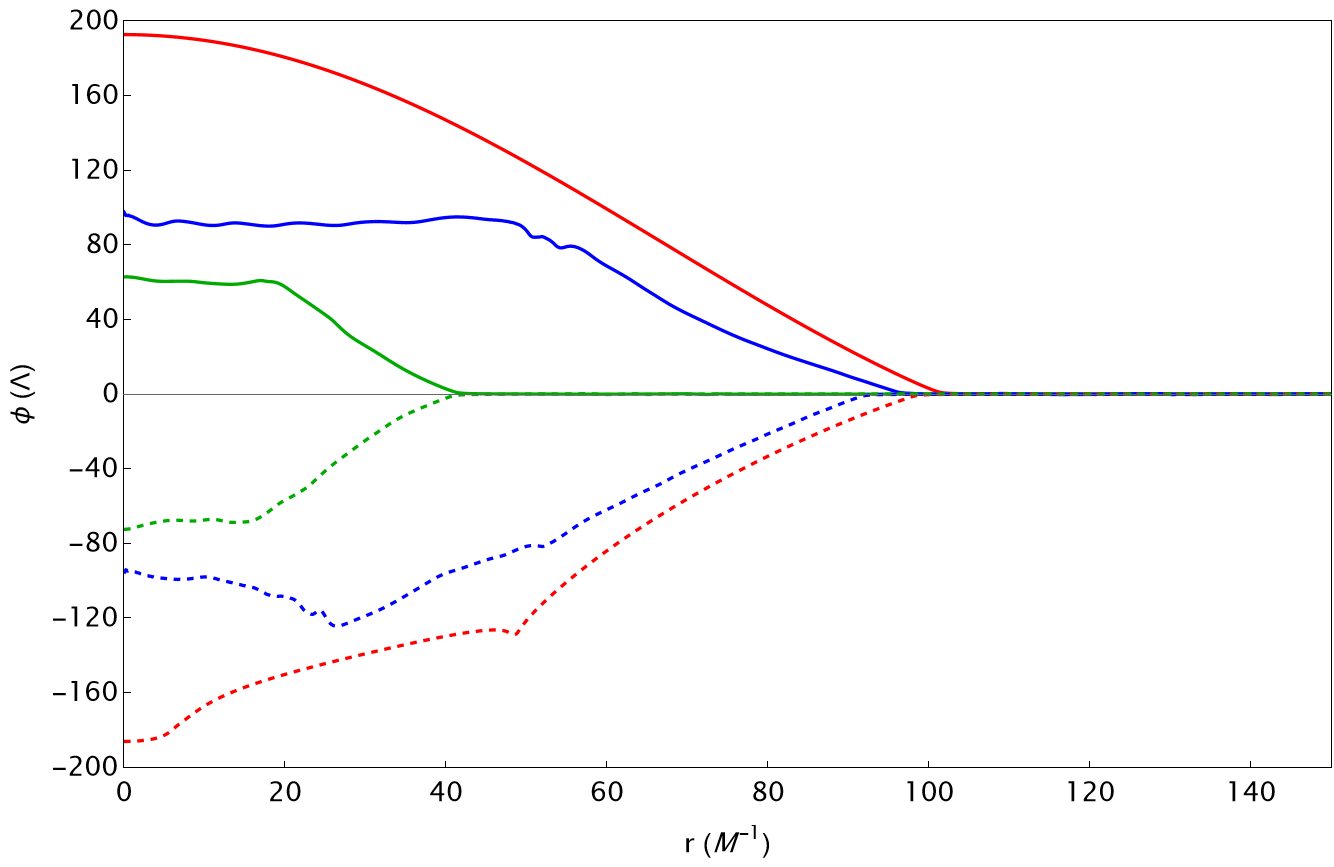}
    \caption{The evolution of the field after initializing with a configuration that is differs significantly from the ansatz of eq.~\eqref{eq:softoscansatz}. We show the field after one period (red), 10 periods (blue) and 50 periods (green). The field naturally moves from a soft oscillon in the dirty regime (blue regime) to one in the clean regime (green profile). This is not surprising since soft oscillons are the unique configurations of size $R_o$ that have a static boundary, and thus don't radiate.}
    \label{fig:attfield}
\end{figure}

There is a large region of parameter space where the existence of the soft oscillon solutions increases the lifetime of localized configurations. The fact that soft oscillons have attractive properties can be seen explicitly by looking at the evolution of the field, starting from away from the soft oscillon ansatz. We show an example of this in Fig.~\ref{fig:attfield}. The initialization (red line) is a solution of the breather profile equation \eqref{eq:effectivebreathermassive} in the limit of $\omega \rightarrow 0$ and $m = 0$. It differs significantly from a soft oscillon as the breather contains only a single cavity mode. We can schematically describe the dynamics in Fig.~\ref{fig:attfield} as follows. The breather is quickly attracted to a soft oscillon of some size $R_o$. Since we started far away from the exact mode distribution necessary in the cavity, we are initially in the dirty regime where the size of the object decays linearly (blue profile). After some time, and below the critical radius we enter the clean regime after which the evolution proceeds exactly as if we started with the soft oscillon ansatz.\\

Figs.~\ref{fig:attractors} and \ref{fig:attfield} are convincing evidence for the attractive properties of soft oscillons in spherical symmetry, both in the clean and dirty regimes. Although we mostly focus on studying their properties in this work, this fact is promising in light of potential applications. We now move on to understanding their decay rate in the different regimes of evolution.

\section{Evaporation}
\label{sec:evaporation}
In the previous section we have described our numerical observations regarding the behavior of the novel soft oscillon ansatz of Eq.~\ref{eq:softoscansatz}. It is remarkable that, as the oscillon radiates away energy through what we called the nonlinear scattering event taking place at the hDW skin of the configuration, the bulk modes are able to arrange themselves in such a way that the configuration transitions through oscillons of ever smaller sizes. The size of the soft oscillons $R_o$ is at the same time a measure of its overall amplitude $\phi_0 \propto R_o$, the amount of trapped modes $n_{max} \approx MR_o/\pi$ and its energy $E\propto R_o^3$. Its adiabatic evolution $R_o(t)$ therefore gives information about all the important properties of the oscillon itself. In this section we will therefore offer an analytic understanding of the adiabatic evolution of $R_o(t)$. In Sec.~\ref{sec:softosc} we highlighted two distinct ways in which the oscillon evolves, namely:
\begin{enumerate}
    \item The clean regime: a deterministic evolution with little to no noise, given by the analytic form $R_o(t) = \sqrt{R_o(t_0)^2 - Ct}$, where $C$ is some constant which can depend on the exact shape of the potential.
    \item The dirty regime: a stochastic decay, relevant after what we call the classical breaking time of the configuration $t_{br}$, where the exact way the configuration reduces its size depends on the exact deviation from the bag model. After taking the mean of multiple different evolutions, we concluded that the size decayed approximately linearly with $R_o(t) = R_o(t_{br}) - v t$. Here, again $v$ is some constant which can be model dependent.
\end{enumerate}
Given these two distinct ways in which the configuration evolves in these regimes, and the fact that $E \propto R_o^3$, it is simple to guess how the energy loss of the oscillon depends on its size. Since $\dot{E} \propto 3 \dot{R_o} R_o^2$, it must be that for the two cases:
\begin{enumerate}
    \item Clean: $\dot{E} \propto R_o$, so that $\dot{R_o} \propto R_o^{-1}$;
    \item Dirty: $\dot{E} \propto R_o^2$, so that $\dot{R_o} \propto v$, with $v$ again a constant.
\end{enumerate}
In this section we offer an analytic understanding of both of these regimes using the cavity and skin picture.

\subsection{Analytics}
To obtain an analytic understanding of the evaporation of soft oscillons, we will make use of the notion that the field behaves as a set of free waves trapped in a cavity, confined by the hDW skin. This perspective provides accurate quantitative predictions for the rate of evaporation for soft oscillons in gapless plateau potentials, regardless of the exact shape of the potential. We remind the reader of the following definition:
\begin{equation}
    \mu \equiv \sqrt{V''_{max}(\phi)}
\end{equation}
Where $V''_{max}(\phi)$ is the maximum value of the potential under consideration. We use this as the height of the confining potential trapping modes within the cavity.\\

One can write a general solution of the field, confined to the cavity, as

\begin{equation}
\phi_{cavity}(r, t) = 
\begin{cases} 
      \sum_n^{n_{max}} \frac{4\sqrt{2} R_o}{n\pi} \sin(n\pi/2) \cdot A_n \frac{\sin(k_n r)}{k_n r} \cos(k_n t + \delta_n) & r< R_o \\
      0 &r\geq R_o \\
   \end{cases}
   \label{eq:bagansatz}
\end{equation}
Where $k_n = n \pi / R_o$ ($n = 1, 2, 3...$) and $n_{max}$ is the largest value of $n$ for which $n\pi/R_o < \mu$. Note that we are working in this section (and will continue to do so) with rescaled dimensionless variables, meaning that spatio-temporal variables are always understood in terms of $M^{-1}$ and field variables in terms of $\Lambda$. For the bag model solution one has $A_n = 1$ and $\delta_n= 0$. We could also have a different distribution of amplitudes and phases, which we will use to interpret the classical breaking of the solution and the dirty regime. First, we will focus on the bag model ansatz. Our numerics suggest that $\dot{E} \propto R_o$ in that case, corresponding to the clean regime.
\subsubsection{Clean regime}
\label{sec:lineardec}
\begin{figure}
    \centering
    \includegraphics[width=0.5\linewidth]{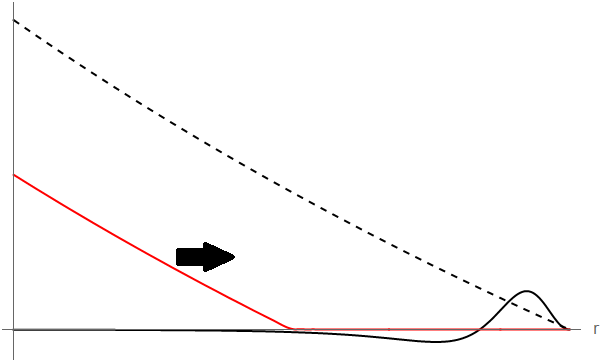}
    \caption{The Bag model can be understood as an outgoing front $|\phi_{front}|$ (red) traveling across a static background $\phi_{bag}(r,0)$ (dashed). The outgoing wave is free until it reaches the boundary at $r = R_o$. There it will scatter off the potential sourced by the background $V''(\phi_{bag}(r, 0))$ (black) after which some of the energy in the wave is transmitted and lost. The typical height of the scattering barrier is set by $\mu$ and is independent of $R_o$.}
    \label{fig:scatterprob}
\end{figure}

It is convenient to think of Eq.~\ref{eq:bagansatz} as the sum of an incoming wavepacket traveling towards the origin at $r = 0$ and an outgoing wavepacket traveling to $r = R_o$. One cycle is completed when the two waves reach their respective destinations and the overall envelope of the solution switches sign. Part of the energy contained in the outgoing wave will be transmitted to infinity, leading to the evaporation of the solution. To calculate the transmission coefficient of the wavepacket, we are required to solve a nonlinear scattering problem, since the scattering potential is generated by the waves themselves. To make progress, it is useful to write the cavity field as the sum of a static background and a traveling front
\begin{equation}
    \phi_{bag}(r, t) = \phi_{bag}(r, 0) + \phi_{front}(r, t)
    \label{eq:scattersep}
\end{equation}
Note that this separation is somewhat trivial as $\phi_{front}(r, t) = \phi_{cavity}(r, t) - \phi_{bag}(r, 0)$, in the clean regime where we assume $A_n = 1$ and $\delta_n = 0$. It is important conceptually however. The equation of motion for the front is then given by
\begin{equation}
    \Ddot{\phi}_{front} - \nabla^2 \phi_{front} + V'(\phi_{front} + \phi_{bag}(r, 0)) = \nabla^2 \phi_{bag}(r,0)
    \label{eq:scatterder}
\end{equation}
To make progress, we approximate $\phi_{front}$ as a small perturbation of the static background, which should be valid at the edges of the traveling front. This sets up the linear scattering problem
\begin{equation}
    \ddot{\phi}_{front} - \nabla^2 \phi_{front} + V''(\phi_{bag}(r, 0))\phi_{front} =\nabla^2 \phi_{bag}(r,0) - V'(\phi_{bag}(r, 0))
    \label{eq:scatterf}
\end{equation}
The r.h.s. of this equation is identically zero if the skin of the cavity is in fact the static hDW. We can thus focus on the l.h.s. of Eq.~\eqref{eq:scatterf}. The flux emitted by the soft oscillons in the clean regime can be estimated as the transmission of $\phi_{front}$ through the static potential barrier $V''(\phi_{bag}(r, 0))$.  A schematic setup of the scattering problem is shown in Fig.~\ref{fig:scatterprob}. 
We need to compute the energy in the wavepacket that is transmitted through the barrier in this simple linear scattering setup. This will provide an estimate for the energy lost by the oscillon every cycle (half a period) since the process each time the front reaches the skin. Note that this way of estimating fluxes is merely an analogue of the full nonlinear system. However, as we'll see the results are accurate enough to confirm this interpretation of the physics at play. \\

We approximate the scattering potential $V''(\phi_{bag}(r,0)) \equiv V(r)$ as,
\begin{equation}
    V(r) = 2 \sqrt{V_{max}''(\phi)} \delta(r - R_o)= 2 \mu \delta(r - R_o), 
\end{equation}
The transmission coefficient of this potential is well known. The ratio of the amplitudes of an incoming wave $I$ and an outgoing wave $T$, depends on its wavenumber $k$ and is given by

\begin{equation}
    t(k) = |T/I|^2 = \frac{1}{1 + \left(\mu/k\right)^2} 
    \label{eq:tcoefficient}
\end{equation}
Waves with $k \gg \mu$ get transmitted completely and won't be confined by the potential. The transmitted part of the outgoing front of the bag model can thus be written as
\begin{equation}
    \phi_{rad}(r, t) = 2 \sqrt{2} \sum_n^{n_{max}}\left(\frac{R_o}{n \pi}\right)^2\frac{\sin\left(k_n(r - t)\right)}{r} \sin(n \pi/2) \frac{1}{\sqrt{1 + \left(\mu/k_n\right)^2}} 
    \label{eq:radbag}
\end{equation}
We can estimate the energy loss per cycle by computing the total energy flux of the radiation field through the hDW-skin at $r = R_o$. This is given by
\begin{equation}
    \Delta E = 4\pi R_o^2 \int_0^{R_o} dt' T_{0r}|_{r=R_o}= 4\pi R_o^2 \int_0^{R_o} dt' \dot{\phi}_{rad}(t', R_o) \phi'_{rad}(t', R_o)
    \label{eq:energylossbag}
\end{equation}
Plugging \eqref{eq:radbag} into \eqref{eq:energylossbag}, ignoring terms of $\mathcal{O}(1/R_o)$ and performing the integration over one cycle, one obtains the rather simple expression
\begin{equation}
    \Delta E = 16 \pi R_o^3 \sum_{n,odd}^{n_{max}} \frac{1}{(n \pi)^2} \frac{1}{1 + \left(\frac{\mu R_o}{ n \pi}\right)^2}
    \label{eq:Elossperc1bag}
\end{equation}
The sum of eq.~\ref{eq:Elossperc1bag}, with $n_{max} = \mu R_o/\pi$ converges in the limit of $R_o \rightarrow \infty$. Namely, in that limit
\begin{equation}
    \sum_{n,odd}^{n_{max}} \frac{1}{(n \pi)^2} \frac{1}{1 + \left(\frac{\mu R_o}{n \pi}\right)^2} \simeq \frac{\tanh{\mu R_o/2}}{4 \mu R_o} \approx \frac{1}{4 \mu R_o} 
\end{equation}
In the large $R_o$ limit the total energy loss per cycle of the soft oscillon can thus be estimated to be
\begin{equation}
    \Delta E \approx 4 \pi R_o^2/\mu
    \label{eq:Elosscyclebag}
\end{equation}
Eq.~\eqref{eq:Elosscyclebag} can be understood as the total energy that the soft oscillon emits per cycle. The average energy loss $\dot{E}$ is given by averaging over the duration of the cycle $\Delta t = R_o$. We finally obtain
\begin{equation}
    \dot{E} = \frac{\Delta E}{\Delta t} \approx 4\pi R_o/\mu
    \label{eq:dEdtbag}
\end{equation}
This linear dependence of $\dot{E}$ on the size of the soft oscillon $R_o$ predicts that its size $R_o$ decays as $\dot{R_o} \propto R_o^{-1}$, as we observed in sec.~\ref{sec:softosc} during the clean regime.\\

Note that this computation was performed using an explicit form of the cavity field of eq.~\eqref{eq:bagansatz}. In particular we chose the various $A_n$ and $\delta_n$ as to reproduce the bag model ansatz in the cavity, which we associate to the clean regime of evolution. In the linear picture, the various modes of the field do not couple, and the distribution of amplitudes and phases do not change over time. However, in reality the scattering event taking place every cycle is nonlinear. It is clear that nonlinearities can in principle lead to a redistribution and dephasing of the modes contained in the cavity (so different values of $A_n$ and $\delta_n$). In what follows we will argue why this happens above a critical size of the oscillon. Then we'll show that this redistribution on average leads to an energy loss that scales with the surface of the oscillon $\dot{E} \propto R_o^2$.


\subsubsection{Critical radius and the dirty regime}
The derivation given in the previous section, which provides an understanding of the linear evaporation rate of the soft oscillon $\dot{E} \propto R_o$ in the clean regime. A crucial prerequisite for the validity of this derivation is that the relative amplitudes of the modes trapped within the oscillon follow the $1/n$ dependence associated to the bag model. There is no guarantee that, starting from the `perfect' soft oscillon ansatz of Eq.~\eqref{eq:softoscansatz}, the modes trapped within the confining skin of the configuration still obey this scaling after the oscillon has completed one cycle. It even seems somewhat surprising that we found soft oscillons that are capable of adiabatically changing their size $R_o$, while retaining the appropriate mode composition in the bulk. In fact, the scattering event coincides with the overall sign flip of the field. Near that time ($t \mod{R_o} \sim 1/2$ assuming $\dot\phi=0$ initialization at $t = 0$), 
$V'(\phi) \approx \mathcal{O}(1)$ everywhere in the bulk and the soft oscillon is not an approximate solution of the equations of motion. In addition to an overall energy loss through the transmission of a wavepacket to spatial infinity, this will also result in a perturbation of the field within the bulk. This perturbation can in part be interpreted as the reflected part of the nonlinear scattering event taking place near $r = R_o$. After one cycle, the oscillon field profile should be written as a soft oscillon with an additional perturbation
\begin{equation}
    \phi(t,r) =  \phi_{soft}(t,r) + \delta(t,r)
    \label{eq:perturbosc}
\end{equation}
Where the perturbation $\delta(t,r)$ obeys the equation of motion
\begin{equation}
    \ddot{\delta} - \nabla^2 \delta + V''(\phi_{soft}) \delta = - V'(\phi_{soft})
    \label{eq:perturbationeom}
\end{equation}
From Eq.~\eqref{eq:perturbationeom} it is evident that perturbations are sourced whenever $V'(\phi_{soft}) \sim \mathcal{O}(1)$, namely during the nonlinear scattering event. A heuristic way to interpret the sourcing of $\delta(t,r)$ once per cycle is to understand its presence as an effective reshuffling of the relative amplitudes of the bulk modes. Schematically, one expects that multiple low frequency modes of the background $\phi_{soft}$ combine in $V'(\phi_{soft})$ to source a perturbation of higher frequency, leading to a general hardening of the relative amplitudes of the bulk modes. In Sec.~\ref{sec:stochevap} we will show that (in the linear approximation), this leads to an increase in the evaporation rate of the oscillon, something that we called the classical breaking of the solution.\\

In Sect.~\ref{sec:softosc}, we observed that classical breaking of the solution typically only occurs when the size of the oscillon lies above a certain threshold, which we called the critical size $R_c$. It is beyond the scope of this work to obtain an accurate prediction of $R_c$. The exact value of $R_c$ is weakly dependent on the initial conditions taken (as long as we're close to a soft oscillon), and will strongly depend on the model under consideration. However, we can intuitively understand why the sourcing of perturbations by soft oscillons of large size necessarily has to lead to a significant reshuffling of the bulk modes.\\

Suppose we want to compute the perturbation sourced by a soft oscillon of size $R_1$. By solving eq.~\eqref{eq:perturbationeom} we can obtain $\delta_1(t,r)$. This is the perturbation living atop of the soft oscillon background after one cycle. This perturbation $\delta_1(t,r)$ should be is a wavepacket of typical wavenumber $k_\delta$ travelling across the bulk. Now, importantly, if $k_\delta \gg \mu$ (the height of the confining potential) the wavepacket will be transmitted perfectly through the skin layer at $r = R_1$. In this scenario, the sourced perturbation is shed in a time $t \sim R_o$. So we conclude there is no significant build-up of perturbations in the bulk as they are shed within one cycle. This is why soft oscillons are able to evolve adiabatically with a linear evaporation rate $\dot{E} \propto R_o$ in the clean regime: the cavity literally remains `clean'. Next, we can try to compute the sourced perturbation for some oscillon of larger size $R_2$. In particular we would like to obtain a solution of the equation
\begin{equation}
    \ddot{\delta_2} - \nabla^2 \delta_2 + V''(\phi_{soft, 2}) \delta_2 = - V'(\phi_{soft,2})
    \label{eq:perturbbigger}
\end{equation}
It is important to note that in the limit of large size $R_o$, the effective source and mass terms evaluated on the soft oscillon ansatz, are for any plateau potential, approximately given by

    \begin{equation}
V_{S}(t,r) \equiv V'(\phi_{soft}(t, r)) \approx 
\begin{cases} 
      V'_{max}(\phi)\delta'(1/2 - t\mod R_o)  & r< R_o \\
      0 &r\geq R_o \\
   \end{cases}
   \label{eq:Vsource}
\end{equation}

\begin{equation}
V_M(t,r) \equiv V''(\phi_{soft}(t, r)) \approx
\begin{cases} 
      V''_{max}(\phi)\delta(1/2 - t\mod R_o)  & r< R_o \\
      0 &r\geq R_o \\
   \end{cases}
   \label{eq:Vmass}
\end{equation}
Where $\delta(x)$ is the Dirac-delta function. There is now an explicit scaling transformation relating the source and mass terms in eq.~\eqref{eq:perturbationeom} for background oscillons of different sizes. Namely,
\begin{equation}
    V_M^2(t, r) = V_M^1(t\cdot(R_1/R_2), r\cdot(R_1/R_2)); \quad V_S^2(t, r) = V_S^1(t\cdot(R_1/R_2), r\cdot(R_1/R_2))
    \label{eq:sourcescalings}
\end{equation}
Where the superscripts $V_{M,S}^1$ and $V_{M,S}^2$ refer to the effective source and mass functions of the two background oscillons of different sizes $R_1$ and $R_2$. Using this scaling we can rewrite \eqref{eq:perturbbigger},

\begin{equation}
    \ddot{\delta_2} - \nabla^2 \delta_2 + V_M^1(t\cdot(R_1/R_2), r\cdot(R_1/R_2)) \delta_2 = - V_S^1(t\cdot(R_1/R_2), r\cdot(R_1/R_2))
\end{equation}
Which, after changing variables $\tilde{t}  = t\cdot(R_1/R_2)$, $\tilde{r}  = r\cdot(R_1/R_2)$ and $\tilde{\delta}_2  = \delta_2\cdot(R_1/R_2)^2$, becomes

\begin{equation}
   \tilde{\ddot{\delta_2}} - \tilde{\nabla}^2 \tilde{\delta}_2 + (R_2/R_1)^2V_M^1(\tilde{t}, \tilde{r}) \tilde{\delta}_2 = - V_S^1(\tilde{t}, \tilde{r})
   \label{eq:rescaledperturb}
\end{equation}
The crux of our argument is that $\delta_1(\tilde{t}, \tilde{r})$ is almost a solution of this equation, up to a correction originating from the shift in the effective mass term.

\begin{figure}[h!]
\includegraphics[width=0.85\textwidth]{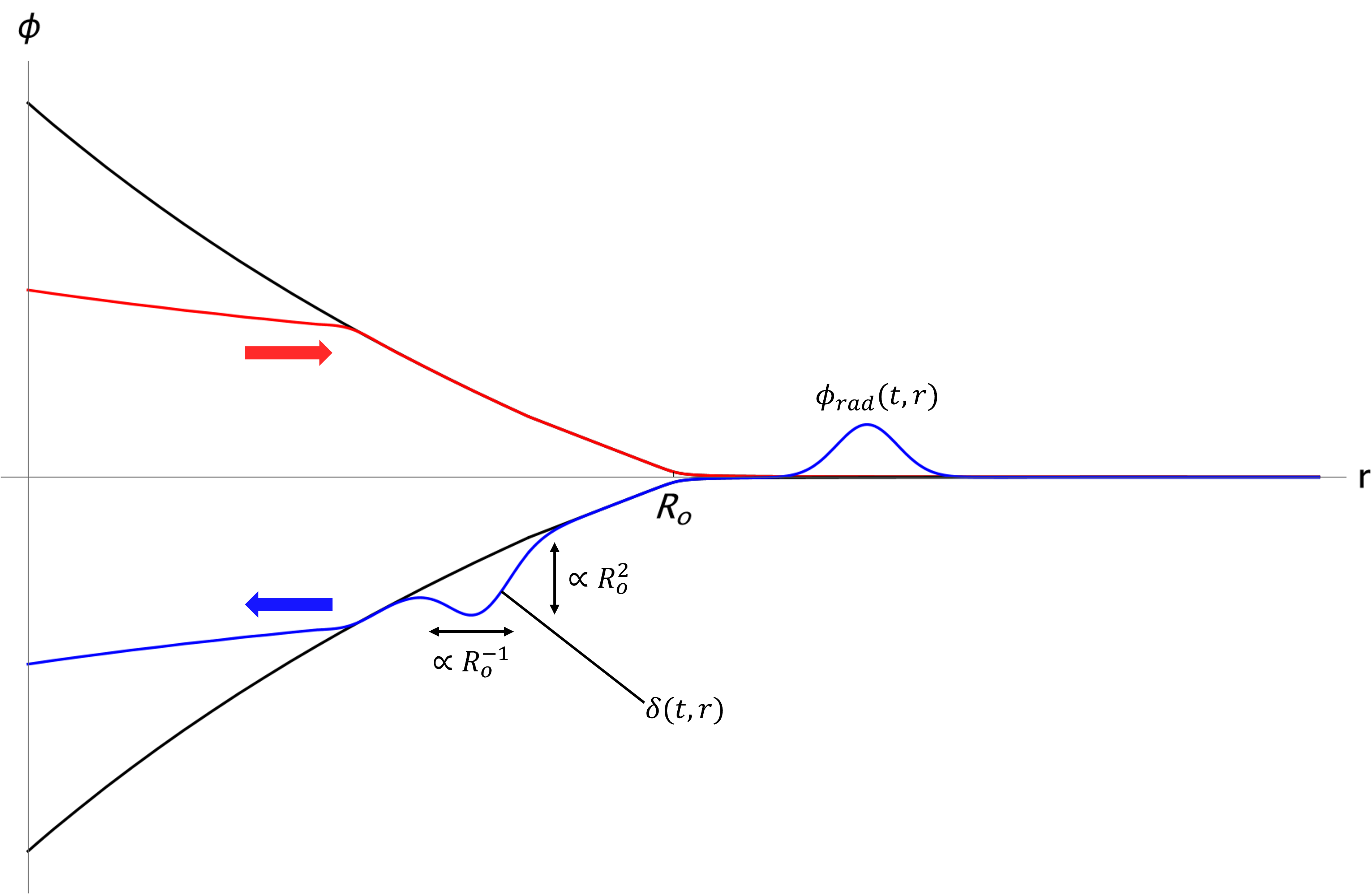}
\centering
\caption{A schematic representation of the nonlinear scattering event. The outgoing front of the bag model travels towards the skin at $R_o$ where it scatters nonlinearly as the whole bulk field switches sign. After the sign flip an ingoing front (in blue) travels towards the origin. However the scattering event has created a transmitted wavepacket $\phi_{rad}$ and a peturbation in the bulk $\delta$. The size and width of $\delta$ depends on the size of the soft oscillon, causing an increased evaporation rate above the critical radius. In black we show the static bag model at $t = 0$ and $t = R_o$, across which the outgoing and ingoing front travel respectively.}
\label{fig:NLscattering}
\end{figure}

The approximate form of the perturbation $\delta_2$ sourced by the soft oscillon of size $R_2$ can be guessed
\begin{equation}
    \delta_2(t,r) \approx A \cdot (R_2/R_1)^2 \delta_1\left(t\cdot(R_1/R_2), r\cdot(R_1/R_2)\right)
    \label{eq:scalingrelationperturbations}
\end{equation}
Where we allow a free parameter $A$ to accommodate for the shift in the effective mass term in \eqref{eq:rescaledperturb}. To make sure our reasoning is correct, we explicitely solved the linearized equations numerically and checked that the approximate scaling relation \eqref{eq:scalingrelationperturbations} holds. In Fig.~\ref{fig:NLscattering} we show the effect of the nonlinear scattering event schematically. Each cycle, the outgoing front (in red) scatters nonlinearly off the hDW. The whole configuration flips sign and the front travels again towards the origin (in blue). However, the scattering event results in a transmitted wavepacket $\phi_{rad}$ and an extra perturbation in the bulk $\delta$.\\

We thus conclude that the wavepacket associated to $\delta_2$ has a similar form as $\delta_1$, albeit with a shifted wavenumber $k_\delta\rightarrow (R_1/R_2) k_\delta$. Moreover the prefactor in eq.~\eqref{eq:scalingrelationperturbations} suggests that its amplitude is also larger. Each cycle of the soft oscillon of size $R_2$ such a wavepacket is created. However there necessarily exists a threshold value of $R_2$ where the typical wavenumber $k_\delta \sim \mu$. At that size, a significant portion of the perturbation does not get transmitted through the hDW skin. The bulk modes are thus unable to relax to the soft oscillon ansatz within one cycle, leading to a build-up of perturbations in the cavity. The fact that the amplitude of the perturbations also grows with $R_o$ also contributes to this fact. It is to be expected that above some value $R_2$ the perturbations in the bulk grow large in $O(1)$ cycles. This threshold value of $R_2$ is what we previously referred to as the critical size $R_c$. Above this size the dynamics of the soft oscillon lead to a significant reshuffling of the relative amplitude of trapped bulk modes, away from the $1/n$ dependence of the bag model. This leads to a larger evaporation rate $\dot{E}$, the classical breaking of the soft oscillon solution.

\label{sec:nse}

\subsubsection*{Evaporation estimate }
\label{sec:stochevap}
Based on the qualitative considerations of the previous section, we will use our understanding of the soft oscillon after classical breaking to get a quantitative understanding of how evaporation proceeds in such a phase. Again, it is useful to start from the general solution of a scalar field contained in a cavity, which we repeat here for clarity,

\begin{equation}
\phi_{cavity}(r, t) = 
\begin{cases} 
      \sum_n^{n_{max}} \frac{4\sqrt{2} R_o}{n\pi} \sin(n\pi/2) \cdot A_n \frac{\sin(k_n r)}{k_n r} \cos(k_n t + \delta_n)  & r< R_o \\
      0 &r\geq R_o \\
   \end{cases}
   \label{eq:gencavityansatz}
\end{equation}

As outlined in sec.~\ref{sec:nse}, we interpret the classical breaktime of the soft oscillon solution as the time after which the relative amplitudes and phases of the various modes deviate significantly from the bag model solution that has $A_n = 1$ and $\delta_n = 0$. This type of deviation from the bag model ansatz is generally expected due to the nonlinear nature of the scattering event. However, the key observation we made in the previous section is that, above the critical size $R_c$, the timescale of relaxation to the bag model solution is larger than the time it takes for a cycle of the solution to complete (and thus larger than the time until the next nonlinear scattering event). This will evidently lead to a significant redistribution of the amplitudes and phases of the various cavity modes. We thus interpret the soft oscillon after classical breaking as a field configuration trapped in a cavity as in Eq.~\ref{eq:gencavityansatz}, but differing significantly from $A_n = 1$ and $\delta_n = 0$. We can try to predict the rate of evaporation $\dot{E}$ for this more generic configuration. \\

First note that the total energy in the free cavity is readily computed and equals,
\begin{equation}
    E_{cavity} = \frac{32}{\pi} R_o^3 \sum_{n,odd}^{n_{max}} \frac{A_n^2}{n^2},
    \label{eq:Ecav}
\end{equation}
independent of the relative phases $\delta_n$. For $A_n = 1$, the sum (for large size $R_o$ and therefore $n_{max}$), converges to $\pi^2/8$, which gives the energy associated to the bag model. We will use this fact later.\\

We are interested in computing the energy loss of a configuration with some generic distribution of $A_n$ and $\delta_n$. Analogously to what was done in Sec.~\ref{sec:lineardec}, we will approximate the radiation field of the cavity modes as the transmitted part of the outgoing field modes. As before,
\begin{equation}
    \phi_{rad}(r, t) = 2 \sqrt{2} \sum_n^{n_{max}}A_n\left(\frac{R_o}{n \pi}\right)^2\frac{\sin\left(k_n(r - t) + \delta_n\right)}{r} \sin(n \pi/2) \frac{1}{\sqrt{1 + \left(\mu/k_n\right)^2}} 
    \label{eq:radcav}
\end{equation}
The energy loss per cycle of this configuration can be computed in a similar manner,
\begin{equation}
    \Delta E = 4 \pi R_o^2 \int_0^{R_o} dt T_{0r}|_{r = R_o} = 16 \pi R_o^3 \sum_{n,odd}^{n_{max}} \frac{A_n^2}{(n\pi)^2} \frac{1}{1 + \left(\frac{\mu R_o}{n\pi}\right)^2}
    \label{eq:Elosscyclgen}
\end{equation}
To get a quantitative prediction for the evaporation rat $\dot{E}$ we need to assume a particular distribution of the modes $A_n$. As observed numerically the process of breaking leads to an unpredictable and somewhat chaotic global evolution of the soft oscillon\footnote{See \cite{Adam:2019xuc,Blaschke:2024uec} for recent discussions of chaotic behaviour in other oscillon models.}. It was observed that a slight perturbation in the initial conditions of order $\mathcal{O}(5\%)$ can lead to substantial differences in the evolution of $R_o(t)$ and thus $\dot{E}$. This can be understood by realizing that a perturbation on the initial condition is equivalent to a particular distribution of $A_n$, differing from $A_n = 1$ by $\mathcal{O}(5\%)$. It seems reasonable that the redistribution of relative amplitudes, caused by the nonlinear scattering event proceeds differently for each individual initialization. Using Eq.~\eqref{eq:Elosscyclgen} it is not surprising that we observe a different global evolution for each instantiation.\\

To make predictions about the ensemble of distributions we will assume the following form for the average distribution of amplitudes
\begin{equation}
    \langle A_n^2\rangle = B  n^p
    \label{eq:averageredist}
\end{equation}
Where $\langle \cdot\rangle$ should be interpreted as an average over many initializations, and $B$ is some constant. The constant $B$ can be fixed by requiring that the total energy in the cavity is equal to the equivalent bag model configuration. Using Eq.~\eqref{eq:Ecav},
\begin{equation}
    B = \pi^2/8 \left(\sum_{n,odd}^{n_{max}} n^{p-2} \right)^{-1}
\end{equation}
Next, we will assume that classical breaking translates into a deposition of low frequency modes into high frequency modes, so that $p > 0$. This is generally what happens when low frequency modes interact nonlinearly. Moreover, we seem to explicitly observe such a hardening in our numerics (see e.g. Fig.~\ref{fig:criticalgrid}). Finally, we will assume that $p < 2$, so that the low frequency remain dominant in the cavity. These modes are associated with the static hDW skin and are required to be dominant in order for the cavity picture to be valid. The average energy loss per cycle is then given by
\begin{equation}
    \langle\Delta E\rangle = 2 \pi R_o^3 \frac{\sum_{n,odd}^{n_{max}} n^{p-2} \frac{1}{1 + \left(\frac{\mu R_o}{n\pi}\right)^2}}{\sum_{n, odd}^{n_{max}} n^{p-2}}
    \label{eq:elosscyclegenf}
\end{equation}
Now, the sum in this expression converges for $n_{max} = \mu R_o/\pi$, taking the limit of $R_o \rightarrow \infty$. Although the sum has no closed form analytic expression, we checked numerically that in this limit the sum converges to, depending on the value of $p$ 

\begin{equation}
\frac{\sum_{n,odd}^{n_{max}} n^{p-2} \frac{1}{1 + \left(\frac{\mu R_o}{n\pi}\right)^2}}{\sum_{n, odd}^{n_{max}} n^{p-2}}\propto 
\begin{cases} 
    K (\mu R_o)^{p-1} & 0<p<1\\
     K \log(\mu R_o)^{-1} & p = 1 \\
      K & p > 1 \\
   \end{cases}
   \label{eq:sumconvergencestoch}
\end{equation}
With $K$ some constant that depends on the value of $p$. We see that for a generic value of $p$ between $0$ and $2$, the sum on the r.h.s. of eq.~\eqref{eq:elosscyclegenf} is constant for a large range of sizes $R_o$. In what follows, we take it to be exactly $K$, which should be accurate for the range of sizes we have considered in this work. The average evaporation rate is then,
\begin{equation}
    \langle \dot{E} \rangle = \frac{\langle \Delta E \rangle}{\Delta t} = 2 \pi K R_o^2
\end{equation}
consistent with the surface evaporation observed for soft oscillons in the dirty regime. This prediction is completely generic: hardening of the cavity modes will typically lead to an evaporation rate that is proportional to the surface of the oscillon. Moreover we expect that, in the large $R_o$ limit, the proportionality constant $K$ is only mildly dependent on the exact shape of the potential. We can get an order of magnitude estimate of its value by evaluating the sum in Eq.~\eqref{eq:sumconvergencestoch} explicitly at $\mu R_o = 200$, for the range of $0<p<2$, and averaging the result. We obtain
\begin{equation}
    K \sim 8\cdot 10^{-2}
\end{equation}
The average of the evolution of the size of the soft oscillon $R_o(t)$ is then,
\begin{equation}
    R_o(t) = R_o(t_{br}) - \frac{K}{18} t = R_o(t_{br}) - v t
\end{equation}
Where $K/18 \equiv v$ to make connection to our previous notation. This matches our expectation of oscillons in the dirty regime.\\

Note that our analysis assumes that the reshuffling of cavity modes is not too severe with respect to the bag model. If for some reason the hardening of the cavity modes is more extreme (e.g. $p > 2$) our estimate for $K$ can be quite far off the true value. Even in those cases our analysis predicts an evaporation rate proportional to the surface of the object, although it will be larger. We expect this to be the case for a generic object of size $R$ whose mode decomposition is not close to any soft oscillon. See sec.~\ref{sec:attractiveness} for a discussion concerning this point.\\

To conclude, we will test our analytic understanding of the evaporation rate of soft oscillons against numerical results.

\subsection{Comparison to numerics}
\label{sec:potcomparisons}
\begin{figure}[h!]
    \centering
    \includegraphics[width=\linewidth]{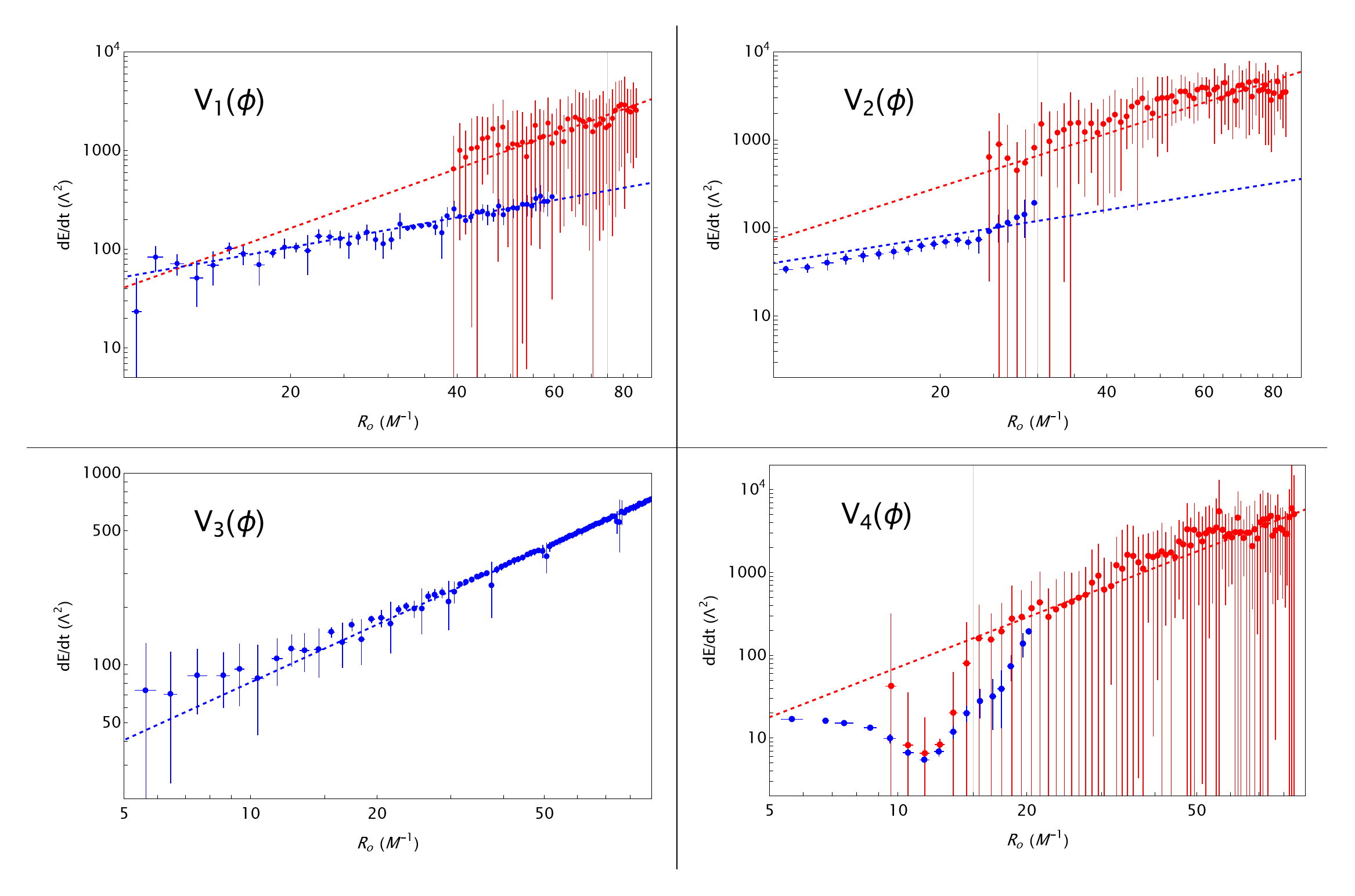}
    \caption{The average evaporation rate soft oscillons of various sizes $R_o$ for the four potentials given in \eqref{eq:testpotentials}. The blue dots correspond to a background oscillon that is initialized with a size below $R_c$ (indicated by the gridline in each plot). In that case $\dot{E} \propto R_o$ (blue dotted line). This is true for all potentials except $V_4(\phi)$ for which the soft oscillon ansatz is not a good solution at small $R_o < R_c$ (see text). The red dots on the other hand correspond to large background oscillons that undergo classical breaking. Each initialization evolves differently in that scenario resulting in large variances of the measurements. On average the evaporation rate scales as $\dot{E} \propto R_o^2$ (red dotted line). Note that the critical radius of $V_3$ is larger than the numerical domain, which is why we have no red dots in that plot.}
    \label{fig:fluxgrid}
\end{figure}

Now that we have built a solid understanding of the mechanisms driving the evaporation of soft oscillons in gapless potentials, it is interesting to see how accurately the naive cavity picture can reproduce the full nonlinear evolution of the system. To do this we simulate soft oscillons of different sizes, for various gapless plateau potentials. In principle, soft oscillons are a valid ansatz for any plateau potential. Based on the considerations outlined in this section, we expect the influence of the potential to be visible through three phenomenological effects
\begin{enumerate}
    \item In the clean regime, the rate of evaporation depends on the height of the scattering potential $\mu$, which will be model dependent.
    \item The critical size $R_c$, above which we expect classical breaking to occur and the soft oscillon enters the dirty regime, can differ for different potentials. This is obvious as the nonlinear scattering event and ensuing redistribution of particles in the bulk depend on the microscopic details of the interactions in the model.
    \item Interestingly the evaporation rate in the dirty regime is only mildly dependent on the microscopic details of the potential. This assumes that the way in which the cavity modes reshuffle is not strongly dependent on the model under consideration (the value of $p$ in the previous section).    
\end{enumerate}
In our comparison we will focus on four distinct gapless potentials to test our understanding,
\begin{equation}
    V_1(\phi) = \frac{\phi^4}{1 + \phi^4}; \quad V_2(\phi) = \frac{\phi^4}{(1 + \phi^2)^2}; \quad V_3(\phi) = \frac{\phi^6}{1 + \phi^6}; \quad V_4(\phi) = \frac{\phi^4}{1 + 10 \phi^2 + \phi^4}
    \label{eq:testpotentials}
\end{equation}
Each of these potentials has a different critical size $R_c$ and value of  $\mu$. We are interested in measuring the evaporation rates of soft oscillons at various sizes $R_o$, both below and above the critical radius. Since above the critical radius, we expect the evaporation rate to be stochastic, we average $\dot{E}(R_o)$ over multiple different instantiations of the field. We perform measurements as follows
\begin{enumerate}
    \item Initialize a soft oscillon at $t = 0$ as in Eq.\eqref{eq:softoscansatz}. To get distinct evolutions of the system we randomly perturb a fixed number of the bulk modes by $\mathcal{O}(5\%)$ with respect to the bag model.
    \item Numerically time evolve the system. At each point in time we measure the size and evaporation rate of the oscillon.
    \item After each cycle (defined as the moment that the bulk average amplitude $\bar\phi$ crosses $0$), we average the size and evaporation rate over the duration of that cycle. We then save that data.
    \item We gather data for $\mathcal{O}(10)$ different instantiations of the soft oscillon, each with random initial perturbations.
    \item By binning over the sizes of the measured data, and averaging the evaporation rate within each bin, we obtain a measurement of $\langle \dot{E}\rangle$ at each size $R_o$.
\end{enumerate}
By a process of trial and error, we first determine the approximate value of $R_c$ for each potential. Based on our previous discussion, we expect that a soft oscillon initialized with a size smaller than this, will evolve similarly, regardless of the exact way in which we initially perturb it. The procedure outlined above should therefore give a value of $\langle\dot{E}\rangle$ with small error bars. The opposite is true above $R_c$, where the evolution is stochastic, and we are only able to understand the average $\langle \dot{E}\rangle$. In Fig.~\ref{fig:fluxgrid}, we show results for the four potentials under consideration. The blue dots correspond to a background oscillon initially smaller than $R_c$, while the opposite is true for the red dots.\\

The two distinct scaling behaviors of $\dot{E}$ are clearly visible in Fig.~\ref{fig:fluxgrid}. For the potential $V_4(\phi)$ however, we don't observe the linear $\dot{E} \propto R_o$ associated to the clean regime. This can be understood by noting that $\mu_4\approx 0.5$. This means that even at moderately large $MR_o \sim 20$, the number of modes trapped by the hDW skin within the bulk is small. This means that the soft oscillon ansatz must necessarily break down. Instead, the configuration is attracted into a different type of solution, which behaves more like a typical single-frequency oscillon. This attractor is similar to the oscillons found in \cite{Dorey:2023sjh}. Strictly speaking the critical size of $V_4(\phi)$ therefore doesn't exist. but we interpret it here as the size for which a soft oscillon is quickly attracted to this type of single-frequency solution. \\

The reason that there are no red dots in the bottom left plot is because the critical radius for $V_3$ is very large. For computational reasons didn't perform a dedicated analysis of soft oscillons in the dirty regime for this potential. However, we did check that classical breaking and the ensuing dirty regime still occur for $R_o > R_c \approx 170 M^{-1}$.\\

The red and blue dotted lines in Fig.~\ref{fig:fluxgrid} are obtained by searching for the best fit parameters of the data in the two scaling regimes $\dot{E} = C R_o$ and $\dot{E} = K R_o^2$ respectively. We previously obtained an estimate of the fit parameters using a linear analysis of the scattering problem. It is interesting to see how close our linear predictions are to the actual data obtained here. We summarize this along with some other characteristics in Table~\ref{tab:resultstable}. Our analytic estimates are remarkably close to the numerical best fit parameters, only differing by an $\mathcal{O}(1)$ factor. This is quite remarkable since the scattering process is highly nonlinear in reality. Also note that the relative difference between the proportionality factor $K$ in the dirty regime is less model dependent than $C$ in the clean regime. However we would need more data to confirm this.

\begin{table}[h!]
\centering

\begin{tabular}{|c|c|c|c|c|c|c|}
\hline
Potential & $\mu/M$ & $MR_c$ & $C$  & $K$ & $|\frac{C - C_{theory}}{C}|$ & $|\frac{K - K_{theory}}{K}|$ \\
\hline
$V_1(\phi)$ & 1.54 & 80 & 5.24 & 0.41 & 0.2 & 0.23 \\
\hline
$V_2(\phi)$ & 0.96 & 30 & 4.01 & 0.74 & 0.66 & 0.32 \\
\hline
$V_3(\phi)$ & 2.1 & 170 & 8.12 & N.A. & 0.62 & N.A. \\
\hline
$V_4(\phi)$ & 0.49 & 15 & N.A. & 0.72 & N.A. & 0.31 \\
\hline
\end{tabular}
\caption{The summary of the main results of this section. It is remarkable that the proportionality constants obtained using an entirely linear analysis, $K_{theory}$ and $C_{theory}$ are so close to the best fit values obtained from our simulations. }
\label{tab:resultstable}
\end{table}

We want to close this section with a short discussion around massive plateau potentials. A priori, it seems that there is no reason to make a distinction between plateau potentials that are and aren't gapped in the vacuum of the theory. In fact, the cavity picture is even more appropriate in massive theories, since there the skin confines modes that have $\omega < m$ perfectly. Following the linear analysis of Sec.~\ref{sec:lineardec}, we'd expect that the soft oscillons in these models radiate very little as no `leakage' of cavity modes is taking place. To check this we also performed numerical simulations of soft oscillons in massive potentials. As the cavity picture suggests, the soft oscillons initially radiate very little and the radius of the configuration stays almost constant for $\mathcal{O}$(10) cycles. However, after this initial period, the solution typically undergoes a severe type of classical breaking, where we even observed situations for which the cavity picture breaks down significantly, meaning that the fundamental $n = 1$ mode of the bag model bacomes of the same order as the higher overtones ($ p>2$ in eq.~\eqref{eq:averageredist}). After this breaking evaporation proceeds through the typical $\langle \dot{E}\rangle \propto R_o^2$, with the proportionality constant $K$ larger than that found for the gapless potentials of this section. This type of severe breaking can be understood heuristically by considering the framework of sec.~\ref{sec:nse}. As for the gapless potentials, the soft oscillons produce a perturbation every cycle that needs to be shed before the next nonlinear scattering event. It is quite intuitive that this relaxation is made impossible if the cavity is perfect as is the case for massive potentials. Since the perturbations produced during the nonlinear scattering event can not be removed from the bulk we expect a large buildup of perturbations causing the breaking of the solution. We have looked into a few different massive potentials and always see this type of breaking occuring. For this reason we don't expect soft oscillons of large size $R_o$ to be a general attractor in gapped theories. Interestingly, the fact that the confining skin of the cavity isn't perfect in gapless theories, increases the lifetime of large objects in spherical symmetry.\\

Having obtained a solid understanding of soft oscillons enforcing spherical symmetry, both in this section and in section~\ref{sec:softosc}, it is interesting how these results are modified if we perturb the system with aspherical perturbations. We will do this in what follows.

\section{Anisotropies}
\label{sec:anisotropies}
So far, we have analyzed the soft oscillons, enforcing exact spherical symmetry on the system. It can be worthwhile to understand how some of their properties are altered when allowing for asphericity. This is important for two distinct reasons. First, although we saw that soft oscillons in spherical symmetry have a moderate to large island of attractiveness, this is not guaranteed to translate to the full three dimensional situation. It'd be interesting to see how asphericity in the initial conditions affects this property. This can tell us whether the soft oscillons as we described them are the generic end products of classical large aspherical wavepackets in plateau potentials, with potential applications in cosmology. Second, even if one preparess a classical spherical field configuration following the soft oscillon ansatz of Eq.~\eqref{eq:softoscansatz}, there will always be an unavoidable aspherical component due to the quantum nature of the field. One typically expects that the spectrum of quantum fluctuations scales with $\langle \hat{\phi}^2_{nlm}\rangle \sim \mathcal{O}(\lambda)$, where $\hat{\phi}_{nlm}$ can be understood as the quantized cavity modes. The presence of these fluctuations has two effects. On the one hand, if there is an aspherical exponential instability of the background soft oscillon, these modes will rapidly become populated and cause a significant deviation from the semiclassical evolution of the system. This type of instability was observed in \cite{vanDissel:2023zva}, causing a rapid aspherical breakup of certain types of oscillons. On the other hand, even when this doesn't happen, the aspherical quantum modes can become of the same order as the lowest occupied cavity modes (see Eq.~\eqref{eq:gencavityansatz}) as the oscillon shrinks. It might thus be expected that the interaction between small aspherical modes and these lowest occupied bag model modes can have an impact on the overall evolution of the configuration.
In this section we will mainly be interested in studying the second of these two issues, namely the effect of small aspherical perturbations on the soft oscillon ansatz. We leave the general attractiveness of these solutions for future work.\\

To investigate this question we simulate soft oscillons on a three dimensional cartesian lattice using a VV2-integrator and implement an absorbing boundary layer at spatial infinity by adding an artificial dissipative term to the equations of motion. Here, we limit ourselves to a few examples of soft oscillons in the potential,
\begin{equation}
    V(\phi) = \frac{\phi^4}{1 + \phi^4},
\end{equation}
which we called $V_1(\phi)$ in Sec.~\ref{sec:evaporation}. The setup of the simulations is as follows. First, we initialize a soft oscillon with some size $R_o$. Next, we add perturbations with an amplitude of $\mathcal{O}(10\%)$ of the background. We do this by adding a random variable to the field in Fourier space (drawn from a Gaussian distribution) and transform back to real space to obtain the perturbed initial field configurations. We then evolve the system and take measurements every cycle.\\

Even with perturbations that are considerable with respect to the background, we report that the asphericity of the initial conditions does not cause a rapid breakup of the configuration as was observed in \cite{vanDissel:2023zva}. This is remarkable and shows that spherically symmetric soft oscillons can exist for a large number of periods in three dimensions.\\

There is a different type of effect that can decrease the lifetime of the soft oscillons with respect to the spherically symmetric solutions. Aspherical modes seem to cause an effective lowering of the critical radius of the soft oscillon. We thus enter the dirty regime of evolution more easily. The evolution of the system proceeds as follows:
\begin{enumerate}
    \item Initially, the oscillon evolves as in spherical symmetry.
    \item After a certain number of cycles the hDW skin of the oscillon starts to deform.
    \item The deformation leads to an increased evaporation rate which is again proportional to the surface of the object $\dot{E} \propto R_o^2$. This is similar to the dirty regime described in earlier sections.
\end{enumerate}

\begin{figure}[h!]
    \centering
    \includegraphics[width=1\linewidth]{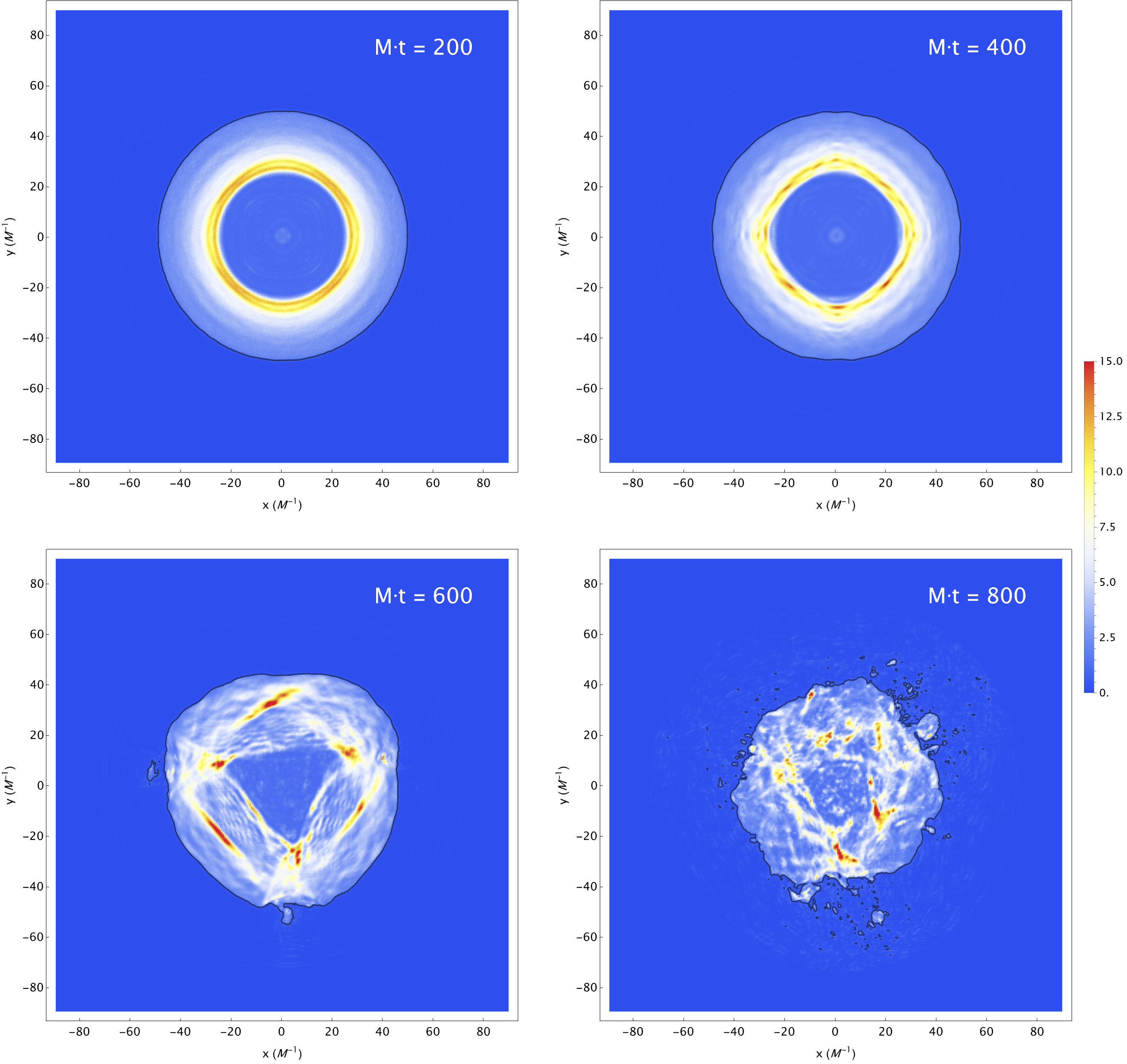}
    \caption{Showing a typical aspherical breaking of the soft oscillon solution. After the hDW skin deform considerably, the evaporation proceeds as we have seen in the case of spherical symmetry above the critical radius, with a rate proportional to the surface of the object $\dot{E}\propto R_o^2$. We put the contour line of $\rho = M^2 \Lambda^2$ in black around the configuration. This broadly delimits the location of the cavity skin.
    }
    \label{fig:asphericalbreaking}
\end{figure}

We show an example in Fig.~\ref{fig:asphericalbreaking} for $MR_o = 50$ (initially). The figures show $\rho(\vec{x}, t)$ in the $z = 0$ plane.\\

It seems that the small aspherical perturbations in the initial conditions eventually start to interact with the spherical modes, causing significant deformation of the hDW skin, speeding up the decay process by emitting surface radiation as in the dirty regime. This is not entirely surprising, since the perturbation $\delta$ created in the bulk during each cycle during the nonlinear scattering event, has a larger phase space to populate in three dimensions.\\

\begin{figure}
    \centering
    \includegraphics[width=1\linewidth]{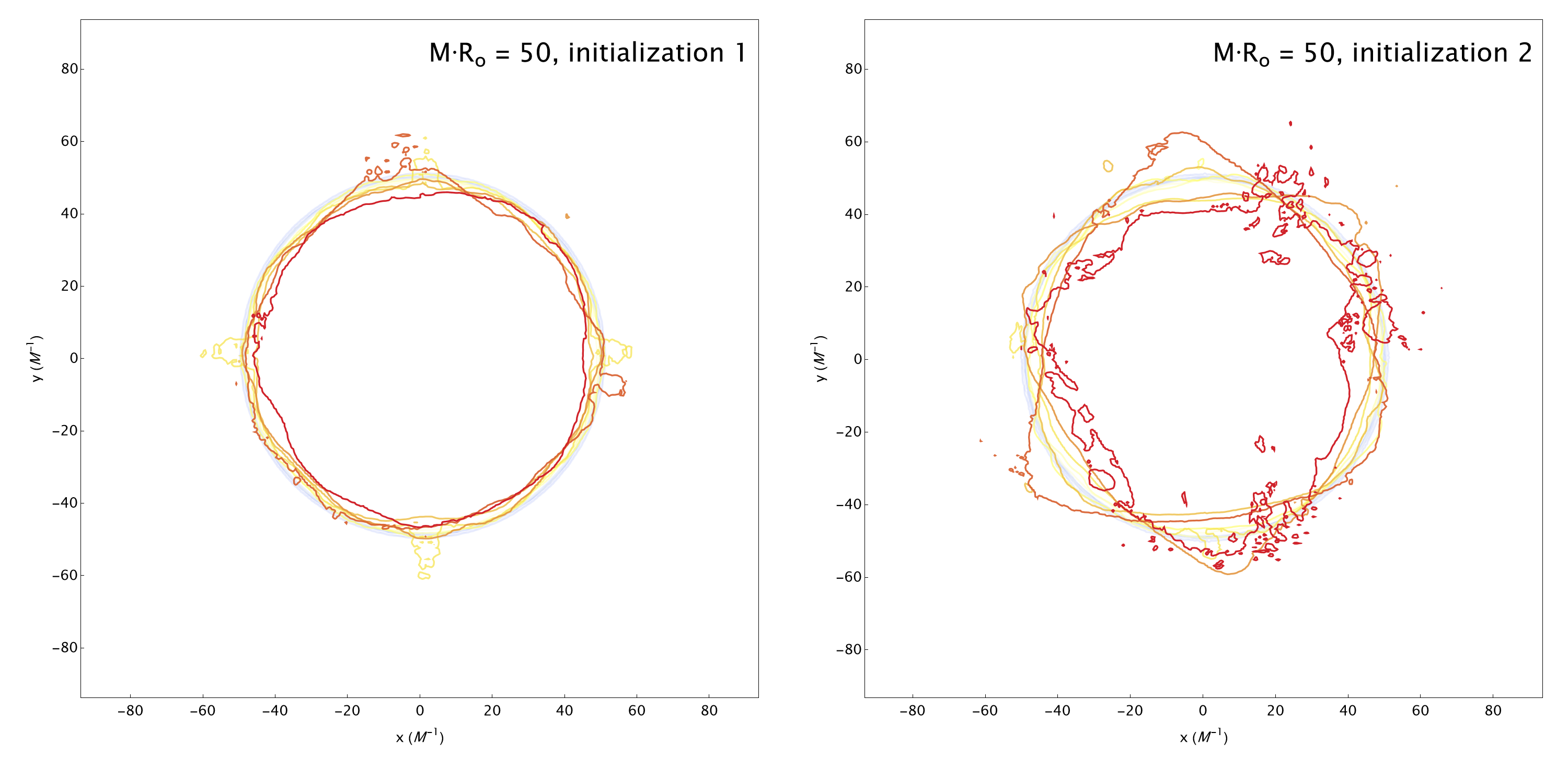}
    \caption{Contours of $\rho(\vec{x}, t) = M^2\Lambda^2$ separated by different cycles of the oscillon. Blue contours correspond to earlier times while red contours correspond to later times. Both cases correspond to an initial soft oscillon of size $MR_o = 50$, but both are initialized with distinct randomized perturbations. Although the initial spherical shapes of the solutions are identical, the randomized initial conditions result in widely different shapes.}
    \label{fig:shapes}
\end{figure}

What's interesting about this new type of breaking is that the microscopic initial conditions of the aspherical modes eventually lead to different shapes of the oscillon, similarly to how the evolution of $R_o(t)$ is stochastic in the dirty regime. In Fig.~\ref{fig:shapes} we show two examples of initially equal macroscopic configurations. The macroscopic object thus retains some type of memory of the initial microstate of the system. This property should translate to the quantum treatment of the cavity modes as well.\\

It's plausible that aspherical perturbations reduce the lifetime of soft oscillons by causing a significant deformation of the cavity skin. Analyzing how this novel type of breaking differs from the spherically symmetric case, and how it depends on the size of the initial perturbation, is beyond the scope of the current work. This is in part because soft oscillons require a large resolution to resolve accurately (simply because of the amount of modes that are contained in them), which is computationally expensive and difficult to attain in larger dimension. We therefore cannot at this point answer exact quantitative questions about soft oscillons with aspherical modes. One important goal for future work will be to understand their effect in more detail.



\section{Conclusion}
\label{sec:concl}

In this work we have presented a new soliton type, termed {\em soft oscillons}, similar to oscillons but with a number of quite distinctive features: 

\begin{itemize}

\item Soft oscillons come in a continuum of sizes $R_o$ (and masses, that scale like $\propto R_o^3$).

\item Their size $R_o$ is unbounded from above (up to gravitational effects), and unrelated to the parameters in the Lagrangian.

\item The oscillation frequency is governed by the size $\omega \sim 1/R_o$. 

\item They have a high harmonic content, with Fourier modes with frequencies $n\pi /R_o$ (with $n$ odd integers) scaling like $ 1/n$. 
This translates into two coherent radiation fronts in the bulk and a pulsating emission of scalar radiation to spatial infinity.

\end{itemize}

These oscillons appear in a broad class of scalar field theories with plateau potentials $V(\phi)$, in the large amplitude regime where the field probes the plateau.
They are longer-lived in `gapless’ models where $V(\phi)$ has a quartic$+$plateau form. Two parameters characterize these potentials: the quartic coupling $\lambda$ and the scale $\Lambda$ where the potential changes shape.  The maximum typical gradient and frequency in the solution is set by a  $M=\sqrt{\lambda} \,\Lambda$. Hence at weak coupling $\lambda\ll1$ our oscillons involve only soft modes, from $1/R \ll M$ to $M\ll\Lambda$.\\

We have further studied the evolution and the scalar radiation of these novel oscillons and found that: i) there are two regimes of evaporation, dubbed clean and dirty, characterized by an emitted power scaling like $R_o$ and $R_o^2$ respectively; ii) the clean (dirty) regimes also correspond to sizes $R_o$ smaller (larger) than a critical radius whose value is model dependent; iii) the lifetime scales with the initial radius $R_i$  typically like $\sim 100 R_{i}$ in the dirty regime and $\sim R_{i}^2$ in the clean regime; iv) soft oscillons exhibit a clear attractor nature to spherical perturbations. They are also stable under small anisotropic deformations, however the lifetime can be reduced. \\

Let us now give a few remarks. \\

The natural interpretation of oscillons in quantum field theory is that they are simply many-particle quasi-bound states of scalar quanta \cite{Dvali:2011aa,Dvali:2012en,Dvali:2018tqi,Olle:2020qqy}. Remarkably, then, oscillons in gapless models correspond to bound states of {\em massless} quanta. This is also the case for potentials without a plateau such as those in \cite{Dorey:2023sjh,Lozanov:2017hjm,Piani:2025dpy} (the only difference is that the radius is fixed). 
Notice that bound states of massless constituent particles are particularly challenging as they are profoundly relativistic. Similar objects in other theories include glueballs in pure Yang Mills and black holes in pure GR \cite{Dvali:2011aa,Dvali:2012en,Dvali:2018tqi}. Oscillons in gapless scalar models represent the equivalent of these interesting objects in a much simpler, spin 0,  theory. This seems an analogy worth exploiting \cite{inprep}. \\

Another byproduct of our work that could be useful to the general understanding of oscillons is the mechanism behind soft oscillons. The peculiar core motion (Fourier mode composition) is understood at once due to the dynamical cavity effect that operates in plateau potentials: large field excursions lead to a cavity inside which the field is effectively massless. This singles out a special solution, the attractor, that i) is static near the surface during most of the evolution (implying small radiation and therefore long lifetime) and ii) acts as a confining cavity capable of trapping many soft modes.\\

This cavity picture emerges (and allows for a controlled understanding) in the asymptotic limit of large field excursion. The perturbative understanding of oscillons in general starting from the operators in the Lagrangian is a challenging task, especially in the large-field and relativistic regime. For perturbative treatments in the (unstable) small amplitude approximation see \cite{Segur:1987mg,Fodor:2008es,Hertzberg:2010yz}. For a recent improvement to large fields in the nonrelativistic limit see \cite{Levkov:2022egq}. In the QFT language \cite{Dvali:2011aa,Dvali:2012gb,Dvali:2017ruz,Dvali:2017eba,Olle:2020qqy,vanDissel:2023zva,Hertzberg:2016tal,Eby:2014fya},  large field oscillons correspond to large occupation number $N\gtrsim 1/\lambda$ with the dimensionless microscopic coupling. For soft oscillons, the occupation number is even higher as it scales with $R_o$. In other words they correspond to large collective coupling. The fact that a simple controllable picture emerges in terms of weakly coupled (`asymptotically free') modes in a cavity is quite welcome and remarkable.\\

This perhaps opens the door to a new aproximation scheme  to understand oscillons in more general models without a plateau or with a mass gap.
In the well-studied sine-Gordon case in 3+1 dimensions,  for example, there is a well known oscillon of amplitude $4\pi$ that has been seen to confine the first harmonic at $3w_o$, see \cite{Hormuzdiar:1998dn,Cyncynates:2021rtf,vanDissel:2023zva}. This oscillon has a remarkably low oscillation frequency, $w_o\simeq 0.55 m$, which also renders it close to a relativistic regime, much like the soft oscillons in the gapless models. \\

Another derivative of the confining cavity effect is the close similarity with the MIT bag model of hadrons \cite{Chodos:1974je}, especially manifest in our bag model discussion in  Sec.~\ref{sec:bagmodel}. At least in the spherical case studied here, it seems fair to say that the soft oscillons arising in plateau potentials realize MIT bags, filled with certain coherent radiation. Inspired by the movable bag of the  MIT bag model, it is also natural to ask whether soft oscillons with a moving cavity or with angular momentum exist in our model.  \\

In this work we've been concerned with describing the properties of soft oscillons as isolated spherical configurations. Such large objects are expected to have an impact on cosmology, as they act as decaying dark matter. Cosmology also provides a natural context for formation of large soft oscillons. It is well known that oscillons can form via parametric resonance fragmentation instability of a homogeneous scalar~\cite{Amin_2010}. This has been explored in gapless and plateau potentials~\cite{Piani:2025dpy, Lozanov:2017hjm} like the ones discussed here. It was observed that short-lived \textit{transients} form, which correspond to objects of marginally large size $R\sim M^{-1}$. Large `soft' objects  seem less likely to form in this mechanism because the most unstable wavelengths have wavenumbers $k \sim M$. There is, however, another formation mechanism that can lead to larger objects. Namely, if our field $\phi$ is a spectator during inflation ($M \,\Lambda \ll H_{\text{inf}}\, M_P$ and $H_{\text{inf}}\gg M$), the typical field displacement away from its vacuum scales with the energy scale of inflation $H_{\text{inf}}$ as $\Delta \phi \sim \sqrt{N_e} \, H_{\text{inf}}$ where $N_e$ is the number of e-folds that inflation lasts. To evade isocurvature bounds one can assume for instance that this is a second stage of inflation, see {\em e.g.} \cite{Faria:2025nlm}. After inflation, the scalar is roughly a random Gaussian variable with field displacements $\Delta\phi\gg \Lambda$. In many Hubble patches, the field will be frozen on the plateau of the potential. How many soft oscillons end up being formed and whith what mass function is a nontrivial question, but an appreciable yield is certainly expected. 
This is an interesting direction for future work.\\

There are yet even more questions that we must leave for the future. A proper account of the impact of anisotropic perturbations, the emission of quantum radiation, the dependence on the shape of the potential, the  effects of gravity, or possible formation mechanisms. We hope to return to these topics soon.

\acknowledgments
We thank Mark Hertzberg, Fabrizio Rompineve, Juan Sebastian Valbuena Bermúdez and Andrzej Wereszczyński for useful discussions and comments. 
We acknowledges the support from the Departament de Recerca i Universitats from Generalitat de Catalunya to the Grup de Recerca `Grup de Fisica Teorica UAB/IFAE' (2021 SGR 00649). 
This publication is part of the R\&D\&i project PID2023-146686NB-C31, funded by MICIU/AEI/10.13039/501100011033/ and by ERDF/EU. IFAE is partially funded by the CERCA program of the Generalitat de Catalunya.  F.~D. acknowledges funding from the ESF under the program Ayudas predoctorales of the Ministerio de Ciencia e Innovación PRE2020-094420. This work is part of the doctoral thesis of F.D. within the framework of the Doctoral Program in Physics at the Autonomous University of Barcelona.

\bibliography{biblio}

\providecommand{\href}[2]{#2}\begingroup\raggedright\begin{thebibliography}{10}

\bibitem{Bogolyubsky:1976nx}
I.~L. Bogolyubsky and V.~G. Makhankov, \emph{{On the Pulsed Soliton Lifetime in Two Classical Relativistic Theory Models}}, {\emph{JETP Lett.} {\bf 24} (1976) 12}.

\bibitem{Bogolyubsky:1976sc}
I.~L. Bogolyubsky and V.~G. Makhankov, \emph{{Dynamics of Heavy Spherically-Symmetric Pulsons}}, {\emph{Pisma Zh. Eksp. Teor. Fiz.} {\bf 25} (1977) 120--123}.

\bibitem{Gleiser:1993pt}
M.~Gleiser, \emph{{Pseudostable bubbles}}, \href{http://dx.doi.org/10.1103/PhysRevD.49.2978}{\emph{Phys. Rev. D} {\bf 49} (1994) 2978--2981}, [\href{https://arxiv.org/abs/hep-ph/9308279}{{\tt hep-ph/9308279}}].

\bibitem{Kolb:1993hw}
E.~W. Kolb and I.~I. Tkachev, \emph{{Nonlinear axion dynamics and formation of cosmological pseudosolitons}}, \href{http://dx.doi.org/10.1103/PhysRevD.49.5040}{\emph{Phys. Rev. D} {\bf 49} (1994) 5040--5051}, [\href{https://arxiv.org/abs/astro-ph/9311037}{{\tt astro-ph/9311037}}].

\bibitem{Copeland:1995fq}
E.~J. Copeland, M.~Gleiser and H.~R. Muller, \emph{{Oscillons: Resonant configurations during bubble collapse}}, \href{http://dx.doi.org/10.1103/PhysRevD.52.1920}{\emph{Phys. Rev. D} {\bf 52} (1995) 1920--1933}, [\href{https://arxiv.org/abs/hep-ph/9503217}{{\tt hep-ph/9503217}}].

\bibitem{Kasuya:2002zs}
S.~Kasuya, M.~Kawasaki and F.~Takahashi, \emph{{I-balls}}, \href{http://dx.doi.org/10.1016/S0370-2693(03)00344-7}{\emph{Phys. Lett. B} {\bf 559} (2003) 99--106}, [\href{https://arxiv.org/abs/hep-ph/0209358}{{\tt hep-ph/0209358}}].

\bibitem{Hertzberg:2010yz}
M.~P. Hertzberg, \emph{{Quantum Radiation of Oscillons}}, \href{http://dx.doi.org/10.1103/PhysRevD.82.045022}{\emph{Phys. Rev. D} {\bf 82} (2010) 045022}, [\href{https://arxiv.org/abs/1003.3459}{{\tt 1003.3459}}].

\bibitem{Amin:2010jq}
M.~A. Amin and D.~Shirokoff, \emph{{Flat-top oscillons in an expanding universe}}, \href{http://dx.doi.org/10.1103/PhysRevD.81.085045}{\emph{Phys. Rev. D} {\bf 81} (2010) 085045}, [\href{https://arxiv.org/abs/1002.3380}{{\tt 1002.3380}}].

\bibitem{Sfakianakis:2012bq}
E.~I. Sfakianakis, \emph{{Analysis of Oscillons in the SU(2) Gauged Higgs Model}},  \href{https://arxiv.org/abs/1210.7568}{{\tt 1210.7568}}.

\bibitem{Amin:2013ika}
M.~A. Amin, \emph{{K-oscillons: Oscillons with noncanonical kinetic terms}}, \href{http://dx.doi.org/10.1103/PhysRevD.87.123505}{\emph{Phys. Rev. D} {\bf 87} (2013) 123505}, [\href{https://arxiv.org/abs/1303.1102}{{\tt 1303.1102}}].

\bibitem{Kawasaki:2015vga}
M.~Kawasaki, F.~Takahashi and N.~Takeda, \emph{{Adiabatic Invariance of Oscillons/I-balls}}, \href{http://dx.doi.org/10.1103/PhysRevD.92.105024}{\emph{Phys. Rev. D} {\bf 92} (2015) 105024}, [\href{https://arxiv.org/abs/1508.01028}{{\tt 1508.01028}}].

\bibitem{Ibe:2019vyo}
M.~Ibe, M.~Kawasaki, W.~Nakano and E.~Sonomoto, \emph{{Decay of I-ball/Oscillon in Classical Field Theory}}, \href{http://dx.doi.org/10.1007/JHEP04(2019)030}{\emph{JHEP} {\bf 04} (2019) 030}, [\href{https://arxiv.org/abs/1901.06130}{{\tt 1901.06130}}].

\bibitem{Fodor:2019ftc}
G.~Fodor, \emph{{A review on radiation of oscillons and oscillatons}}.
\newblock PhD thesis, Wigner RCP, Budapest, 2019.
\newblock \href{https://arxiv.org/abs/1911.03340}{{\tt 1911.03340}}.

\bibitem{Olle:2020qqy}
J.~Olle, O.~Pujolas and F.~Rompineve, \emph{{Recipes for oscillon longevity}}, \href{http://dx.doi.org/10.1088/1475-7516/2021/09/015}{\emph{JCAP} {\bf 09} (2021) 015}, [\href{https://arxiv.org/abs/2012.13409}{{\tt 2012.13409}}].

\bibitem{Zhang:2020bec}
H.-Y. Zhang, M.~A. Amin, E.~J. Copeland, P.~M. Saffin and K.~D. Lozanov, \emph{{Classical Decay Rates of Oscillons}}, \href{http://dx.doi.org/10.1088/1475-7516/2020/07/055}{\emph{JCAP} {\bf 07} (2020) 055}, [\href{https://arxiv.org/abs/2004.01202}{{\tt 2004.01202}}].

\bibitem{Cyncynates:2021rtf}
D.~Cyncynates and T.~Giurgica-Tiron, \emph{{Structure of the oscillon: The dynamics of attractive self-interaction}}, \href{http://dx.doi.org/10.1103/PhysRevD.103.116011}{\emph{Phys. Rev. D} {\bf 103} (2021) 116011}, [\href{https://arxiv.org/abs/2104.02069}{{\tt 2104.02069}}].

\bibitem{Levkov:2022egq}
D.~G. Levkov, V.~E. Maslov, E.~Y. Nugaev and A.~G. Panin, \emph{{An Effective Field Theory for large oscillons}}, \href{http://dx.doi.org/10.1007/JHEP12(2022)079}{\emph{JHEP} {\bf 12} (2022) 079}, [\href{https://arxiv.org/abs/2208.04334}{{\tt 2208.04334}}].

\bibitem{vanDissel:2023zva}
F.~van Dissel, O.~Pujolas and E.~I. Sfakianakis, \emph{{Oscillon spectroscopy}}, \href{http://dx.doi.org/10.1007/JHEP07(2023)194}{\emph{JHEP} {\bf 07} (2023) 194}, [\href{https://arxiv.org/abs/2303.16072}{{\tt 2303.16072}}].

\bibitem{Dorey:2023sjh}
P.~Dorey, T.~Romanczukiewicz, Y.~Shnir and A.~Wereszczynski, \emph{{Oscillons in gapless theories}}, \href{http://dx.doi.org/10.1103/PhysRevD.109.085017}{\emph{Phys. Rev. D} {\bf 109} (2024) 085017}, [\href{https://arxiv.org/abs/2312.05308}{{\tt 2312.05308}}].

\bibitem{Blaschke:2024dlt}
F.~Blaschke, T.~Roma\'nczukiewicz, K.~S\l{}awi\'nska and A.~Wereszczy\'nski, \emph{{Oscillons from Q-Balls through Renormalization}}, \href{http://dx.doi.org/10.1103/PhysRevLett.134.081601}{\emph{Phys. Rev. Lett.} {\bf 134} (2025) 081601}, [\href{https://arxiv.org/abs/2410.24109}{{\tt 2410.24109}}].

\bibitem{Zhou:2024mea}
S.-Y. Zhou, \emph{{Non-topological solitons and quasi-solitons}}, \href{http://dx.doi.org/10.1088/1361-6633/adc69e}{\emph{Rept. Prog. Phys.} {\bf 88} (2025) 046901}, [\href{https://arxiv.org/abs/2411.16604}{{\tt 2411.16604}}].

\bibitem{Amin:2010dc}
M.~A. Amin, R.~Easther and H.~Finkel, \emph{{Inflaton Fragmentation and Oscillon Formation in Three Dimensions}}, \href{http://dx.doi.org/10.1088/1475-7516/2010/12/001}{\emph{JCAP} {\bf 12} (2010) 001}, [\href{https://arxiv.org/abs/1009.2505}{{\tt 1009.2505}}].

\bibitem{Amin:2011hj}
M.~A. Amin, R.~Easther, H.~Finkel, R.~Flauger and M.~P. Hertzberg, \emph{{Oscillons After Inflation}}, \href{http://dx.doi.org/10.1103/PhysRevLett.108.241302}{\emph{Phys. Rev. Lett.} {\bf 108} (2012) 241302}, [\href{https://arxiv.org/abs/1106.3335}{{\tt 1106.3335}}].

\bibitem{Fonseca:2019ypl}
N.~Fonseca, E.~Morgante, R.~Sato and G.~Servant, \emph{{Axion fragmentation}}, \href{http://dx.doi.org/10.1007/JHEP04(2020)010}{\emph{JHEP} {\bf 04} (2020) 010}, [\href{https://arxiv.org/abs/1911.08472}{{\tt 1911.08472}}].

\bibitem{Gleiser:2009ys}
M.~Gleiser and D.~Sicilia, \emph{{A General Theory of Oscillon Dynamics}}, \href{http://dx.doi.org/10.1103/PhysRevD.80.125037}{\emph{Phys. Rev. D} {\bf 80} (2009) 125037}, [\href{https://arxiv.org/abs/0910.5922}{{\tt 0910.5922}}].

\bibitem{Farhi:2005rz}
E.~Farhi, N.~Graham, V.~Khemani, R.~Markov and R.~Rosales, \emph{{An Oscillon in the SU(2) gauged Higgs model}}, \href{http://dx.doi.org/10.1103/PhysRevD.72.101701}{\emph{Phys. Rev. D} {\bf 72} (2005) 101701}, [\href{https://arxiv.org/abs/hep-th/0505273}{{\tt hep-th/0505273}}].

\bibitem{Graham:2006vy}
N.~Graham, \emph{{An Electroweak oscillon}}, \href{http://dx.doi.org/10.1103/PhysRevLett.98.101801}{\emph{Phys. Rev. Lett.} {\bf 98} (2007) 101801}, [\href{https://arxiv.org/abs/hep-th/0610267}{{\tt hep-th/0610267}}].

\bibitem{Zhou:2013tsa}
S.-Y. Zhou, E.~J. Copeland, R.~Easther, H.~Finkel, Z.-G. Mou and P.~M. Saffin, \emph{{Gravitational Waves from Oscillon Preheating}}, \href{http://dx.doi.org/10.1007/JHEP10(2013)026}{\emph{JHEP} {\bf 10} (2013) 026}, [\href{https://arxiv.org/abs/1304.6094}{{\tt 1304.6094}}].

\bibitem{Fodor:2008es}
G.~Fodor, P.~Forgacs, Z.~Horvath and A.~Lukacs, \emph{{Small amplitude quasi-breathers and oscillons}}, \href{http://dx.doi.org/10.1103/PhysRevD.78.025003}{\emph{Phys. Rev. D} {\bf 78} (2008) 025003}, [\href{https://arxiv.org/abs/0802.3525}{{\tt 0802.3525}}].

\bibitem{Hiramatsu:2020obh}
T.~Hiramatsu, E.~I. Sfakianakis and M.~Yamaguchi, \emph{{Gravitational wave spectra from oscillon formation after inflation}}, \href{http://dx.doi.org/10.1007/JHEP03(2021)021}{\emph{JHEP} {\bf 03} (2021) 021}, [\href{https://arxiv.org/abs/2011.12201}{{\tt 2011.12201}}].

\bibitem{VanDissel:2020umg}
F.~Van~Dissel and E.~I. Sfakianakis, \emph{{Symmetric multifield oscillons}}, \href{http://dx.doi.org/10.1103/PhysRevD.106.096018}{\emph{Phys. Rev. D} {\bf 106} (2022) 096018}, [\href{https://arxiv.org/abs/2010.07789}{{\tt 2010.07789}}].

\bibitem{Zhang:2020ntm}
H.-Y. Zhang, \emph{{Gravitational effects on oscillon lifetimes}}, \href{http://dx.doi.org/10.1088/1475-7516/2021/03/102}{\emph{JCAP} {\bf 03} (2021) 102}, [\href{https://arxiv.org/abs/2011.11720}{{\tt 2011.11720}}].

\bibitem{Antusch:2019qrr}
S.~Antusch, F.~Cefal\`a and F.~Torrent\'\i{}, \emph{{Properties of Oscillons in Hilltop Potentials: energies, shapes, and lifetimes}}, \href{http://dx.doi.org/10.1088/1475-7516/2019/10/002}{\emph{JCAP} {\bf 10} (2019) 002}, [\href{https://arxiv.org/abs/1907.00611}{{\tt 1907.00611}}].

\bibitem{Antusch:2017flz}
S.~Antusch, F.~Cefala, S.~Krippendorf, F.~Muia, S.~Orani and F.~Quevedo, \emph{{Oscillons from String Moduli}}, \href{http://dx.doi.org/10.1007/JHEP01(2018)083}{\emph{JHEP} {\bf 01} (2018) 083}, [\href{https://arxiv.org/abs/1708.08922}{{\tt 1708.08922}}].

\bibitem{Amin:2018xfe}
M.~A. Amin, J.~Braden, E.~J. Copeland, J.~T. Giblin, C.~Solorio, Z.~J. Weiner et~al., \emph{{Gravitational waves from asymmetric oscillon dynamics?}}, \href{http://dx.doi.org/10.1103/PhysRevD.98.024040}{\emph{Phys. Rev. D} {\bf 98} (2018) 024040}, [\href{https://arxiv.org/abs/1803.08047}{{\tt 1803.08047}}].

\bibitem{Antusch:2017vga}
S.~Antusch, F.~Cefala and S.~Orani, \emph{{What can we learn from the stochastic gravitational wave background produced by oscillons?}}, \href{http://dx.doi.org/10.1088/1475-7516/2018/03/032}{\emph{JCAP} {\bf 03} (2018) 032}, [\href{https://arxiv.org/abs/1712.03231}{{\tt 1712.03231}}].

\bibitem{Lozanov:2022yoy}
K.~D. Lozanov and V.~Takhistov, \emph{{Enhanced Gravitational Waves from Inflaton Oscillons}}, \href{http://dx.doi.org/10.1103/PhysRevLett.130.181002}{\emph{Phys. Rev. Lett.} {\bf 130} (2023) 181002}, [\href{https://arxiv.org/abs/2204.07152}{{\tt 2204.07152}}].

\bibitem{Levkov:2023ncb}
D.~G. Levkov and V.~E. Maslov, \emph{{Analytic description of monodromy oscillons}}, \href{http://dx.doi.org/10.1103/PhysRevD.108.063514}{\emph{Phys. Rev. D} {\bf 108} (2023) 063514}, [\href{https://arxiv.org/abs/2306.06171}{{\tt 2306.06171}}].

\bibitem{Lozanov:2023knf}
K.~D. Lozanov, M.~Sasaki and V.~Takhistov, \emph{{Universal gravitational waves from interacting and clustered solitons}}, \href{http://dx.doi.org/10.1016/j.physletb.2023.138392}{\emph{Phys. Lett. B} {\bf 848} (2024) 138392}, [\href{https://arxiv.org/abs/2309.14193}{{\tt 2309.14193}}].

\bibitem{Lozanov:2023aez}
K.~D. Lozanov, M.~Sasaki and V.~Takhistov, \emph{{Universal gravitational wave signatures of cosmological solitons}}, \href{http://dx.doi.org/10.1088/1475-7516/2025/01/094}{\emph{JCAP} {\bf 01} (2025) 094}, [\href{https://arxiv.org/abs/2304.06709}{{\tt 2304.06709}}].

\bibitem{Pirvu:2023plk}
D.~P\^\i{}rvu, M.~C. Johnson and S.~Sibiryakov, \emph{{Bubble velocities and oscillon precursors in first-order phase transitions}}, \href{http://dx.doi.org/10.1007/JHEP11(2024)064}{\emph{JHEP} {\bf 11} (2024) 064}, [\href{https://arxiv.org/abs/2312.13364}{{\tt 2312.13364}}].

\bibitem{Cotner:2018vug}
E.~Cotner, A.~Kusenko and V.~Takhistov, \emph{{Primordial Black Holes from Inflaton Fragmentation into Oscillons}}, \href{http://dx.doi.org/10.1103/PhysRevD.98.083513}{\emph{Phys. Rev. D} {\bf 98} (2018) 083513}, [\href{https://arxiv.org/abs/1801.03321}{{\tt 1801.03321}}].

\bibitem{Olle:2019kbo}
J.~Oll\'e, O.~Pujol\`as and F.~Rompineve, \emph{{Oscillons and Dark Matter}}, \href{http://dx.doi.org/10.1088/1475-7516/2020/02/006}{\emph{JCAP} {\bf 02} (2020) 006}, [\href{https://arxiv.org/abs/1906.06352}{{\tt 1906.06352}}].

\bibitem{Kawasaki:2019czd}
M.~Kawasaki, W.~Nakano and E.~Sonomoto, \emph{{Oscillon of Ultra-Light Axion-like Particle}}, \href{http://dx.doi.org/10.1088/1475-7516/2020/01/047}{\emph{JCAP} {\bf 01} (2020) 047}, [\href{https://arxiv.org/abs/1909.10805}{{\tt 1909.10805}}].

\bibitem{Arvanitaki:2019rax}
A.~Arvanitaki, S.~Dimopoulos, M.~Galanis, L.~Lehner, J.~O. Thompson and K.~Van~Tilburg, \emph{{Large-misalignment mechanism for the formation of compact axion structures: Signatures from the QCD axion to fuzzy dark matter}}, \href{http://dx.doi.org/10.1103/PhysRevD.101.083014}{\emph{Phys. Rev. D} {\bf 101} (2020) 083014}, [\href{https://arxiv.org/abs/1909.11665}{{\tt 1909.11665}}].

\bibitem{Kitajima:2020rpm}
N.~Kitajima, J.~Soda and Y.~Urakawa, \emph{{Nano-Hz Gravitational-Wave Signature from Axion Dark Matter}}, \href{http://dx.doi.org/10.1103/PhysRevLett.126.121301}{\emph{Phys. Rev. Lett.} {\bf 126} (2021) 121301}, [\href{https://arxiv.org/abs/2010.10990}{{\tt 2010.10990}}].

\bibitem{Kitajima:2021inh}
N.~Kitajima, K.~Kogai and Y.~Urakawa, \emph{{New scenario of QCD axion clump formation. Part I. Linear analysis}}, \href{http://dx.doi.org/10.1088/1475-7516/2022/03/039}{\emph{JCAP} {\bf 03} (2022) 039}, [\href{https://arxiv.org/abs/2111.05785}{{\tt 2111.05785}}].

\bibitem{Dvali:2023xfz}
G.~Dvali, \emph{{Saturon Dark Matter}},  \href{https://arxiv.org/abs/2302.08353}{{\tt 2302.08353}}.

\bibitem{Hindmarsh:2007jb}
M.~Hindmarsh and P.~Salmi, \emph{{Oscillons and domain walls}}, \href{http://dx.doi.org/10.1103/PhysRevD.77.105025}{\emph{Phys. Rev. D} {\bf 77} (2008) 105025}, [\href{https://arxiv.org/abs/0712.0614}{{\tt 0712.0614}}].

\bibitem{Braden:2015vza}
J.~Braden, J.~R. Bond and L.~Mersini-Houghton, \emph{{Cosmic bubble and domain wall instabilities II: Fracturing of colliding walls}}, \href{http://dx.doi.org/10.1088/1475-7516/2015/08/048}{\emph{JCAP} {\bf 08} (2015) 048}, [\href{https://arxiv.org/abs/1505.01857}{{\tt 1505.01857}}].

\bibitem{Bond:2015zfa}
J.~R. Bond, J.~Braden and L.~Mersini-Houghton, \emph{{Cosmic bubble and domain wall instabilities III: The role of oscillons in three-dimensional bubble collisions}}, \href{http://dx.doi.org/10.1088/1475-7516/2015/09/004}{\emph{JCAP} {\bf 09} (2015) 004}, [\href{https://arxiv.org/abs/1505.02162}{{\tt 1505.02162}}].

\bibitem{Vaquero:2018tib}
A.~Vaquero, J.~Redondo and J.~Stadler, \emph{{Early seeds of axion miniclusters}}, \href{http://dx.doi.org/10.1088/1475-7516/2019/04/012}{\emph{JCAP} {\bf 04} (2019) 012}, [\href{https://arxiv.org/abs/1809.09241}{{\tt 1809.09241}}].

\bibitem{Gorghetto:2020qws}
M.~Gorghetto, E.~Hardy and G.~Villadoro, \emph{{More axions from strings}}, \href{http://dx.doi.org/10.21468/SciPostPhys.10.2.050}{\emph{SciPost Phys.} {\bf 10} (2021) 050}, [\href{https://arxiv.org/abs/2007.04990}{{\tt 2007.04990}}].

\bibitem{Blanco-Pillado:2020smt}
J.~J. Blanco-Pillado, D.~Jim\'enez-Aguilar and J.~Urrestilla, \emph{{Exciting the domain wall soliton}}, \href{http://dx.doi.org/10.1088/1475-7516/2021/01/027}{\emph{JCAP} {\bf 01} (2021) 027}, [\href{https://arxiv.org/abs/2006.13255}{{\tt 2006.13255}}].

\bibitem{Kitajima:2022jzz}
N.~Kitajima, F.~Kozai, F.~Takahashi and W.~Yin, \emph{{Power spectrum of domain-wall network, and its implications for isotropic and anisotropic cosmic birefringence}}, \href{http://dx.doi.org/10.1088/1475-7516/2022/10/043}{\emph{JCAP} {\bf 10} (2022) 043}, [\href{https://arxiv.org/abs/2205.05083}{{\tt 2205.05083}}].

\bibitem{Dvali:2011aa}
G.~Dvali and C.~Gomez, \emph{{Black Hole's Quantum N-Portrait}}, \href{http://dx.doi.org/10.1002/prop.201300001}{\emph{Fortsch. Phys.} {\bf 61} (2013) 742--767}, [\href{https://arxiv.org/abs/1112.3359}{{\tt 1112.3359}}].

\bibitem{Dvali:2012en}
G.~Dvali and C.~Gomez, \emph{{Black Holes as Critical Point of Quantum Phase Transition}}, \href{http://dx.doi.org/10.1140/epjc/s10052-014-2752-3}{\emph{Eur. Phys. J. C} {\bf 74} (2014) 2752}, [\href{https://arxiv.org/abs/1207.4059}{{\tt 1207.4059}}].

\bibitem{Dvali:2012gb}
G.~Dvali and C.~Gomez, \emph{{Landau\textendash{}Ginzburg limit of black hole's quantum portrait: Self-similarity and critical exponent}}, \href{http://dx.doi.org/10.1016/j.physletb.2012.08.019}{\emph{Phys. Lett. B} {\bf 716} (2012) 240--242}, [\href{https://arxiv.org/abs/1203.3372}{{\tt 1203.3372}}].

\bibitem{Dvali:2017ruz}
G.~Dvali and S.~Zell, \emph{{Classicality and Quantum Break-Time for Cosmic Axions}}, \href{http://dx.doi.org/10.1088/1475-7516/2018/07/064}{\emph{JCAP} {\bf 07} (2018) 064}, [\href{https://arxiv.org/abs/1710.00835}{{\tt 1710.00835}}].

\bibitem{Dvali:2017eba}
G.~Dvali, C.~Gomez and S.~Zell, \emph{{Quantum Break-Time of de Sitter}}, \href{http://dx.doi.org/10.1088/1475-7516/2017/06/028}{\emph{JCAP} {\bf 06} (2017) 028}, [\href{https://arxiv.org/abs/1701.08776}{{\tt 1701.08776}}].

\bibitem{Eby:2014fya}
J.~Eby, P.~Suranyi, C.~Vaz and L.~C.~R. Wijewardhana, \emph{{Axion Stars in the Infrared Limit}}, \href{http://dx.doi.org/10.1007/JHEP11(2016)134}{\emph{JHEP} {\bf 03} (2015) 080}, [\href{https://arxiv.org/abs/1412.3430}{{\tt 1412.3430}}].

\bibitem{Guth:2014hsa}
A.~H. Guth, M.~P. Hertzberg and C.~Prescod-Weinstein, \emph{{Do Dark Matter Axions Form a Condensate with Long-Range Correlation?}}, \href{http://dx.doi.org/10.1103/PhysRevD.92.103513}{\emph{Phys. Rev. D} {\bf 92} (2015) 103513}, [\href{https://arxiv.org/abs/1412.5930}{{\tt 1412.5930}}].

\bibitem{Lozanov:2017hjm}
K.~D. Lozanov and M.~A. Amin, \emph{{Self-resonance after inflation: oscillons, transients and radiation domination}}, \href{http://dx.doi.org/10.1103/PhysRevD.97.023533}{\emph{Phys. Rev. D} {\bf 97} (2018) 023533}, [\href{https://arxiv.org/abs/1710.06851}{{\tt 1710.06851}}].

\bibitem{Piani:2025dpy}
M.~Piani, J.~Rubio and F.~Torrenti, \emph{{Ephemeral Oscillons in Scalar-Tensor Theories: The Higgs-like case}},  \href{https://arxiv.org/abs/2501.14869}{{\tt 2501.14869}}.

\bibitem{Chodos:1974je}
A.~Chodos, R.~L. Jaffe, K.~Johnson, C.~B. Thorn and V.~F. Weisskopf, \emph{{A New Extended Model of Hadrons}}, \href{http://dx.doi.org/10.1103/PhysRevD.9.3471}{\emph{Phys. Rev. D} {\bf 9} (1974) 3471--3495}.

\bibitem{Kaplan:2015fuy}
D.~E. Kaplan and R.~Rattazzi, \emph{{Large field excursions and approximate discrete symmetries from a clockwork axion}}, \href{http://dx.doi.org/10.1103/PhysRevD.93.085007}{\emph{Phys. Rev. D} {\bf 93} (2016) 085007}, [\href{https://arxiv.org/abs/1511.01827}{{\tt 1511.01827}}].

\bibitem{inprep}
F.~van Dissel and O.~Pujolàs, \emph{{in preparation}}, .

\bibitem{Bezrukov:2007ep}
F.~L. Bezrukov and M.~Shaposhnikov, \emph{{The Standard Model Higgs boson as the inflaton}}, \href{http://dx.doi.org/10.1016/j.physletb.2007.11.072}{\emph{Phys. Lett. B} {\bf 659} (2008) 703--706}, [\href{https://arxiv.org/abs/0710.3755}{{\tt 0710.3755}}].

\bibitem{DeSimone:2008ei}
A.~De~Simone, M.~P. Hertzberg and F.~Wilczek, \emph{{Running Inflation in the Standard Model}}, \href{http://dx.doi.org/10.1016/j.physletb.2009.05.054}{\emph{Phys. Lett. B} {\bf 678} (2009) 1--8}, [\href{https://arxiv.org/abs/0812.4946}{{\tt 0812.4946}}].

\bibitem{Alexeeva:2023rfi}
N.~V. Alexeeva, I.~V. Barashenkov, A.~A. Bogolubskaya and E.~V. Zemlyanaya, \emph{{Understanding oscillons: Standing waves in a ball}}, \href{http://dx.doi.org/10.1103/PhysRevD.107.076023}{\emph{Phys. Rev. D} {\bf 107} (2023) 076023}, [\href{https://arxiv.org/abs/2304.05911}{{\tt 2304.05911}}].

\bibitem{Arodz:2007jh}
H.~Arodz, P.~Klimas and T.~Tyranowski, \emph{{Compact oscillons in the signum-Gordon model}}, \href{http://dx.doi.org/10.1103/PhysRevD.77.047701}{\emph{Phys. Rev. D} {\bf 77} (2008) 047701}, [\href{https://arxiv.org/abs/0710.2244}{{\tt 0710.2244}}].

\bibitem{Dvali:2018tqi}
G.~Dvali, M.~Michel and S.~Zell, \emph{{Finding Critical States of Enhanced Memory Capacity in Attractive Cold Bosons}}, \href{http://dx.doi.org/10.1140/epjqt/s40507-019-0071-1}{\emph{EPJ Quant. Technol.} {\bf 6} (2019) 1}, [\href{https://arxiv.org/abs/1805.10292}{{\tt 1805.10292}}].

\bibitem{Dvali:2020wqi}
G.~Dvali, \emph{{Entropy Bound and Unitarity of Scattering Amplitudes}}, \href{http://dx.doi.org/10.1007/JHEP03(2021)126}{\emph{JHEP} {\bf 03} (2021) 126}, [\href{https://arxiv.org/abs/2003.05546}{{\tt 2003.05546}}].

\bibitem{Dvali:2021tez}
G.~Dvali, O.~Kaikov and J.~S.~V. Berm\'udez, \emph{{How special are black holes? Correspondence with objects saturating unitarity bounds in generic theories}}, \href{http://dx.doi.org/10.1103/PhysRevD.105.056013}{\emph{Phys. Rev. D} {\bf 105} (2022) 056013}, [\href{https://arxiv.org/abs/2112.00551}{{\tt 2112.00551}}].

\bibitem{Dvali:2021rlf}
G.~Dvali and O.~Sakhelashvili, \emph{{Black-hole-like saturons in Gross-Neveu}}, \href{http://dx.doi.org/10.1103/PhysRevD.105.065014}{\emph{Phys. Rev. D} {\bf 105} (2022) 065014}, [\href{https://arxiv.org/abs/2111.03620}{{\tt 2111.03620}}].

\bibitem{Dvali:2023qlk}
G.~Dvali, O.~Kaikov, F.~K\"uhnel, J.~S. Valbuena-Bermudez and M.~Zantedeschi, \emph{{Vortex Effects in Merging Black Holes and Saturons}}, \href{http://dx.doi.org/10.1103/PhysRevLett.132.151402}{\emph{Phys. Rev. Lett.} {\bf 132} (2024) 151402}, [\href{https://arxiv.org/abs/2310.02288}{{\tt 2310.02288}}].

\bibitem{Adam:2019prh}
C.~Adam, K.~Oles, T.~Romanczukiewicz and A.~Wereszczynski, \emph{{Kink-antikink scattering in the $\phi^4$ model without static intersoliton forces}}, \href{http://dx.doi.org/10.1103/PhysRevD.101.105021}{\emph{Phys. Rev. D} {\bf 101} (2020) 105021}, [\href{https://arxiv.org/abs/1909.06901}{{\tt 1909.06901}}].

\bibitem{GarciaMartin-Caro:2025zkc}
A.~Garc\'\i{}a Mart\'\i{}n-Caro, J.~Queiruga and A.~Wereszczynski, \emph{{Feshbach resonances and dynamics of BPS solitons}}, \href{http://dx.doi.org/10.1103/PhysRevD.111.096002}{\emph{Phys. Rev. D} {\bf 111} (2025) 096002}, [\href{https://arxiv.org/abs/2501.02589}{{\tt 2501.02589}}].

\bibitem{Adam:2019xuc}
C.~Adam, K.~Oles, T.~Romanczukiewicz and A.~Wereszczynski, \emph{{Spectral Walls in Soliton Collisions}}, \href{http://dx.doi.org/10.1103/PhysRevLett.122.241601}{\emph{Phys. Rev. Lett.} {\bf 122} (2019) 241601}, [\href{https://arxiv.org/abs/1903.12100}{{\tt 1903.12100}}].

\bibitem{Blaschke:2024uec}
F.~Blaschke, T.~Roma\'nczukiewicz, K.~S\l{}awi\'nska and A.~Wereszczy\'nski, \emph{{Amplitude modulations and resonant decay of excited oscillons}}, \href{http://dx.doi.org/10.1103/PhysRevE.110.014203}{\emph{Phys. Rev. E} {\bf 110} (2024) 014203}, [\href{https://arxiv.org/abs/2403.00443}{{\tt 2403.00443}}].

\bibitem{Segur:1987mg}
H.~Segur and M.~D. Kruskal, \emph{{Nonexistence of Small Amplitude Breather Solutions in $\phi^4$ Theory}}, \href{http://dx.doi.org/10.1103/PhysRevLett.58.747}{\emph{Phys. Rev. Lett.} {\bf 58} (1987) 747--750}.

\bibitem{Hertzberg:2016tal}
M.~P. Hertzberg, \emph{{Quantum and Classical Behavior in Interacting Bosonic Systems}}, \href{http://dx.doi.org/10.1088/1475-7516/2016/11/037}{\emph{JCAP} {\bf 11} (2016) 037}, [\href{https://arxiv.org/abs/1609.01342}{{\tt 1609.01342}}].

\bibitem{Hormuzdiar:1998dn}
J.~N. Hormuzdiar and S.~D.~H. Hsu, \emph{{Pion breather states in QCD}}, \href{http://dx.doi.org/10.1103/PhysRevC.59.889}{\emph{Phys. Rev. C} {\bf 59} (1999) 889--893}, [\href{https://arxiv.org/abs/hep-ph/9805382}{{\tt hep-ph/9805382}}].

\bibitem{Amin_2010}
M.~A. Amin, R.~Easther and H.~Finkel, \emph{Inflaton fragmentation and oscillon formation in three dimensions}, \href{http://dx.doi.org/10.1088/1475-7516/2010/12/001}{\emph{Journal of Cosmology and Astroparticle Physics} {\bf 2010} (Dec., 2010) 001–001}.

\bibitem{Faria:2025nlm}
M.~Faria, R.~Z. Ferreira and F.~Rompineve, \emph{{Stupendously Large Primordial Black Holes from the QCD axion}},  \href{https://arxiv.org/abs/2504.07890}{{\tt 2504.07890}}.

\end{thebibliography}\endgroup

\end{document}